\documentclass[%
 reprint,
superscriptaddress,
nofootinbib,
 amsmath,amssymb,
 aps,
showkeys
]{revtex4-1}

\usepackage{graphicx}
\usepackage{dcolumn}
\usepackage{bm}
\usepackage{lipsum}
\usepackage{amsmath,amssymb}
\usepackage{booktabs}
\usepackage[utf8]{inputenc}
\usepackage[T1]{fontenc}
\usepackage[normalem]{ulem} 
\usepackage{verbatim}
\usepackage[left=2cm, right=2cm, top=2cm]{geometry}

\usepackage{epigraph}
\usepackage{siunitx}
\usepackage[english]{babel}
\usepackage{listings}
\usepackage{graphicx}
\usepackage[utf8]{inputenc}
\usepackage{listings}
\usepackage{matlab-prettifier}
\usepackage{amsfonts}
\usepackage{amsmath}
\usepackage{amssymb}
\usepackage{amsthm}
\usepackage{array}
\newcolumntype{P}[1]{>{\centering\arraybackslash}p{#1}}
\newcolumntype{M}[1]{>{\centering\arraybackslash}m{#1}}
\usepackage{braket}
\usepackage{xcolor}
\usepackage{enumitem}   
\usepackage{tabularx}
\usepackage{booktabs}

\usepackage{booktabs}
\usepackage{multirow}

\usepackage[colorlinks = true,
            linkcolor = blue,
            urlcolor  = blue,
            citecolor = blue,
            anchorcolor = blue]{hyperref}
\usepackage{comment}
\usepackage{hyperref}
\usepackage{color}

\newcommand{\totder}[2]{\frac{\mathrm{d} #1}{\mathrm{d} #2}}
\renewcommand{\Re}{\mathrm{Re}\,} 
\renewcommand{\Im}{\mathrm{Im}\,} 

\newcommand{\eps}{\epsilon}

\begin{document}

\preprint{APS/123-QED}

\title{Accuracy Requirements: Assessing the Importance of First Post-Adiabatic Terms for Small-Mass-Ratio Binaries}

\author{Ollie Burke}
\email[]{ollie.burke@l2it.in2p3.fr}
\affiliation{Laboratoire des 2 Infinis -- Toulouse (L2IT-IN2P3), Universitè de Toulouse, CNRS, UPS, F-31062 Toulouse Cedex 9, France}
\author{Gabriel Andres Piovano}
\email[]{gabriel.piovano@ucd.ie}
\affiliation{School of Mathematics and Statistics, University College Dublin, Belfield, Dublin 4, Ireland}
  \author{Niels Warburton}
 \affiliation{School of Mathematics and Statistics, University College Dublin, Belfield, Dublin 4, Ireland}   
\author{Philip Lynch}
\affiliation{Max Planck Institute for Gravitational Physics (Albert Einstein Institute) Am M\"uhlenberg 1, 14476 Potsdam, Germany}
\author{Lorenzo Speri}
\affiliation{Max Planck Institute for Gravitational Physics (Albert Einstein Institute) Am M\"uhlenberg 1, 14476 Potsdam, Germany}
\author{Chris Kavanagh}
\affiliation{School of Mathematics and Statistics, University College Dublin, Belfield, Dublin 4, Ireland}
\author{Barry Wardell}
\affiliation{School of Mathematics and Statistics, University College Dublin, Belfield, Dublin 4, Ireland}
\author{Adam Pound}
\affiliation{School of Mathematical Sciences and STAG Research Centre,
University of Southampton, Southampton, SO17 1BJ, United Kingdom}
\author{Leanne Durkan}
\affiliation{Center of Gravitational Physics, University of Texas at Austin, Austin, Texas, USA, 78712}
\author{Jeremy Miller}
\affiliation{Shamoon College of Engineering, 56 Bialik Street, Be'er Sheva 8410802, Israel.}

\begin{abstract}
We investigate the impact of post-adiabatic (1PA) terms on parameter estimation for extreme and intermediate mass-ratio inspirals using state-of-the-art waveform models.
Our analysis is the first to employ Bayesian inference to assess systematic errors for 1PA waveforms.
We find that neglecting 1PA terms 
introduces significant biases for the (small) mass ratio $\eps \gtrsim 10^{-5}$ for quasi circular orbits in Schwarzschild spacetime, which can
be mitigated with resummed 3PN expressions at 1PA order.
Moreover, we show that the secondary spin is strongly correlated with the other intrinsic parameters, and it can not be constrained for $\eps \lesssim  10^{-5}$.  
Finally, we highlight the need for addressing eccentric waveform systematics in the small-mass-ratio regime, as they yield stronger biases than the circular limit in both intrinsic and extrinsic parameters.
\end{abstract}
\pacs{Valid PACS appear here}
\maketitle

\section{Introduction}

One of the most promising sources for the Laser Interferometer Space Antenna (LISA), the first planned space-based gravitational-wave detector, are extreme-mass-ratio inspirals (EMRIs). An EMRI is the slow inspiral of a stellar-mass compact object (CO) of mass $\mu \sim 10^{0-2} M_{\odot}$ into a massive black hole (MBH) with mass $M\sim 10^{5-7}M_{\odot}$. The successful detection of an EMRI within the LISA data stream will be a groundbreaking achievement, offering outstanding tests of general relativity and unique insights into the fundamental nature of the central MBH~\cite{barack2007using,gair2013testing,Maselli:2020zgv,Maselli:2021men,Barsanti:2022ana}. Extracting the maximum information from LISA will only be possible with sufficiently high-fidelity waveform models. The source modeling of the gravitational-wave (GW) signal is exceptionally complicated, requiring sophisticated mathematical techniques arising from black hole perturbation theory and gravitational self-force (GSF) theory~\cite{barack2018self,Pound:2021qin}. Work in these areas dates back to the late 1950s, but, to this day, the mathematical details for generic orbits (eccentric and precessing inspirals) into rotating BHs, when accounting for GW emission, are still being developed.
Recent progress in this waveform-modeling program includes rapid generation of waveforms~\cite{Speri:2023jte,katz2021fast,chua2021rapid,VanDeMeent:2018cgn,McCart:2021upc, Lynch:2021ogr, Lynch:2023gpu}, the inclusion of transient resonances~\cite{Isoyama:2021jjd,Gupta:2022fbe,speri2021assessing,Lynch2022} and of the small companion's spin~\cite{Akcay:2019bvk,Piovano:2021iwv,Skoupy:2021asz,Mathews:2021rod,Drummond:2022xej,Drummond:2022efc,Skoupy:2023lih}, generation of adiabatic waveforms for generic orbits around a Kerr BH~\cite{Hughes:2021exa,Isoyama:2021jjd} and of some specific post-adiabatic corrections on generic orbits~\cite{vandeMeent:2017bcc}, and generation of waveforms including all effects at first post-adiabatic order in the case of quasicircular, nonspinning binaries~\cite{Wardell:2021fyy}. 

Not only are EMRIs challenging to model, but they prove to be one of the hardest problems in LISA data analysis~\cite{MockLISADataChallengeTaskForce:2009wir,Chua:2019wgs,chua2022nonlocal}. Due to the rich structure of the waveform and sheer volume of parameter space, grid-based searches such as those used by LIGO will be impossible~\cite{gair2004event}. Furthermore, EMRI signals are typically long-lived and may remain within the LISA sensitivity band for multiple years. The number of cycles scales with the inverse of the (small) mass ratio, $\eps = \mu/M$, implying that one could observe hundreds of thousands of orbits. Thus, we expect to constrain parameters to sub-percent level precision \cite{Babak:17aa, burke2021extreme}.

The detection and eventual characterization of an EMRI signal require waveform models that are faithful to the GW signal buried within the noise of the instrument. A detailed knowledge of the GSF is required, which involves solving the Einstein field equations order by order in powers of the small perturbative variable $\eps\ll1$. Using a two-timescale expansion~\cite{Hinderer:2008dm,Pound:2021qin}, it has been shown that the CO's orbital phase has an expansion of the form\footnote{In this expansion, we have neglected the effect of resonances that appear at fractional powers of the mass ratio $\mathcal{O}(\eps^{-1/2})$. For the class of orbits considered in this work, resonant orbits do not exist and thus will be neglected.} 
\begin{equation}\label{eq:expansion_orbital_phase}
    \phi = \underbrace{\eps^{-1} \phi_{0}}_{\text{0PA}} + \underbrace{\phi_{1}}_{\text{1PA}} +\underbrace{\eps\phi_{2}}_{\text{2PA}}  + O(\eps^2).
\end{equation}
The leading order in this expansion represents the \emph{adiabatic} (0PA) evolution, which can be understood as a slow evolution of geodesic orbital parameters due to the orbit-averaged dissipative piece of the first-order term in the self-force (arising from the $\sim \eps$ term in the metric perturbation). 
The forcing functions that drive this evolution have been computed at 0PA order for generic bound orbits~\cite{Fujita:2009us,Flanagan:2012kg,Isoyama:2021jjd,Hughes:2021exa}. 
However, it is known that we also require the \emph{first post--adiabatic} (1PA) contribution, $\phi_1$, to ensure sufficiently accurate waveform models~\cite{Hinderer:2008dm,LISA:2022kgy,LISA:2022yao}. This 1PA term involves not only the conservative and dissipative-oscillatory corrections at first order in the mass ratio, but also the orbit-averaged dissipative effects of the second-order self-force (arising from the $\sim \eps^2$ term in the metric perturbation). Finally, linear effects from the secondary spin and evolving spin and mass of the primary feed into this 1PA term~\cite{Mathews:2021rod,Miller:2020bft}. Since the \emph{second post-adiabatic} (2PA) corrections induce a contribution to the orbital phase which scales linearly with the mass ratio, they are deemed sufficiently small to be unnecessary for parameter estimation of EMRIs. Thus, knowledge of the second-order self-force corrections (and linear corrections from the secondary spin) is thought to be both necessary \emph{and sufficient} for EMRI data analysis. 
In 2021, a major breakthrough in accuracy was achieved: the second-order metric perturbation was computed, allowing the construction of the first complete 1PA waveforms for quasicircular orbits in Schwarzschild spacetimes~\cite{Wardell:2021fyy} (building on Refs.~\cite{Pound:2019lzj,Miller:2020bft,Warburton:2021kwk}). This waveform showed remarkable agreement with numerical relativity (NR) waveforms, even for $\eps \sim 10^{-1}$, far outside the EMRI regime; we refer to Ref.~\cite{Albertini:2022rfe} for a thorough accuracy analysis. 

It is crucial that EMRI waveforms are not only accurate but also fast and computationally efficient, as typical Bayesian analysis requires $\sim 10^{5-6}$ waveform evaluations to infer the parameters that govern the underlying true model.
The vast parameter space of EMRIs (given by fourteen parameters, excluding the small companion spin) further adds to the complexity and computational burden of Bayesian inference.
In order to develop approaches to tackle the huge parameter space, and because self-force models were not available, several fast-to-evaluate but approximate ``kludge'' models ~\cite{barack2004lisa, babak2007kludge, chua2017augmented} were developed. Despite not being as faithful as self-force waveforms, kludge models were essential for scoping out early LISA science.
These models have been commonly employed in parameter estimation studies in conjunction with very efficient, but unfortunately non-robust systematic tests. Such systematic studies include the Lindblom criteria~\cite{lindblom2008model}, mismatches between waveforms, and the requirement that the orbital dephasing between two different trajectories is $\lesssim$ 1 radian. 
More sophisticated data analysis studies on EMRIs employed Fisher Matrix-based estimates to quote precision measurements of parameters~\cite{vallisneri2008use, Babak:17aa, Piovano:2021iwv, burke2021extreme, Piovano:2022ojl, gupta2022modeling, gair2013testing, moore2017gravitational,maselli2022detecting} and biases on parameters arising from waveform modeling errors~\cite{cutler2007lisa, speri2021assessing}. Fisher matrices are cheap to compute, requiring a tiny number of waveform evaluations compared to Bayesian inference. However, they are prone to severe numerical instabilities~\cite{vallisneri2008use}, and they assume that the underlying probability distribution of the parameters is approximately Gaussian. The use of Bayesian inference is not a silver bullet either: un-converged posteriors can result in false conclusions, which can turn out to be quite dangerous when forecasting LISA science. 
However, if performed with care, the results can be interpreted as more robust than other systematic tests discussed in this paragraph.

Waveform models based on self-force calculations are now emerging. 
Typically, these require an expensive off\-line step involving the calculation of the self-force or other post-adiabatic effects, and a more rapid online step that computes the inspiral trajectory and the associated waveform.
The first waveform models that contained partial post-adiabatic phasing information took minutes to hours to evaluate \cite{Osburn:2015duj,Warburton:2017sxk,Drummond:2023loz}, but more recently near-identity averaging \cite{VanDeMeent:2018cgn,Lynch:2021ogr,Lynch:2023gpu} and two-timescale approaches \cite{Miller:2020bft} have reduced the calculation of the inspiral trajectory to a few seconds.
Rapidly generating waveforms that include the full mode content then requires additional optimizations \cite{katz2021fast,chua2021rapid}.

The work presented here is a first of its kind. It is the first waveform accuracy study to assess the requirement of post-adiabatic terms for LISA-focused studies based entirely on Bayesian inference. We employ the state-of-the-art \texttt{FastEMRIWaveforms} (FEW) framework~\cite{katz2021fast,chua2021rapid}, which exploits the EMRI multiscale structure to quickly generate sub-second waveforms with accuracy suitable for LISA data analysis. 
We also incorporate state-of-the-art second-order GSF results into the FEW framework. Moreover, our analysis includes the latest LISA response function~\cite{katz2022assessing}, yielding second-generation Time-Delay Interferometry (TDI) variables with realistic orbits generated by the European Space Agency~\cite{martens2021trajectory}. Finally, we also use the most recent instrumental noise model given by the LISA mission requirements team~\cite{LISAsr:18aa}. All computations presented in this work exploit Graphical Processing Units (GPUs) to accelerate the waveform and LISA response evaluation time to $\lesssim$ 1 second, making Bayesian inference possible. Thus, we employ the most accurate EMRI waveforms present in the literature to date. Armed with these tools, our goal is to understand the importance of the first post-adiabatic corrections when performing parameter estimation. Finally, we extended our analysis for a class of intermediate mass-ratio inspirals (IMRIs), in particular assuming the primary is a massive black hole (MBH) with mass $M\sim 10^{5-7}M_{\odot}$ (like an EMRI) while the smaller companion has mass $\mu \sim 10^{3} M_{\odot}$\footnote{We choose these masses for an IMRI to ensure that the corresponding GW signal is within the LISA band.
In general, both components of an intermediate mass-ratio binary may be much lighter~\cite{mandel2009can, mandel2008rates}.}.

The paper is organized as follows: in Sec.~\ref{sec:setup} we describe the approximate and exact waveform models used throughout this work; Sec.~\ref{sec:data_analysis} outlines the data analysis schemes. The results are presented in Sec~\ref{sec:results}, more specifically Sec.~\ref{sec:mismodelling} on mismodeling post-adiabatic templates and Sec.~\ref{sec:secondary_spin} on constraining the spin of the smaller companion. Finally, we discuss the impact of mismodeling adiabatic templates for eccentric orbits in Sec.~\ref{sec:eccentricity}. A comprehensive summary of the results is given in Sec.\ref{sec:summary}. Finally, we present a discussion alongside the scope for future work in Sec.~\ref{discussion} and Sec.~\ref{sec:Future_Work} respectively.

Expert readers who do not wish to dig through the details of the paper can instead skip straight to the main plots, namely Fig.~\ref{fig:Summary_all_mass_ratios_spin}, circular orbit corner plots Figs.~\ref{fig:corner_q_1e5_spin}--\ref{fig:corner_q_1e3_spin} and eccentric orbit corner plots Figs.~\ref{fig:circular_9PN_result_e0_0p0}--\ref{fig:eccentric_9PN_result_e0_0p2}. We use $G=c=1$ units throughout the paper unless otherwise stated. 

\section{Waveform models}\label{sec:setup}
We assume that the central body is a spinless BH described by the Schwarzschild metric with line element
\begin{align}
    ds^2 &= - \bigg(1- \frac{2M}{r}\bigg) dt^2 + \bigg(1- \frac{2M}{r}\bigg)^{-1} dr^2  \nonumber \\
         &\quad +  r^2 \left(d\theta^2 + \sin \theta^2 d\phi^2\right) \,.
\end{align}
The binary's smaller companion is a generic CO endowed with spin; we do not make any assumption on its internal composition.

To conform with Ref.~\cite{Wardell:2021fyy}, we find it convenient in our computations to use the symmetric mass ratio $\nu = \mu/(M + \mu)^2 =  \eps/(1 + \eps)^2 = \eps + \mathcal{O}(\eps^2)$. The symmetric mass ratio ranges from $0<\nu\leq1/4$ and has been shown to provide a better agreement for binding energy, fluxes, and waveforms when compared with numerical relativity simulations for comparable mass binaries~\cite{Pound:2019lzj,Warburton:2021kwk,Wardell:2021fyy,vandeMeent:2020xgc}. However, it does not materially affect accuracy at the mass ratios we consider here. 
We implement in FEW two 1PA waveforms specialized to quasicircular orbits. The first one is a hybrid, approximate waveform, where the 0PA fluxes were computed exactly (up to negligible numerical error) using linear BH perturbation theory while the post-adiabatic second-order self-force corrections are obtained by resumming a third-Post-Newtonian-order (3PN) expansion.
We label this model as cir0PA+1PA-3PN.
The second 1PA waveform is a state-of-the-art model that includes all relevant 1PA terms: the second-order self-force fluxes and binding energy corrections, computed in Ref.~\cite{Warburton:2021kwk} and~Ref.~\cite{Pound:2019lzj}, respectively, and the contributions from the secondary spin, given in Refs.~\cite{Piovano:2021iwv, Akcay:2019bvk}. This second waveform model is labeled as cir1PA. Finally, to allow for comparison between 0PA and 1PA waveform models, we implement an adiabatic template that we call cir0PA. This is identical to the cir1PA model but with the post-adiabatic terms removed and setting the spin of the secondary to zero.

To assess the potential impact of eccentricity on EMRI data analysis, we additionally compare two other adiabatic models: the fully relativistic 0PA waveform available in the FEW package of the BHPToolkit~\cite{BHPToolkit} and an approximate waveform that includes GW fluxes known at 9PN \cite{Munna:2020juq}. We label the former as ecc0PA and the latter as ecc0PA-9PN. 
The two models have the same waveform amplitudes and geodesic orbital frequencies and differ only in their expressions for the energy and angular momentum fluxes that drive the inspiral. 
For more details on the ecc0PA model, we refer the reader to Refs.~\cite{Chua:2020stf,Katz:2021yft}, while we provide the evolution equations for the ecc0PA-9PN model in Sec.~\ref{sec:0PAecc}.

In the following subsections, we summarize the cir1PA model and describe the  approximations introduced in cir0PA+1PA-3PN and ecc0PA-9PN.
For all waveform models, we follow the convention adopted in Ref.~\cite{Katz:2021yft} for the source reference frame, namely  the orbital angular momentum is set to be aligned to the z-axis of a Cartesian frame centered on the MBH. In this way, the motion is confined to the equatorial plane $\theta = \pi/2$, and the orbital plane does not precess.

\subsection{Quasicircular inspirals with spinning secondary to 1PA order }\label{sec:1PA}

The equations of motion for EMRIs can be conveniently expanded in the small mass-ratio $\eps$. At zeroth order, the motion of the binary components is approximated by a free-falling point-particle in a fixed background spacetime. Post-geodesic corrections arise from the self-force (see Refs.~\cite{Pound:2021qin,Harte:2014wya} for a comprehensive review) and the coupling between the curvature and spin of the smaller companion, called spin-curvature force~\cite{Tanaka:1996ht, Dixon:1964NCim, Dixon:1970I,Dixon:1970II}. The model we consider is specifically based on a two-timescale formulation of this small-$\eps$ expansion. See Refs.~\cite{Hinderer:2008dm,Miller:2020bft} and Ref.~\cite{Pound:2021qin} for a comprehensive review. As the name suggests, a two-timescale expansion assumes the existence of two disparate timescales. The short one is the orbital timescale, i.e. the orbital period of the secondary. The long timescale is the radiation-reaction timescale. During the inspiral, orbital quantities like the frequency, radius, and waveform amplitudes evolve on the slow timescale, at a rate of order $\mathcal{O}(\eps)$. By contrast, orbital phases evolve on the fast timescale, at a rate of order $\mathcal{O}(1)$.

In this section we summarize the orbital evolution in this formulation through 1PA order, dividing the discussion into conservative corrections and dissipative evolution.

\subsubsection{Conservative corrections to orbital motion}\label{sec:orbital_dynamics}

We first summarize the conservative corrections to the (slowly varying) constants of motion in the case of quasicircular, equatorial orbits in the Schwarzschild spacetime. For more details on the interplay between the self-force and spin-curvature force, see Ref.~\cite{Mathews:2021rod}, while for more details on the dynamics of spinning particles in circular equatorial orbits, see~\cite{Tanaka:1996ht,Piovano:2020zin}.

In general, if dissipation is neglected, a spinning particle in Schwarzschild spacetime admits four conserved quantities, which are the (normalized) energy $\check E = E/\mu$ and the components of the (normalized) total angular momentum $\vec{J}/(\mu M) =(J_x,J_y,J_z)/(\mu M)$~\cite{Rudiger:1981a, Rudiger:1981b}.
With our choice of reference frame, $\vec{J}/(\mu M) = (0,0,J_z)/(\mu M)$, we define accordingly $\check J = J_z/(\mu M)$. As initial conditions, we choose $\theta = \pi/2$ and set the secondary spin (anti\nobreakdash-)aligned to the $z$-axis, which implies that $\chi>0$ is aligned ($\chi <0$ is anti-aligned) to $\vec{J}$, where $\chi$ is the secondary's dimensionless spin parameter. Such conditions ensure that neither the orbital plane nor the secondary spin precess~\cite{Tanaka:1996ht, Witzany:2023bmq}. 

Hereafter, hatted quantities refer to dimensionless variables normalized by $M$ (as opposed to quantities with checks, such as $\check{E}$, which are normalized with $\mu$). For example, $\widehat \Omega_\phi = M \Omega_\phi$ and $\hat r = r/ M$. The binding energy $\check{E}$ is the only first integral we need to model the inspiral at 1PA order for our quasicircular orbital configurations.
It is convenient to parameterize the orbit in terms of its orbital frequency $\widehat{\Omega}_{\phi}$. The orbital radius of the particle, $\hat r_p$, can then be written to linear order in $\nu$, at fixed frequency $\widehat \Omega_\phi$, as
\begin{equation}
 \hat r_p = \hat r + \nu \chi \delta \hat r^{\chi}(\widehat \Omega_\phi) + \nu \delta \hat r^{\text{SF}}(\widehat \Omega_\phi).
\end{equation}
The leading-order term is the \emph{geodesic} orbital radius, $\hat r = \widehat \Omega_\phi^{-2/3}$, while the corrections $\delta \hat r^{\chi}(\widehat \Omega_\phi)$ and  $\delta \hat r^{\text{SF}}(\widehat \Omega_\phi)$ are the linear shifts due to the secondary spin and to the conservative first-order self-force, given by
\begin{equation}
\delta \hat r^{\chi}  = -\frac{1}{\sqrt{\hat r}}
\end{equation}
and by Eq.~(A8) of Ref.~\cite{Miller:2020bft}, respectively. We can similarly expand $\check E$ to first order in $\nu$ at fixed frequency:
\begin{align}
\check E &= \check E_0 + \nu \chi \check E_1^{\chi} + \nu \check E_1^{\text{SF}} \ ,
\end{align}
where $\check E_0 $ is the geodesic binding energy and $\check E_1^{\chi}$ is the shift induced by the secondary spin~\cite{Jefremov:2015gza}, given by
\begin{align}
\check E_0 = \frac{\hat r-2}{\sqrt{\hat r}\sqrt{\hat r-3}}-1 \ , \qquad  \check E_1^{\chi} = -\frac{1}{\hat r^2\sqrt{\hat r-3}} \ .
\end{align}
Finally, $\check E_1^{\text{SF}}$ is the correction to the binding energy induced by the conservative piece of the first-order self-force~\cite{LeTiec:2011dp} (see Refs.~\cite{Pound:2019lzj,Albertini:2022rfe} for discussion).

\subsubsection{Evolution equations}\label{sec:radiation_reaction_equations}
We now consider the binary evolution through 1PA order, accounting for the slow evolution of due to dissipation. 

In our models, we neglect the evolution of the primary mass and spin and second-order GSF horizon fluxes, the latter of which are currently unknown. While these effects appear formally at 1PA order, they have been shown to have a numerically small effect compared to the overall 1PA contribution to the inspiral \cite{Martel:2003jj, Hughes:2018qxz, Albertini:2022rfe}, so we can safely neglect them.

The azimuthal orbital phase $\Phi_\phi$ and the leading-order orbital radius $\hat r$ evolve according to the following coupled set of equations: 
\begin{align}
\totder{\Phi_\phi}{\hat t} &= \widehat \Omega_\phi (\hat{r} )  \ , \label{eq:phase_ev} \\
\totder{\hat r}{\hat t} &= - \nu \big [F_0(\hat{r} )+ \nu F_1(\hat{r} )\big]  \label{eq:radius_ev}  \ .
\end{align}
Here, $F_0(\hat r)$ is the leading-order, adiabatic forcing term given by $F_0 (\hat r) =  \big( \partial \check E_0 / \partial \hat r\big)^{-1} \mathcal{F}_0(\hat r)$, where $\mathcal{F}_0(\hat r)$ is the leading-order energy flux. The subleading, 1PA term  $F_1(\hat r)$ is what differs between our two circular orbit models. For the complete 1PA model, cir1PA, we have
\begin{equation}
F_1(\hat r) = F_1^{\text{SF}}(\hat r)+\chi F_1^{\chi}(\hat r).
\end{equation}
where
\begin{align}
F_1^{\text{SF}}(\hat r) &=  a(\hat r) \mathcal{F}_1(\hat{r}) + a(\hat r)^2 \bigg(\frac{\partial \check E_1^{\text{SF}}}{\partial \hat r} \bigg) \mathcal F_0(\hat r) \ , \label{eq:fix_freq_force_term}\\
F_1^{\chi}(\hat r)&= a(\hat r) \mathcal F_1^{\chi}(\hat r) + a(\hat r)^2 \bigg(\frac{\partial \check E_1^{\chi}}{\partial \hat r} \bigg) \mathcal F_0(\hat r) \label{eq:spin_fix_freq_force_term} \ ,
\end{align}
and  $a(\hat r) = \big( \partial \check E_0 / \partial \hat r \big)^{-1}$. The post-adiabatic corrections to the fluxes $\mathcal{F}_1^{\text{SF}}(\hat r)$ and $\mathcal F_1^{\chi}(\hat r)$ are, respectively, generated by nonlinear (quadratic) terms in the field equations~\cite{Warburton:2021kwk} and by the secondary spin~\cite{Piovano:2021iwv}. 

In the cir0PA+1PA-3PN model, $F_1(\hat r)$ is approximated as
\begin{equation}\label{eq:PN_Waveform}
F_1(\hat r) = F_1^{3\text{PN}}(\hat r)+\chi F_1^{\chi}(\hat r),
\end{equation}
where $F_1^{3\text{PN}}$ has the same functional form as Eq.~\eqref{eq:fix_freq_force_term}, but it is constructed using analytic post-Newtonian expressions for the fluxes at 0PA and 1PA order, including the self-force correction to the binding energy, in which we retain relative 3PN order accuracy for each. The adiabatic and post-adiabatic fluxes can be extracted by taking the leading and next-to-leading-order-in-$\nu$ terms in the 4PN fluxes recently computed in Ref.~\cite{Blanchet:2023sbv}.  The binding energy is computed using post-Newtonian self-force expansions~\cite{Fujita:2012cm}. Explicitly, these are given in App.~\ref{appendix:PN}. It is crucially important that we treat the 3PN flux terms and binding energy as polynomial approximants when used in Eq.~\eqref{eq:fix_freq_force_term}, without further expanding the geodesic quantity $a(\hat{r})$. This is because $a(\hat{r})$ is divergent at the lightring, rendering its large-radius Taylor expansion extremely inaccurate in the strong field, which would corrupt the entire model.

The evolution equations for our adiabatic model are obtained by simply setting the forcing function $F_1(\hat r)$ to zero in Eqs.~(\ref{eq:phase_ev}-\ref{eq:radius_ev}). We label this model as cir0PA. Finally, we remark that for all mass-ratios considered in this work, each inspiral terminates far (in radial coordinate distance) from where the transition to plunge begins~\cite{ori2000transition, o2003transition, kesden2011transition, apte2019exciting, burke2020transition, compere2020transition,compere2021self}, $\hat{r} - \hat{r}_{\text{isco}} \sim \eps^{2/5}$. This is important since our two-timescale evolution equations cease to accurately describe the trajectory around that point~\cite{Albertini:2022rfe}.

\subsection{Non-spinning eccentric 0PA inspirals} \label{sec:0PAecc}

In the eccentric case, the orbital radius $\hat r$ oscillates  on the orbital timescale between a maximum value $\hat r_{\text{max}}$ and a minimum value $\hat r_{\text{min}}$, meaning it is not an ideal quantity to evolve directly. Instead, we parameterize the eccentric orbit using the slowly evolving semi-latus rectum $p$ and eccentricity $e$, which are defined in terms of the slowly evolving maximum and minimum values of the orbital radius:
\begin{align}
p = \frac{2 \hat r_{\text{max}} \hat r_{\text{min}}}{(\hat r_{\text{max}} + \hat r_{\text{min}})} \ , \qquad 
e = \frac{ \hat r_{\text{max}}  - \hat r_{\text{min}}}{\hat r_{\text{max}} + \hat r_{\text{min}}} \ .
\end{align}
Note that in the circular orbit limit, $e \rightarrow 0$ and $p \rightarrow \hat r$.

Due to the orbital-timescale radial motion, the waveform picks up an additional phase $\Phi_r$ which evolves with the Boyer-Lindquist fundamental radial frequency $\widehat \Omega_r$. 
Expressions for the fundamental frequencies $\widehat\Omega_\phi(p,e)$ and $\widehat\Omega_r (p,e)$  for eccentric geodesic orbits in Schwarzschild spacetime can be found in Ref.~\cite{Cutler:1994pb}. In practice we make use of the semi-analytic expressions in terms of elliptic integrals in Refs.~\cite{Fujita:2009bp} which have been implemented in the \texttt{KerrGeodesics} package \cite{KerrGeodesicsPackage}. 

The eccentric equations of motion at adiabatic order can be written as
\begin{align} \label{eq:Eccentric_EOM1}
\begin{split}
  \totder{\Phi_\phi}{\hat t} &= \widehat \Omega_\phi (p,e )  \ , 
\end{split} \\
\begin{split}
   \totder{\Phi_r}{\hat t} &= \widehat \Omega_r (p,e )  \ ,
\end{split} \\
\begin{split}
\totder{p}{\hat t} &= - \eps \left( \frac{\partial p}{\partial \check E_0} \mathcal{F}_0^E(p,e) + \frac{\partial p}{\partial \check J_0} \mathcal{F}_0^J(p,e)\right) ,
\end{split}\\
\begin{split}
\totder{e}{\hat t} &= - \eps \left( \frac{\partial e}{\partial \check E_0} \mathcal{F}_0^E(p,e) + \frac{\partial e}{\partial \check J_0} \mathcal{F}_0^J(p,e) \right)  ,
\end{split} \label{eq:Eccentric_EOM2}
\end{align}
where $\mathcal{F}_0^E$ and $\mathcal{F}_0^J$ are the leading-order total fluxes of energy and angular momentum, respectively. We switch back to the small mass ratio $\eps$ in the above equations since both the vanilla FEW and ecc0PA-9PN models are implemented in terms of $\eps$ instead of the symmetric mass ratio $\nu$.
By employing the geodesic relations for $\check E_0(p,e)$ and $\check J_0(p,e)$, i.e., \cite{Cutler:1994pb}
\begin{align}
    \check E_0(p,e) &= \sqrt{\frac{(p-2)^2-4 e^2}{p (p -e^2 - 3)}} \ , \\
    \check J_0(p,e) &= \frac{p}{\sqrt{p -e^2 - 3}} \ ,
\end{align}
one can obtain the various partial derivatives in Eqs.~(\ref{eq:Eccentric_EOM1} - \ref{eq:Eccentric_EOM2}) by constructing a Jacobian and inverting so that
\begin{align}
 \begin{pmatrix}
  \frac{\partial p}{\partial \check E_0} &  \frac{\partial p }{\partial \check J_0} \\ 
 \frac{ \partial e }{\partial \check E_0} & \frac{\partial e}{ \partial \check J_0}
\end{pmatrix} &=
 \begin{pmatrix}
  \frac{\partial \check E_0 }{\partial p } &  \frac{\partial \check E_0 }{\partial e} \\ 
 \frac{ \partial \check J_0 }{\partial p} & \frac{\partial \check J_0}{ \partial e}
\end{pmatrix} ^{-1}  \\
&= \frac{1}{\frac{\partial \check E_0 }{\partial p } \frac{\partial \check J_0}{ \partial e} - \frac{\partial \check E_0 }{\partial e} \frac{ \partial \check J_0 }{\partial p} } \begin{pmatrix}
  \frac{\partial \hat J_0}{ \partial e} &  - \frac{\partial \check E_0 }{\partial e} \\ 
 - \frac{ \partial \hat J_0 }{\partial p} &  \frac{\partial \check E_0 }{\partial p }
\end{pmatrix} \ .
\end{align}
Explicit expressions are available in Ref.~\cite{Hughes:2021exa}.
Note that this procedure will give a singular expression for the rate of change of $e$ in the circular orbit limit, $e = 0$. However, it is well established that for adiabatic inspirals \cite{Kennefick:1995za}, quasicircular orbits remain quasicircular and so we can simply use $d e / d \hat{t} = 0$ in this special case.

The only difference between the two eccentric models is the expressions for the total energy and angular momentum fluxes. For the ecc0PA model, we numerically solve the Teukolsky equations for the energy and angular momentum fluxes to infinity and down the horizon generated by a point particle on a geodesic orbit with a given value of $p$ and $e$. We do this for multiple points in the $(p,e)$ parameter space and interpolate using splines so that the fluxes can be rapidly evaluated when solving the eccentric equations of motion.  For more details, see Sec. IV A of Ref.~\cite{katz2021fast}.  

By contrast, the ecc0PA-9PN model uses analytic PN expansions of the fluxes of energy and angular momentum to infinity and through the horizon \cite{Munna:2020juq,Munna:2023vds}. These are valid to leading order in the mass ratio and are readily available in the \texttt{PostNewtonianSelfForce} package of the BHPToolkit \cite{PostNewtonianSelfForcePackage}. The PN series contain expansions in eccentricity to at least order $e^{10}$, with some PN components having even higher order expansions up to $e^{30}$. In this work we use the terms in the series through relative 9PN order.

We highlight that the cir0PA model and ecc0PA model will not agree in the limit as $e \rightarrow 0$ because the former is expanded in $\nu$ whereas the latter is expanded in $\eps$. For our purposes, this is not a problem: we will only directly compare waveforms expanded in the same small perturbative variable, e.g., cir0PA with cir1PA ($\nu$) and ecc0PA with ecc0PA-9PN ($\eps$). We checked that recovery of a cir0PA model with an ecc0PA model (at $e = 0$) yields a bias on the secondary mass $\mu$. In turn, such bias results in a biased mass ratio $\eps_{\text{bias}}$ given by
\begin{equation}
    \eps_{\text{bias}} \approx \eps_{\text{true}} - 2\eps_{\text{true}}^{2}.
\end{equation}
Since $\nu_{\text{true}} =  \eps_{\text{true}}/(1+\eps_{\text{true}})^2 \sim \eps_{\text{true}} - 2\eps_{\text{true}}^{2} + \mathcal O(\eps^3)$, we recovered the symmetric mass-ratio, as expected. If the 1PA components were known for eccentric orbits, then it would not matter whether the equations of motion were expanded in $\eps$ or $\nu$ (for the range of mass ratios we consider).     

\subsection{Waveform models}
In a multiscale expansion of the Einstein field equations, the gravitational waveform detected by an observer at infinity in the Schwarzschild spacetime can be written as~\cite{Hughes:2021exa}
\begin{align}
 h_+- i h_\times &= \frac{\mu}{D_{\rm S}} \displaystyle \sum_{\ell,m,n} \mathcal{A}_{\ell m n}(p(t),e(t))Y_{\ell m}(\vartheta, \varphi) e^{-i \Phi_{mn}(t)}\,, \label{eq:Teukolskywave}
\end{align}
where $D_{\rm S}$ is the source's luminosity distance from the detector, $\mathcal{A}_{\ell m n} \equiv 2\hat {Z}^{\infty}_{\ell m n}/\hat \omega^2_{mn}$, and $\hat Z^{\infty}_{\ell m n} = M^2 Z^{\infty}_{\ell m n}$, the latter being the inhomogenous solution of the radial Teukoslky equation in the limit $\hat r \to \infty$. The functions $Y_{\ell m}(\vartheta, \varphi)$ are the spin-weighted spherical harmonics for spin weight $s=-2$~\cite{Goldberg:1967}, while the angles $(\vartheta,\varphi)$ identify the direction of the detector in the source reference frame. Here the GW 
frequency $\hat \omega_{mn}$ is defined as $\hat \omega_{mn} = m \widehat \Omega_\phi + n \widehat \Omega_r$.
In practice, the mode amplitudes $\mathcal{A}_{\ell m n}$ are interpolated across the $p-e$ space using a neural network~\cite{chua2021rapid,katz2021fast}. To determine the waveform mode content, the modes are cumulatively summed in order of decreasing power. This summation is truncated at a threshold value of $(1-\kappa)$ times the total power of the modes, with $\kappa$ a tunable parameter. Only the modes that pass this threshold are included in the waveform computation (for further details see \cite{chua2021rapid,katz2021fast}).

For our circular orbit runs, we have checked that the mode content of injected waveforms and model template waveforms are identical. This is to ensure consistency between waveforms and to make sure that biases are not a feature of missing important modes between waveform models in the analysis. When generating waveforms, we use the default mode selection parameter $\kappa = 10^{-5}$ in FEW, resulting in the same 13 $\ell m$-modes \emph{for all} circular orbit-based waveforms present in this work. The number of modes scales considerably with increasing eccentricity (see Fig.~4 in \cite{katz2021fast}); this will be discussed in Sec.~\ref{sec:eccentricity}.

Since our waveform represents an evolving source, we do not have a decomposition into discrete frequencies, but a multi-voice decomposition where the evolving ``voices'' are found by solving the equations of motion for the phases $\Phi_\phi(t)$ and $\Phi_r(t)$ and summing together as $\Phi_{mn}(t) = m \Phi_\phi(t) + n \Phi_r(t)$. 

Due to the LISA constellation's motion, the sources' sky positions are measured with respect to a suitable solar system barycentric (SSB) frame attached to the ecliptic~\cite{Barack:2003fp}. 
We adopt the same convention as Ref.~\cite{Katz:2021yft} and label the binary's sky position and the direction of binary's angular momentum as $(\theta_S, \phi_S)$ and $(\theta_K, \phi_K)$, respectively. The viewing angle $\varphi$ is set to $\varphi=-\pi/2$, which implies that  $\cos\vartheta$ can then be written  in terms of the constant angles $(\theta_S, \phi_S)$ and $(\theta_K, \phi_K)$ as
\begin{equation}
\cos \vartheta = -\cos \theta_S \cos \theta_K - \sin \theta_S \sin \theta_K \cos(\phi_S - \phi_K)\ .
\end{equation}
The initial phases of the waveform are then entirely determined by the angles $\Phi_{\phi_0}$ and $\Phi_{r_0}$.
Notice that the angles $(\theta_S, \phi_S)$ and $(\theta_K, \phi_K)$ also appear in the LISA response function~\cite{katz2022assessing} and ~\cite{katz2021fast}.
\subsubsection{1PA waveforms for circular orbits} 
In the circular orbit case, we no longer have a radial phase and its corresponding harmonic contribution, so the waveform model simplifies to
\begin{align}
 h_+- i h_\times &= \frac{\mu}{D_{\rm S}} \displaystyle \sum_{\ell,m} \mathcal{A}_{\ell m}(\hat{r}(t))Y_{\ell m}(\vartheta, \varphi) e^{-i m \Phi_{\phi}(t)} \ . \label{eq:Teukolskycircwave}
\end{align}

Note that at 1PA order, there will also be order-mass-ratio corrections to the waveform amplitudes $\mathcal{A}_{\ell m} = \mathcal{A}_{\ell m}^{(0)} + \eps \mathcal{A}_{\ell m}^{(1)}+ \mathcal{O}(\eps^2)$. GW interferometers are much more sensitive to fluctuations in frequency, rather than amplitude, especially for asymmetric binaries with $\eps \ll 1$. As such, including sub-leading corrections to the orbital phase is more important than to the amplitudes. In Appendix~\ref{appendix:importance_amplitudes}, we show that we require SNRs $\gtrsim 1/\epsilon$ in order for the 1PA amplitudes to noticeably affect the waveform phase. In other words, the 1PA amplitudes become important in the comparable mass-ratio case, rather than the small mass-ratio cases considered here $\epsilon < 10^{-3}$. Thus, we can safely neglect 1PA corrections to the amplitudes in our parameter estimation studies.
For this reason, we employed Eq.~\eqref{eq:Teukolskycircwave} for our all circular orbit models: cir1PA, cir0PA+1PA-3PN and cir0PA with amplitudes computed at adiabatic order. 

\section{LISA data analysis}\label{sec:data_analysis}
GW data analysis relies on three crucial ingredients: GW waveforms, a description of the time-evolving instrument response to the incoming waves, and an accurate characterization of the noise. In this section, we describe our model for the LISA response function and the noise process, ultimately leading to the Whittle-likelihood.

\subsection{The data stream of LISA}
The projection of the polarised GW signals onto the LISA arms depends on the geometry of the instrument, which changes in time due to the spacecraft's motion. The LISA response function is then a dynamical quantity in both the time and frequency domain, and clearly much more complicated than the ``static'' response of ground-based detectors. Such a feature introduces severe complexities to the accurate modeling and efficient computation of the LISA response function. By projecting the incoming GWs onto the arm-lengths of the detector, one can model the deformations across the six LISA links between the individual craft. It is then possible to construct a first set of time-shifted second-generation TDI variables, which massively suppress the laser noise (by $\sim 8$ orders of magnitude).  These variables can be linearly combined to construct a further set of TDI variables $(A, E, T)$~\cite{armstrong1999time, Tinto:2004wu}, which are uncorrelated in their noise properties. In our work, we use  the TDI variables $(A, E, T)$ ~\cite{Tinto:2004wu,vallisneri2005synthetic}. 

The data streams can then be written as 
\begin{equation}\label{eq:data_stream}
   d^{(X)}(t) = h_{\rm e}^{(X)}(t;\boldsymbol{\theta}_{\text{tr}}) + n^{(X)}(t)\, ,\quad X = \{A , E, T\},
\end{equation} where $h^{(X)}_{\rm e}$ denotes the true deterministic signal with true parameters $\boldsymbol{\theta}_{\text{tr}}$ for the $X$ TDI observable. We use \texttt{lisa-on-gpu}, a framework to compute the response of the LISA instrument to the incoming GWs available at~\cite{katz2022assessing}, which generates the three TDI variables. We employ the most realistic orbits of the LISA crafts generated by the European Space Agency (ESA)~\cite{martens2021trajectory}. These numerically computed orbits take into account gravitational attractions from relevant celestial bodies in our solar system and are built to minimize the ``breathing'' of the LISA interferometer arms. Such orbits have \emph{approximately} equal and constant arm-lengths, resulting in minor correlations between noise components~\cite{vallisneri2012non} that we neglect in favor of computational efficiency~\cite{katz2022assessing}. If these correlations are mismodeled, then the injected mismodeled noise realizations will impact the statistical errors of the recovered parameters~\cite{burke2021extreme, talbot2021inference, edy2021issues}.  This is not a problem in our analysis because, as we will explain later on, we consider zero-noise injections.
Furthermore, the signal-to-noise ratio is weakly impacted when mismodeling the noise process\footnote{This was demonstrated in the case of gaps, where correlations between noise components are severe and cannot be neglected. For more information, see Chap.~8 in~\cite{burke2021extreme}.}.

\subsection{Noise process and likelihood}
We assume that the noise in each channel is a zero-mean, weakly stationary, ergodic, Gaussian random process with noise covariance matrix (in the domain of positive frequencies)~\cite{wiener1930generalized,khintchine1934korrelationstheorie}
\begin{equation}\label{eq:Wiener-Khintchine-Theorem}
\langle \tilde{n}^{(X)}(f)[\tilde{n}^{(X)}(f')]^{\star} \rangle = \frac{1}{2}\delta(f - f')S^{(X)}_{n}(f') \, , 
\end{equation}
with tilded quantities representing Fourier transforms of the time-domain data, i.e.,
\begin{equation}
    \tilde{a}(f) = \int_{-\infty}^{\infty}\text{d}t\, a(t)e^{-2\pi i f t}\,.
\end{equation}
In Eq.~\eqref{eq:Wiener-Khintchine-Theorem}, $\langle \cdot \rangle$ denotes an ensemble averaging process, $\delta$ is the Dirac delta function and $S^{(X)}_{n}(f)$ is the (one-sided) power spectral density (PSD) of the instrumental noise process for each channel $X = \{A, E, T\}$. In our analysis, we use the latest \texttt{SciRDv1} model noise PSD~\cite{LISAScienceRequirementsDocument} and second-generation TDI variables with a \texttt{python} implementation available in~\cite{KatzTools}. Finally, we assume that the PSDs for each channel $X= \{A, E, T\}$ are all known and fixed. 

Furthermore, we neglect noise correlations between each TDI channel, i.e., $\langle \tilde{n}^{X}(f)[\tilde{n}^{Y}(f')]^{\star}\rangle = 0$ for $X \neq Y$~\cite{armstrong1999time, vallisneri2005synthetic, katz2022assessing}. Under our assumptions, we can write the Whittle-likelihood for a known PSD $S^{(X)}_{n}(f)$ as~\cite{whittle:1957, finn1992detection} 
\begin{equation}\label{eq:log-likelihood-function}
    \log p(d|\boldsymbol{\theta}) \propto -\frac{1}{2}\sum_{X = \{A, E, T\}} (d - h_{m}|d - h_{m})_{(X)}
\end{equation}
with approximate model templates $h^{(X)}_{m}$ and noise-weighted inner product $(a|b)_{(X)}$ given by~\cite{finn1992detection}
\begin{equation}\label{eq:Inner_Prod}
(a|b)_{X} = 4\mathrm{Re}\int_{0}^{\infty}\frac{\tilde{a}^{(X)}(f)(\tilde{b}^{(X)})^{\star}(f)}{ S^{(X)}_{n}(f) }\, \text{d}f. 
\end{equation}

Given a model template $h_{m}$, we define the \emph{effective} SNR with respect to the true signal $h_{e}$ as
\begin{equation}\label{eq:eff_SNR}
\rho^{\text{eff}}_{AET}(\boldsymbol{\theta}) = \left[\sum_{X = \{A, E, T\}}\frac{(h_{e}|h_{m}(\boldsymbol{\theta}))^{2}_{(X)}}{(h_{m}(\boldsymbol{\theta})|h_{m}(\boldsymbol{\theta}))_{(X)}}\right]^{1/2},
\end{equation}
with exact templates $h_{e}$ evaluated at the true parameters $\boldsymbol{\theta}_{\text{tr}}$. The maximum of Eq.~\eqref{eq:eff_SNR} is given when $\boldsymbol{\theta} = \boldsymbol{\theta}_{\text{tr}}$ and there are no mismodeling errors, i.e., $h_{m} \equiv h_{e}$ for all $\boldsymbol{\theta}$. This maximum is the \emph{optimal} matched filtering SNR, given by~\cite{woodward2014probability} 
\begin{equation}\label{eq:SNR_AET}
    \rho^{\text{opt}}_{AET} = \left[\sum_{X = \{A, E, T\}}(h_{e}|h_{e})_{(X)}\right]^{1/2},
\end{equation}
where $h_{e}$ is evaluated at the true parameters $\boldsymbol{\theta}_{\text{tr}}$. The optimal SNR over each TDI stream $X = \{A , E, T\}$ denotes the average power of the signal when compared to the root-mean-square average of the noise floor. The greater the SNR of the signal $h_{e}$ in Eq.~\eqref{eq:data_stream}, the greater the likelihood of claiming detection. Previous works ~\cite{Babak:17aa, babak2009algorithm} set, rather arbitrarily, the SNR threshold to claim detection as $\rho_{AET} \gtrsim 20$. The sources selected in this work are well above this threshold, with $\rho_{AET} \sim 70$ in the EMRI regime and $\rho_{AET} \sim 340$ in the IMRI regime. 

\subsection{Bayesian parameter estimation}\label{sec:PE_algorithm}
Parameter estimation in GW astronomy is typically performed using Bayesian inference. At the heart of Bayesian statistics lies Bayes' theorem, which, up to a normalization factor, is given by 
\begin{equation}\label{eq:Bayes_Theorem}
\log p(\boldsymbol{\theta}|d) \propto  \log p(d|\boldsymbol{\theta}) + \log p(\boldsymbol{\theta}) \, .
\end{equation}
On the right-hand side, $p(d | \boldsymbol{\theta})$ is the likelihood function, a probability distribution that describes the probability of observing the data stream given the parameters. Under our assumptions about the noise, we can use the Whittle-likelihood in Eq.~\eqref{eq:log-likelihood-function}. The density $p(\boldsymbol{\theta})$ is the prior probability distribution representing a-priori beliefs on the parameter set $\boldsymbol{\theta}$ before observing the data stream $d$. We opt for uninformative, uniform prior distributions for $\boldsymbol{\theta}$. Finally, the quantity $p(\boldsymbol{\theta}|d)$ is the sought-for posterior distribution reflecting our beliefs on the parameters after our observations. The goal is then to generate auto-correlated samples from the posterior density $p(\boldsymbol{\theta}|d)$ in order to compute summary statistics.

Our analysis employs Markov-Chain Monte-Carlo (MCMC) techniques to sample from the posterior distribution. In particular, we use the \texttt{eryn} sampler~\cite{Karnesis:2023ras, michael_katz_2023_7705496}, which is  based on the \texttt{emcee}~\cite{Foreman-Mackey:2013} code. Our proposal distribution is the default stretch proposal~\cite{goodman2010ensemble}. MCMC algorithms for EMRI inference are non-trivial, so we describe in more details our codes in App.~\ref{appendix:MCMC}. In Sec.~\ref{app:eryn_emcee} we give a brief overview of the \texttt{eryn} and \texttt{emcee} samplers, discussing their strengths and weaknesses  for the simulations presented in this work.

The initial samples are chosen so that $\boldsymbol{\theta}^{(0)} \approx \boldsymbol{\theta}_{\text{tr}}$ since our goal is to identify potential biases close to source parameters, rather than perform a search. More details on our choice of starting coordinates and prior choices (see Eq.~\eqref{eq:prior_distribution}) are given in App.~\ref{app:mcmc_start_prior_burnin}. Given a chain of samples $\boldsymbol{\theta}^{(i)} \sim p(\boldsymbol{\theta}|d)$, we define our ``best-fit parameters'' as the \emph{maximum a posteriori} (MAP) point estimate $ \boldsymbol{\theta}_{\text{bf}} = \text{argmax}_{\boldsymbol{\theta}}\{p(\boldsymbol{\theta}|d)\}$. The \emph{best-fit} parameters give the best ``match'' between the model template and the observed data stream $d$. In other words, when using non-informative uniform priors, $\boldsymbol{\theta}_{\text{bf}}$ maximizes both the likelihood function and posterior density. If the recovered parameters $\boldsymbol{\theta}_{\text{bf}} \neq \boldsymbol{\theta}_{\text{tr}}$ then the MAP estimate $\boldsymbol{\theta}$ gives a biased point estimate of the true source parameters $\boldsymbol{\theta}_{\text{tr}}$. 

We neglect the instrumental noise $n^{(X)}(t)$ present in Eq.~\eqref{eq:data_stream}, which implies that the Whittle-likelihood reduces to
\begin{equation}\label{eq:likelihood_no_noise}
    \log p(d|\boldsymbol{\theta}) \propto -\frac{1}{2}\sum_{X = \{A, E, T\}} (h_{e} - h_{m}|h_{e} - h_{m})_{(X)}.
\end{equation}
In this way, we focus on the impact of biases on the parameters arising due to waveform mismodeling, which may otherwise be obfuscated by nuisance statistical fluctuations given by noise realizations. We also neglect the confusion noise sourced by galactic dwarf binaries~\cite{Cornish_2017} since it has not been implemented yet in the latest PSD $S^{(X)}_{n}$ ~\cite{KatzTools,LISAScienceRequirementsDocument}. If the white-dwarf background were included, the SNR of the signals would be lower, resulting in wider posteriors on the parameters.
Nevertheless, the impact of the white-dwarf background on the SNR is marginal, hence it would not significantly affect our key results.

\subsection{Waveform systematics and detection}
We now outline our systematic tests used to compare waveform models. We begin first by describing our main systematic test defined in a Bayesian framework. We will then describe alternate statistical tests that are present in the literature.

For a probability density $p(\boldsymbol{\theta}|d)$, we define the 68\% credible set, $C_{p(\boldsymbol{\theta}|d)}$, of the samples as the probability $P(\boldsymbol{\theta}\in C_{p(\boldsymbol{\theta}|d)}) = 0.68$. The choice of the credible set $C_{p(\boldsymbol{\theta}|d)}$ is not unique, so we will choose the \textit{highest posterior density interval} (HPDI) in our study. In other words, the smallest possible credible set $C_{p(\boldsymbol{\theta}|d)}$ such that $P(\boldsymbol{\theta}\in C_{p(\boldsymbol{\theta}|d)}) = 0.68$. Whenever we refer to credible sets, we will be referring to the highest posterior density interval. Let $\tilde{p}(\boldsymbol{\theta}|d)$ represent an approximate model posterior, generated through inference using approximate model waveforms. We then define the systematic test

\begin{equation}\label{eq:systematic_ratio}
    \mathcal{C}[i] =     \begin{cases}
        1 & \theta_{\text{tr}}^{i} \in \hat{C}_{\tilde{p}(\theta^{i}|d)} , \\
        0 & \text{otherwise},
    \end{cases}
\end{equation}
where $\hat{C}_{\tilde{p}(\boldsymbol{\theta}^{i}|d)}$ is the estimated \emph{marginalized} HPDI for parameter $\theta^{i}$. Equation~\eqref{eq:systematic_ratio} generalizes the Cutler-Vallisneri CV criterion~\cite{cutler2007lisa}, since it accounts for non-Gaussian features in the posterior that are not captured by a Fisher Matrix-based approach. If $\mathcal{C}[i] = (0)1$ for all recovered parameters $\theta^{i}_{\text{bf}}$, then a waveform model is (un)suitable for parameter estimation. Put simply, if the true parameter is not contained within the 68$\%$ HPDI generated using approximate waveforms, then the waveform model is not suitable for statistical inference. An example is given in Fig.~\ref{fig:easy_example_biases} in App.~\ref{appendix:MCMC}. We stress that Eq.~\eqref{eq:systematic_ratio} is SNR dependent, since the size of the interval will increase (decrease) with a decrease (increase) of SNR. Since these analyses are computationally expensive, we adopted astrophysically motivated SNRs. One can deduce, as a very rough approximation, how the criterion ~\eqref{eq:systematic_ratio} changes for brighter or dimmer sources by reducing the size of the HPDI as the SNR increases.  

We now describe further quantities that are used to draw comparisons between waveform models. The overlap function between two waveforms $h^{(X)}_{1}$ and $h^{(X)}_{2}$ is defined as 
\begin{equation}
   \mathcal{O}(h^{(X)}_{1},h^{(X)}_{2}) = \frac{(h_{1}|h_{2})_{(X)}}{\sqrt{(h_{1}|h_{1})_{(X)} (h_{2}|h_{2})_{(X)}}}.
\end{equation}
Then, over all channels $X = \{A , E, T\}$, the total mismatch between two waveform models $h_{1}$ and $h_{2}$ is
\begin{align*}
    \mathcal{M}_{AET}(h_{1},h_{2}) = 1 - \sqrt{\frac{1}{3}\sum_{X = \{A, E, T\}}\mathcal{O}^{2}(h_1,h_2)}\,,
\end{align*}%
where $\mathcal{M} =  0\ (\mathcal{M} = 1)$ indicates that $h_{1}$ and $h_{2}$ are identical (orthogonal). In a similar way, we define 
\begin{align}
\mathcal{M}^{\text{(inj)}} &= \mathcal{M}_{AET}(h_e(\boldsymbol{\theta}_{\text{tr}}),h_m(\boldsymbol{\theta}_{\text{tr}}))  \label{eq:mismatch_injected} \ ,\\
\mathcal{M}^{(\text{bf})} &= \mathcal{M}_{AET}(h_e(\boldsymbol{\theta}_{\text{tr}}),h_m(\boldsymbol{\theta}_{\text{bf}}))   \ . \label{eq:mismatch_recovered_params}
\end{align}
We remark here that \eqref{eq:mismatch_recovered_params} is the usual \emph{fitting factor}, computed after stochastically identifying parameters $\boldsymbol{\theta}_{\text{bf}}$ that minimize the mismatch function.

It is usual to perform systematic studies between waveform models by analyzing the difference of orbital phase, called dephasing, between two trajectories of the CO. We define two types of dephasing in  $\Phi_{\phi}$: one between two trajectories at the injected parameters $\boldsymbol{\theta}_{\text{tr}}$ (Eq.~\eqref{eq:dephasing_injected}); and one between the two trajectories at inferred parameters $\boldsymbol{\theta}_{\text{bf}}$ (Eq.~\eqref{eq:dephasing_recovered_params}):
\begin{align}
\Delta \Phi^{\text{(inj)}} &= \text{Max}\{\Phi^{\text{exact}}_{\phi}\}_{\boldsymbol{\theta} = \boldsymbol{\theta}_{\text{tr}}} - \text{Max}\{\Phi^{\text{model}}_{\phi}\}_{\boldsymbol{\theta} = \boldsymbol{\theta}_{\text{tr}}} \label{eq:dephasing_injected} \, ,\\
\Delta \Phi^{\text{(bf)}} &= \text{Max}\{\Phi^{\text{exact}}_{\phi}\}_{\boldsymbol{\theta} = \boldsymbol{\theta}_{\text{tr}}} - \text{Max}\{\Phi^{\text{model}}_{\phi}\}_{\boldsymbol{\theta} = \boldsymbol{\theta}_{\text{bf}}} \ .\label{eq:dephasing_recovered_params}
\end{align}
Eq.~\eqref{eq:dephasing_injected} is the usual quantity used to estimate the accuracy requirements of EMRI waveforms. We comment that the two equations above are evaluated over the same time of observation.

Finally, we refer to the accumulated SNR normalised by the optimal SNR as the quantity
\begin{align}
\rho^{\text{(inj)}}/\rho^{\text{(opt)}} &= \rho_{AET}^{\text{eff}}(\boldsymbol{\theta}_{\text{tr}}) / \rho_{AET}^{\text{opt}}\, , \label{eq:accum_SNR_true}\\
\rho^{\text{(bf)}}/\rho^{\text{(opt)}} &= \rho_{AET}^{\text{eff}}(\boldsymbol{\theta}_{\text{bf}}) / \rho_{AET}^{\text{opt}}\, \label{eq:accum_SNR_bf},
\end{align}
for $\rho^{\text{eff}}_{AET}(\boldsymbol{\theta})$ and $\rho^{\text{opt}}$ defined in Eq.~\eqref{eq:eff_SNR} and Eq.~\eqref{eq:SNR_AET}, respectively. The equations above represent the fraction of SNR (normalised between 0 and 1) accumulated throughout the inspiral. The quantity in Eq.~\eqref{eq:accum_SNR_bf} is useful to determine whether a reference true waveform model is detectable by an approximate model. The quantities in Eqs.~(\ref{eq:mismatch_injected} - \ref{eq:accum_SNR_bf}) will be useful to compare against our Bayesian inference results.   

\section{Results}\label{sec:results}

We now present our results on the Bayesian parameter inference with complete circular 1PA and mismodeled eccentric 0PA waveforms. In Sec.~\ref{sec:mismodelling}, we will investigate the impact of mismodeling EMRI templates by neglecting some (or all) of the post-adiabatic corrections. In Sec.~\ref{sec:secondary_spin}, we will then focus our attention on constraining the secondary spin parameter. Finally, in Sec.~\ref{sec:eccentricity} we will describe the impact of mismodeling 0PA eccentric templates.

\begin{figure}
    \includegraphics[width = 0.48\textwidth]{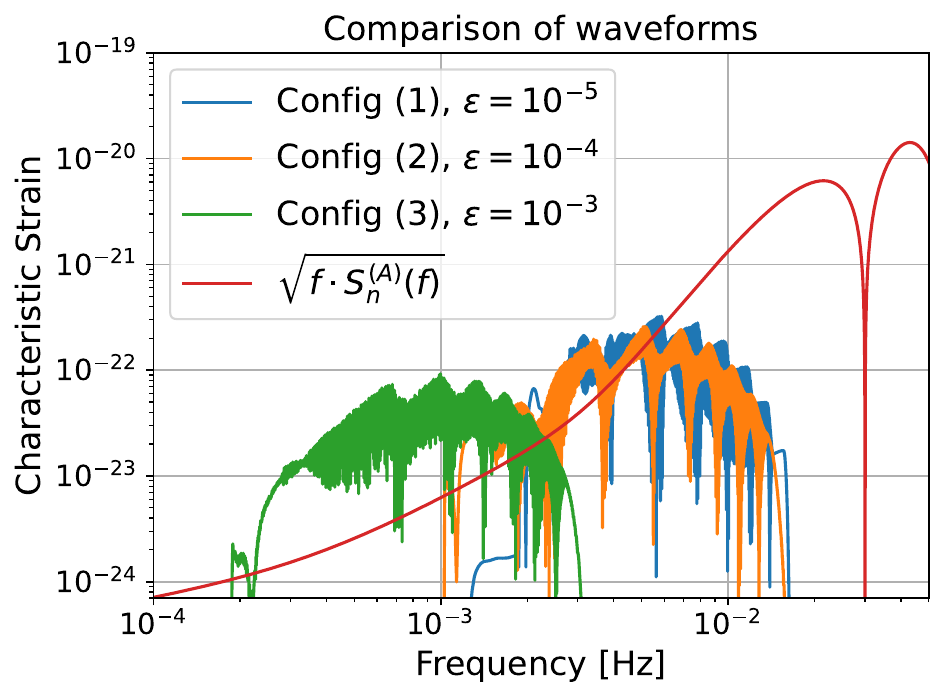}
    \caption{New: The blue, orange and green curves are characteristic strains $h_{c} = |f\tilde{h}(f)|$ of each small mass-ratio binary with parameters given in Table \ref{ref:EMRI_parameters}. The red curve is the characteristic strain of the noise in the A channel using second generation TDI variables. The two EMRIs with $\epsilon = 10^{-5}$ and $\epsilon = 10^{-4}$ respectively have $\sim 70,000$, and $\sim 10,000$ cycles. The IMRI with $\epsilon = 10^{-3}$ has $\sim 4,000$ cycles. As one can see here, the small mass-ratio binaries presented in this paper are inherently different.  }
    \label{fig:summary_waveforms}
\end{figure}

The inferred parameters are the following: the red-shifted masses $M$ and $\mu$,  the initial radial coordinate and initial phase $r_{0}/M$ and $\Phi_{\phi_{0}}$, respectively (both defined at time $t=0$), the luminosity distance to the source $D_{\rm S}$, the source sky position $(\theta_{\rm S}, \phi_{\rm S})$ and the orientation of the orbital angular momentum $(\theta_{\rm K}, \phi_{\rm K})$. Waveforms are generated in the source frame and transformed to the SSB frame via the response function. For the sake of clarity, if a model includes the spin of the secondary as a parameter it will be denoted ``w/ spin'' and otherwise ``w/o spin''. In Sec.~\ref{sec:mismodelling}, we will only infer the secondary spin $\chi$ using the cir1PA w/ spin model, whereas in Sec.~\ref{sec:secondary_spin} we will infer the secondary spin with all approximate models. For sections~\ref{sec:mismodelling} and \ref{sec:secondary_spin}, configurations for each of the mass ratios and the relative SNR can be found in Table~\ref{ref:EMRI_parameters}. Given the parameters of the small mass-ratio binaries, we plot their characteristic strains $| f \tilde{h}(f) | $  in Fig.~\ref{fig:summary_waveforms} alongside the characteristic strain of the noise $(f S_{n}^{(A)}(f))^{1/2}$ corresponding to second generation TDI variables ~\cite{moore2014gravitational, babak2021lisa}. As one can see, each of the small mass-ratio binaries are different and we do not wish to directly compare the three cases. We chose the three mass-ratio cases presented in Fig.~\ref{fig:summary_waveforms} since they represent prototypical small-mass ratio binaries.
We remark here that since we are using second generation TDI variables, the PSD $\propto$ constant as $f \rightarrow 0$. Note that this sensitivity curve is different from the usual curve displayed when using the low frequency and sky/polarisation/inclination-averaged approximation to LISA response (see Fig.(1) in both ~\cite{robson2019construction,LISAScienceRequirementsDocument}). There is no red noise component for second generation TDI since as this has been carefully subtracted via the TDI configurations~\cite{babak2021lisa}, see figure (4). If, on the other hand, we had plotted first generation TDI variables we would see a plot that would look qualitatively similar to figure (1) in ~\cite{robson2019construction} with the usual ``bucket" representing an interval in frequency where LISA is most sensitive to GWs. We chose, instead, to use second generation TDI variables since these variables will most likely be used in practice for LISA data analysis. For further information, see~\cite{bayle2019simulation}.  

\begin{table}[tbp]
\begin{tabular}{c|c|c|c|c|c|c}
\hline \hline
Config.              & $\eps$                        & $M\,[M_{\odot}]$                 & $r_{0}/M$                   & $D_{\rm S}\,[\text{Gpc}]$ & $T_{\text{obs}}\,[\text{yrs}]$ & $\rho_{AET}$         \\ \hline
\multirow{2}{*}{(1)} & \multirow{2}{*}{$10^{-5}$} & \multirow{2}{*}{$10^{6}$}        & \multirow{2}{*}{$10.6025$}  & \multirow{2}{*}{1.0}      & \multirow{2}{*}{2.0}    & \multirow{2}{*}{82}  \\
                     &                            &                                  &                             &                           &                         &                      \\
\multirow{2}{*}{(2)} & \multirow{2}{*}{$10^{-4}$} & \multirow{2}{*}{$10^{6}$}        & \multirow{2}{*}{$15.7905$}  & \multirow{2}{*}{2.0}      & \multirow{2}{*}{1.5}    & \multirow{2}{*}{70}  \\
                     &                            &                                  &                             &                           &                         &                      \\
\multirow{2}{*}{(3)} & \multirow{2}{*}{$10^{-3}$} & \multirow{2}{*}{$5\cdot 10^{6}$} & \multirow{2}{*}{$16.8123$} & \multirow{2}{*}{1.0}      & \multirow{2}{*}{1.0}    & \multirow{2}{*}{340} \\
                     &                            &                                  &                             &                           &                         &    \\ \hline\hline                 
\end{tabular}
\caption{(Update on SNRs). Here we tabulate injection EMRI/IMRI parameters for all waveforms in sections~\ref{sec:mismodelling} and \ref{sec:secondary_spin}. The IMRI configuration is given by the last row in the table. For each case, we use identical extrinsic parameters: $\theta_{\rm S} = \pi/3,\ \phi_{\rm S} = \pi/4,\ \theta_{\rm K} = 2, \phi_{\rm K} = 5, \Phi_{\phi_{0}} = 1.5$. The secondary spin is fixed to the fiducial value $\chi = 0.5$. }
 \label{ref:EMRI_parameters}
\end{table}
Details of our sampling algorithm can be found in App.~\ref{appendix:MCMC}, including starting points and prior bounds. In this analysis, we will use the \texttt{emcee} sampler due to the simplicity of the likelihood structure in the case of circular orbits.

\subsection{Systematic biases --- missing 1PA terms  for quasicircular orbits}\label{sec:mismodelling}

We first consider the impact of using mismodeled waveform templates on LISA science when attempting to extract full 1PA waveforms within the data stream. 

\begin{figure*}
    \centering
    \includegraphics[width = \textwidth]{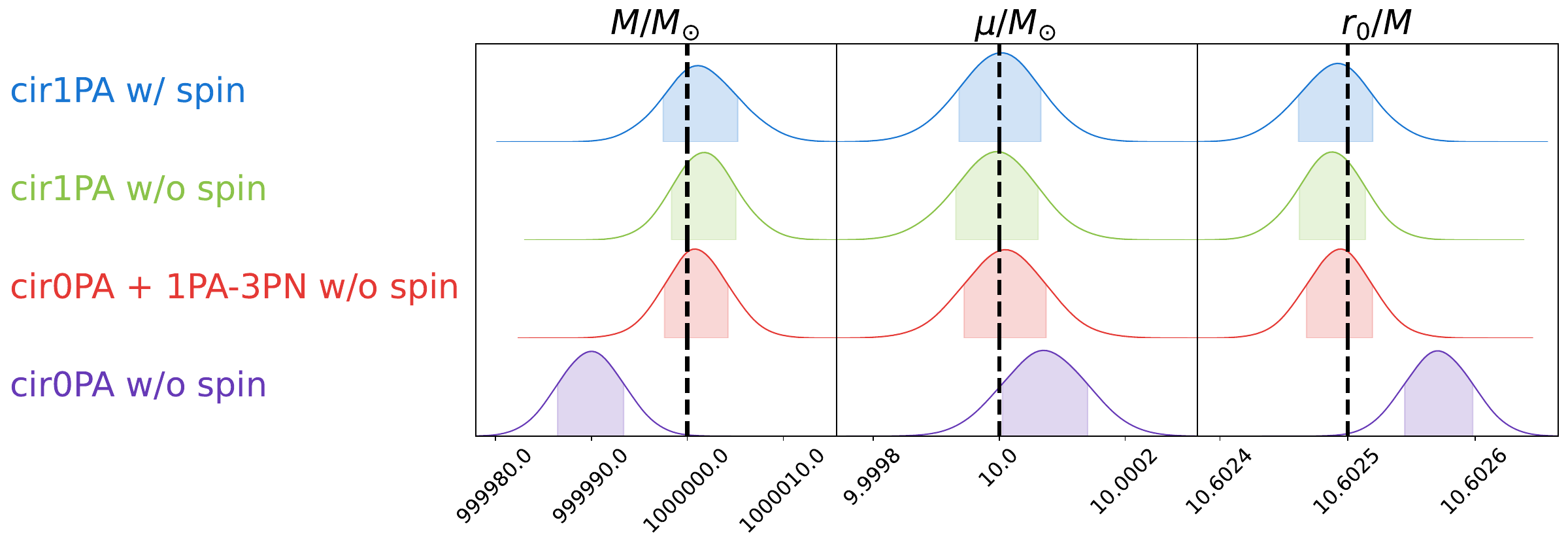}
    \includegraphics[width = \textwidth]{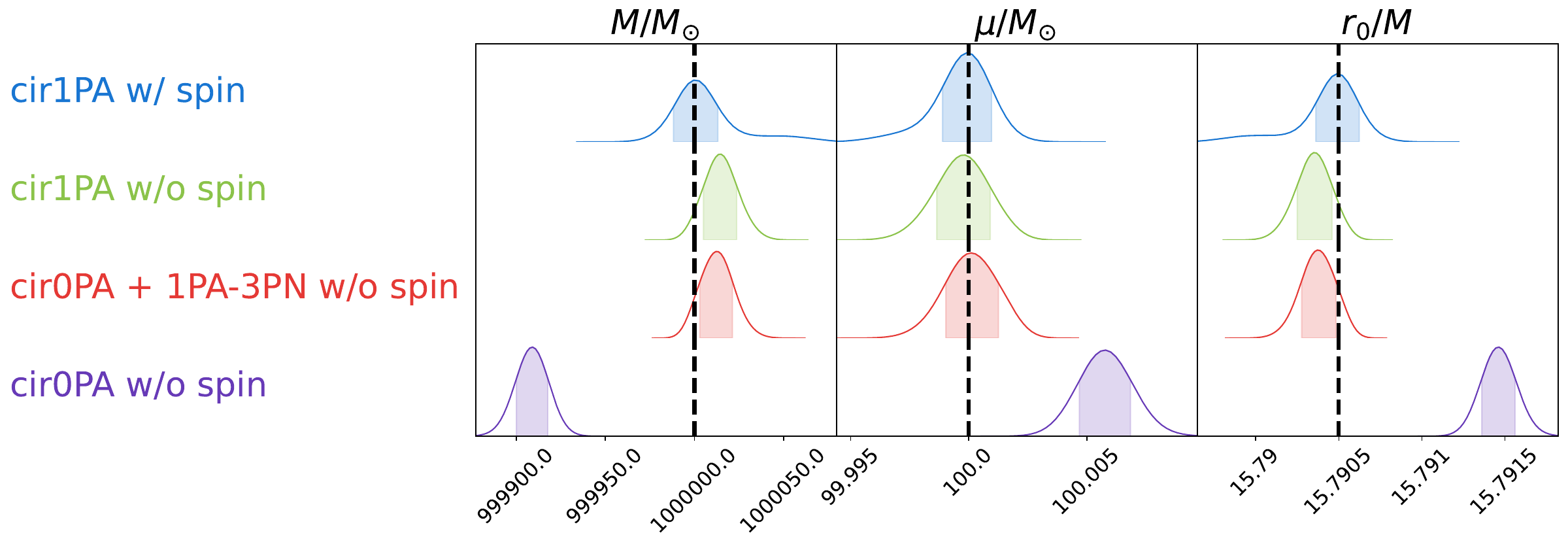}
    \includegraphics[width = \textwidth]{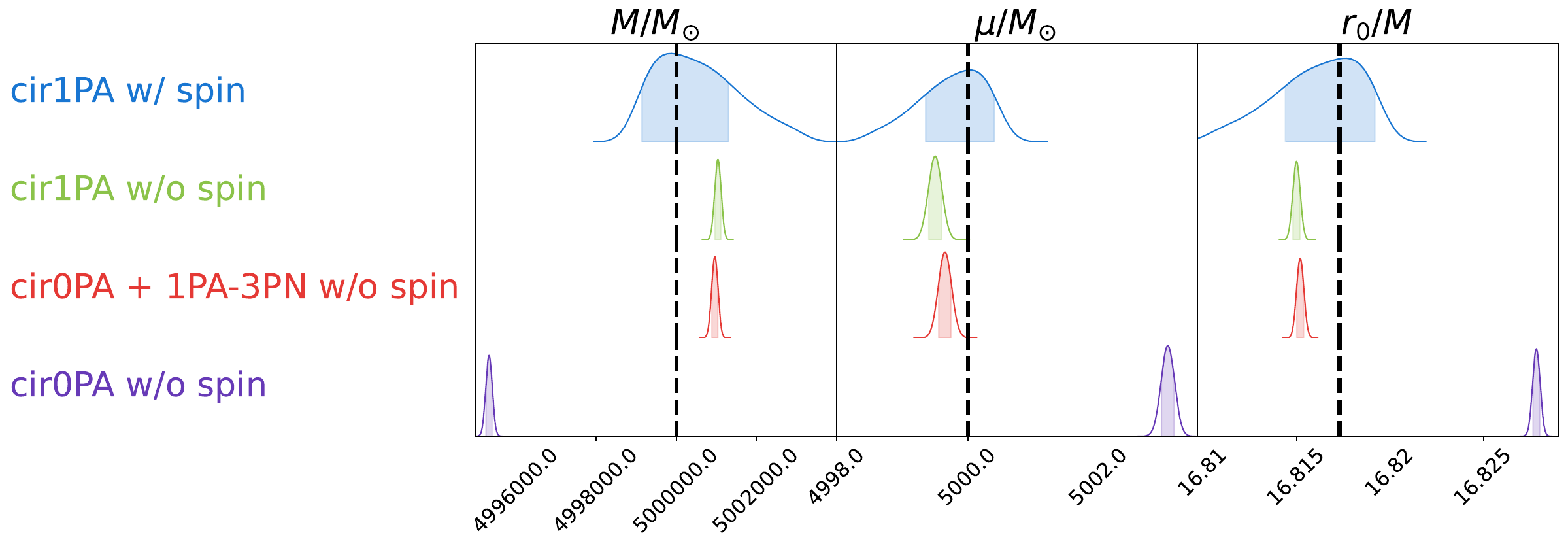}
    
    \caption{Marginalized posteriors with shaded 68\% credible intervals generated by injecting a true reference model cir1PA and recovering using different models with different mass ratios. The top panel is with $\eps = 10^{-5}$, middle panel $\eps = 10^{-4}$ and bottom panel $\eps = 10^{-3}$. (\textbf{Blue:}) recovery with the true reference model cir1PA and sampling over the secondary spin parameter. (\textbf{Green:})  recovery using the cir1PA without spin. (\textbf{Red:}) recovery with cir0PA+1PA-3PN without spin. (\textbf{Purple:}) recovery with cir0PA (purely adiabatic) waveform. The black vertical dashed line indicates the true parameters. These simulations used parameters given by Table \ref{ref:EMRI_parameters} for Configs. (1), (2) and (3) are given by the top row, middle row, and bottom row respectively. Orbital dephasings, mismatches, accumulated SNRs, and maximum log-likelihood values can be found in Table~\ref{Tab:Summary_Stats_MCMC}. Last row for case $\epsilon = 10^{-3}$ is new.}
    \label{fig:Summary_all_mass_ratios_spin}
\end{figure*}

For each mass ratio $\eps = \{10^{-5}, 10^{-4}, 10^{-3}\}$ with respective parameters given by Table ~\ref{ref:EMRI_parameters}, we perform four parameter estimation simulations. We inject a true reference signal cir1PA with (w) spin and recover with the following models each without (w/o) secondary spin
\begin{equation}\label{eq:various_cases_no_spin}
h_{m} =
    \begin{cases}
        \text{cir1PA w/o spin}, \\
        \text{cir0PA+1PA-3PN w/o spin}, \\
        \text{cir0PA w/o spin}.
    \end{cases}
\end{equation}
The results for each studied mass ratio $\eps=\{10^{-5}, 10^{-4}, 10^{-3}\}$ are displayed (from top to bottom respectively) in Fig.~\ref{fig:Summary_all_mass_ratios_spin}. In each of the three panels, the top rows (blue posteriors) are exact marginalized posteriors, generated when injecting and recovering with the exact model cir1PA w/ spin where the spin on the secondary is sampled over. We do not present the posteriors on the extrinsic parameters as they display near-to-zero bias with respect to the true parameters. The non-Gaussian features and shifts to the true posterior are a feature of the secondary spin, which will be discussed in Sec.~\ref{sec:secondary_spin}. 

We present posterior medians and 68\% HPDIs in Tables~\ref{tab:table_eps_1e-5}, \ref{tab:table_eps_1e-4} and \ref{tab:table_eps_1e-3} for the intrinsic parameters. We report summary statistics of the secondary spin when sampling over that parameter in the cir1PA w/ spin model. The parameter posterior distribution of the secondary spin will be discussed in section \ref{sec:secondary_spin}.  We begin by discussing the case with the smallest mass ratio, $\eps = 10^{-5}$.

Referring to the top panel of Fig.~\ref{fig:Summary_all_mass_ratios_spin}, we see that both the approximate models cir1PA w/o spin (green) and cir0PA + 1PA-3PN w/o spin (red) are suitable for parameter estimation of full 1PA waveforms. Each posterior shows statistically insignificant biases at $\rho_{AET} \sim 82$ with Eq.~\eqref{eq:systematic_ratio} resulting in $\mathcal{C} = 1$ for all parameters. he true parameters $\boldsymbol{\theta}_{0}$ lie within the 68\% HPDI given by the first row and final column in Table~\ref{tab:table_eps_1e-5}.  Remarkably, the parameters of the exact cir1PA w/ spin model can be correctly inferred, with statistically insignificant biases, by the cir0PA+1PA-3PN model. We remind the reader that the latter contains 0PA information with a 1PA term approximated by a resummed 3PN expansion. The last row of the top panel in Fig.~\ref{fig:Summary_all_mass_ratios_spin} shows a cir1PA w/ spin model recovered with our least faithful model, cir0PA w/o spin. Making reference to Table~\ref{tab:table_eps_1e-5}, we see that none of the true intrinsic parameters are contained within the 68\% HPDIs. According to Eq.~\eqref{eq:systematic_ratio}, the intrinsic parameters show statistically significant biases with $\mathcal{C} = 0$. We do remark, that the recovered parameters are very similar to the true ones as indicated by the posterior medians in Table \ref{tab:table_eps_1e-5}.  For example, given the true primary mass, $M = 10^{6}M_{\odot}$, our recovered parameter is only $\sim 10M_{\odot}$ away in magnitude. This quantitatively confirms that adiabatic models would be fine for detection purposes, but not for statistical inference. 

\begin{table*}[tbp]
\begin{tabular}{l|c|c|c|c|c|c|c|c|c}
\multicolumn{1}{c|}{$\eps$}                        & \textbf{Model Waveform} & $\Delta \Phi^{(\text{inj})}$ & $\Delta \Phi^{(\text{bf})}$ & $\mathcal{M}^{(\text{inj})}$ & $\mathcal{M}^{(\text{bf})}$ & $\rho^{(\text{inj})}/\rho^{(\text{opt})}$ & $\rho^{(\text{bf})}/\rho^{(\text{opt})}$ & $\log\mathcal{L}^{(\text{inj})}$ & $\log\mathcal{L}^{(\text{bf})}$ \\ \hline\hline
\multirow{6}{*}{$10^{-5}$}                      & \multicolumn{1}{l|}{}   & \multicolumn{1}{l|}{}        & \multicolumn{1}{l|}{}       & \multicolumn{1}{l|}{}        & \multicolumn{1}{l|}{}       & \multicolumn{1}{l|}{}                     & \multicolumn{1}{l|}{}                    & \multicolumn{1}{l|}{}            & \multicolumn{1}{l}{}            \\
                                                & Cir1PA w/o spin         & $0.779$                      & $0.0165$                    & $0.143$                      & $4.497\times 10^{-5}$       & $83.4\%$                                  & $99.9\%$                                 & $-846$                           & -0.250                          \\
                                                & \multicolumn{1}{l|}{}   & \multicolumn{1}{l|}{}        & \multicolumn{1}{l|}{}       & \multicolumn{1}{l|}{}        & \multicolumn{1}{l|}{}       & \multicolumn{1}{l|}{}                     & \multicolumn{1}{l|}{}                    & \multicolumn{1}{l|}{}            & \multicolumn{1}{l}{}            \\
                                                & Cir0PA 1PA-3PN w/o spin & $0.786$                      & $0.00179$                   & $0.163$                      & $4.293 \times 10^{-6}$      & $81.5\%$                                  & $99.8\%$                                 & $-943$                           & -0.0324                         \\
                                                & \multicolumn{1}{l|}{}   & \multicolumn{1}{l|}{}        & \multicolumn{1}{l|}{}       & \multicolumn{1}{l|}{}        & \multicolumn{1}{l|}{}       & \multicolumn{1}{l|}{}                     & \multicolumn{1}{l|}{}                    & \multicolumn{1}{l|}{}            & \multicolumn{1}{l}{}            \\
                                                & Cir0PA w/o spin         & $3.002$                      & $0.00532$                   & $0.889$                      & $2.412 \times 10^{-6}$      & $6.4\%$                                   & $99.8\%$                                 & $-4800$                          & -0.0234                         \\ \hline \hline
\multicolumn{1}{c|}{\multirow{6}{*}{$10^{-4}$}} &                         &                              &                             &                              &                             &                                           &                                          &                                  &                                 \\
\multicolumn{1}{c|}{}                           & Cir1PA w/o spin         & 3.994                        & 0.00702                     & 0.511                        & $8.601 \times 10^{-6}$      & $30.3\%$                                  & $99.9\%$                                 & -5019                            & -0.336                          \\
\multicolumn{1}{c|}{}                           &                         &                              &                             &                              &                             &                                           &                                          &                                  &                                 \\
\multicolumn{1}{c|}{}                           & Cir0PA 1PA-3PN w/o spin & 4.310                        & 0.0179                      & 0.486                        & $1.26 \times 10^{-4}$       & $34.2\%$                                  & $99.9\%$                                 & -4799                            & -0.441                          \\
\multicolumn{1}{c|}{}                           &                         &                              &                             &                              &                             &                                           &                                          &                                  &                                 \\
\multicolumn{1}{c|}{}                           & Cir0PA w/o spin         & 13.093                       & 0.0354                      & 0.653                        & $2.573 \times 10^{-5}$      & $19.0\%$                                  & $99.9\% $                                & -5506                            & -0.122                          \\ \hline \hline
\multirow{6}{*}{$10^{-3}$}&                         & \multicolumn{1}{l|}{}        & \multicolumn{1}{l|}{}       & \multicolumn{1}{l|}{}        & \multicolumn{1}{l|}{}       & \multicolumn{1}{l|}{}                     & \multicolumn{1}{l|}{}                    & \multicolumn{1}{l|}{}            & \multicolumn{1}{l}{}            \\
                                                & Cir1PA w/o spin         & 4.518                        & 0.00559                     & 0.922                        & $3.643 \times 10^{-6}$      & $3.3 \%$                                  & $99.9\%$                                 & -112938                          & -0.226                          \\
                                                &                         &                              &                             &                              &                             &                                           &                                          &                                  &                                 \\
                                                & Cir0PA 1PA-3PN w/o spin & 4.882                        & 0.0218                      & 0.949                        & $3.443 \times 10^{-5}$      & $3.4 \%$                                  & $99.9\%$                                 & -112827                          & -2.132                          \\
                                                &                         &                              &                             &                              &                             &                                           &                                          &                                  &                                 \\
                                                & Cir0PA w/o spin         & 14.958                       & 0.0421                       & 0.938                        & $2.7368 \times 10^{-5}$      & $4.9\%$                                   & $99.9\%$                                 & -122173                          & -1.757                       \\
\multicolumn{1}{c|}{}                           &                         &                              &                             &                              &                             &                                           &                                          &                                  &        
\end{tabular}
\caption{Here we present a summary of computed statistics for various mass ratios $\eps = \{10^{-5}, 10^{-4}, 10^{-3}\}$ (first column) when  comparing an injected cir1PA waveform and approximate model templates (second column).  We compute the orbital dephasings~(Eqs.~\ref{eq:dephasing_injected}-\ref{eq:dephasing_recovered_params})  (third and fourth columns); mismatch (Eqs.~\ref{eq:mismatch_injected}-\ref{eq:mismatch_recovered_params})  (fifth and sixth columns);  accumulated SNRs (Eqs.~\ref{eq:accum_SNR_true}-\ref{eq:accum_SNR_bf}) (seventh and eighth columns); and, finally the log-likelihood function, Eq.~\eqref{eq:likelihood_no_noise}, at the injected/recovered parameters. The top, middle, and bottom panels of this table correspond to the top, middle, and bottom panels of Fig.~\ref{fig:Summary_all_mass_ratios_spin}, respectively. Last row for case $\epsilon = 10^{-3}$ is new.}
\label{Tab:Summary_Stats_MCMC}
\end{table*}


We now discuss our simulations for $\eps \sim 10^{-4}$, given by the middle panel in Fig.~\ref{fig:Summary_all_mass_ratios_spin} and Table \ref{tab:table_eps_1e-4}. The marginalized Gaussians in the top row of the middle panel all exhibit heavy tails. This is due to correlations between the intrinsic parameters and secondary spin, which will be discussed in Sec.~\ref{sec:secondary_spin}. Since the approximate models do not contain spin, there are no such correlations and the resultant posteriors resemble their familiar Gaussian shapes for high SNRs. With all approximate model templates, we see significant biases ($\mathcal{C} = 0$) across the intrinsic parameters when neglecting post-adiabatic terms at $\rho_{AET} \sim 70$. The most dramatic bias arises when using the cir0PA w/o spin model to recover the exact cir1PA w/ spin reference model.  We conclude here that our approximate models summarized in Eq.~\eqref{eq:various_cases_no_spin}, without secondary spin, are unsuitable for parameter estimation.

Finally, we considered an IMRI with $\eps = 10^{-3}$, given by the bottom panel in Fig.~\ref{fig:Summary_all_mass_ratios_spin} and the final row of Table~\ref{tab:table_eps_1e-3}. There are clear differences between the top row (inject/recover with exact model) and the bottom three rows where we recover the reference model with approximations \eqref{eq:various_cases_no_spin} absent of spin. In the top row of the bottom panel, the marginalized posteriors are not Gaussian and are non-trivially skewed. This is further reflected by the 68\% HPDIs given in the first row and final column of Table~\ref{tab:table_eps_1e-3}.  This is a result of strong correlations between the intrinsic parameters and the secondary spin. The approximate waveform models  representing model templates cir1PA w/o spin and cir0PA + 1PA-3PN w/o spin, and cir0PA w/o spin each yield biased Gaussian distributions with roughly an order of magnitude more precision than the cir1PA w/ spin case. This is a consequence of neglecting the important correlations with the secondary spin. Interestingly, the recovered parameters are contained within the approximate posterior 68\% HPDI of the marginalized true posterior distribution, implying that they are consistent with the true parameter distribution\footnote{It is for this reason we generalise the Cutler-Valisneri formalism~\cite{cutler2007lisa} in order to account for non-Gaussian features. According to the usual CV formalism, one may incorrectly infer that the cir0PA model is a suitable waveform model for parameter estimation since the recovered parameter is ``consistent" with the true posterior distribution given by the cir1PA w/ spin model. }. This is alarming: in this situation we have not only incorrectly estimated the true parameters, but our confidence that they are the ``correct'' ones is largely inflated due to the tightness of the posterior distributions.  other words, the approximate models recover parameters with a detrimental loss of accuracy but with a dangerous increase in precision. In light of Eq.~\eqref{eq:systematic_ratio}, we conclude that all models are unsuitable for parameter inference of the cir1PA w/ spin model at $\rho_{AET} \sim 340$. For this small mass-ratio binary case, we require full knowledge of the second-order self force to perform accurate parameter estimation. 

In Table \ref{Tab:Summary_Stats_MCMC}, we give a summary of details regarding the individual MCMC simulations for each small-mass-ratio binary configuration presented in Table \ref{ref:EMRI_parameters}. The details of the specific computations can be found in the caption of the table. One of the main features of this table is the small mismatch, and accumulated SNR normalized by the optimal SNR. Finally, we remark from Table \ref{Tab:Summary_Stats_MCMC} that all unbiased results \emph{satisfy} the condition $\Delta \Phi^{\text{(inj)}} \lesssim 1$ radian. Our simulations indicate that the criterion $\Delta\Phi \lesssim 1$ for unbiased parameter estimation is a reasonable benchmark for exploratory studies. However, more sophisticated methods like Bayesian inference should be used to make more conclusive statements regarding the accuracy of waveform models. 

\subsection{Constraining the secondary spin}\label{sec:secondary_spin}

\begin{figure}[hbpt!]
    \includegraphics[width = \textwidth]{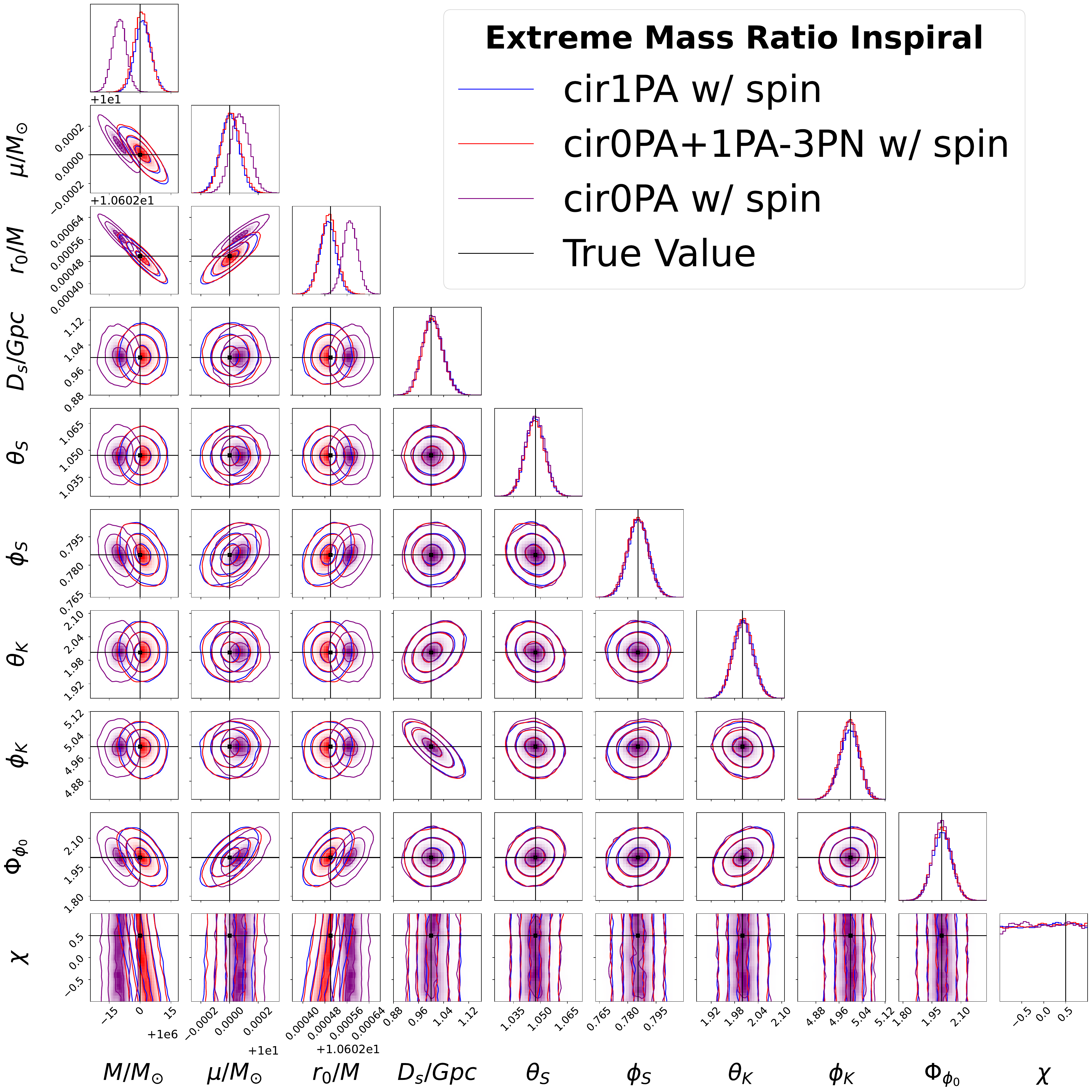}
    \caption{\textbf{(Mass ratio $\eps = 10^{-5}$):} Here we inject an EMRI waveform with \emph{spinning} CO on a circular orbit with parameters $M = 10^{6}M_{\odot}$, $\mu = 10M_{\odot}$, $r_{0}/M = 10.6025$, $D_{\rm S} = 1\,$Gpc and extrinsic parameters given in the caption of Table \ref{ref:EMRI_parameters}. The magnitude of the spinning secondary is $\chi = 0.5$, the SNR is $\rho_{AET} \sim 82$ and the time of observation is $T_{\text{obs}} = 2$ years. The blue, red, and purple parameter posteriors are generated when recovering with a cir1PA w/ spin model, cir0PA + 1PA-3PN w/ spin model, and cir0PA w/ spin model, respectively. The black vertical lines indicate the true parameters. The take-home message is that the spin of the secondary \emph{cannot} be constrained for $\rho_{AET} = 70$ and the models considered here.
    }
    \label{fig:corner_q_1e5_spin}
\end{figure}

\begin{figure}[hbpt!]
    \centering
    \includegraphics[width = \textwidth]{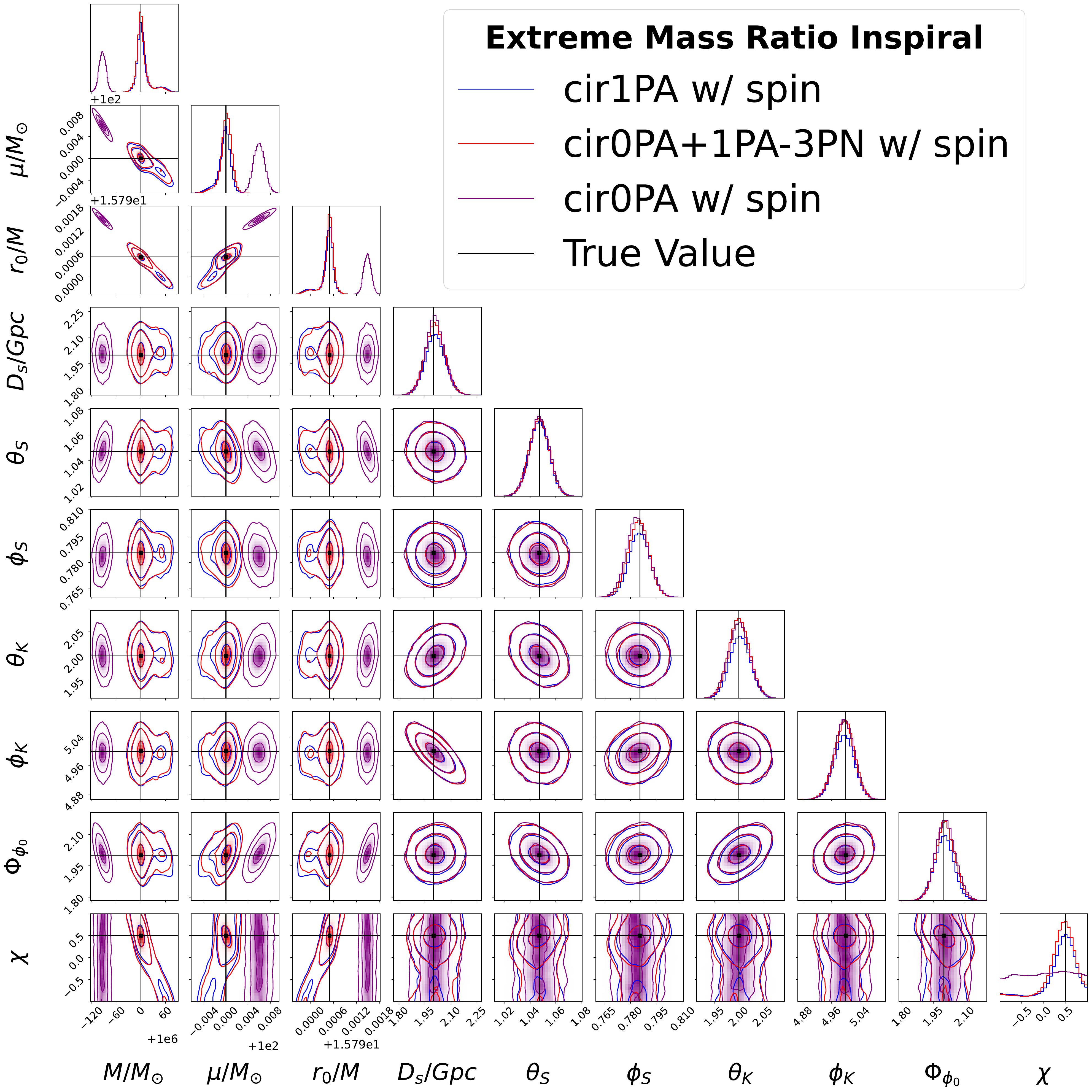}
    \caption{\textbf{(Mass ratio $\eps = 10^{-4}$):} Here we inject an EMRI waveform with \emph{spinning} CO on a circular orbit with parameters $M = 10^{6}M_{\odot}$, $\mu = 100M_{\odot}$, $r_{0}/M = 15.7905$, $D_{\rm S} = 2\,$Gpc and extrinsic parameters given in the caption of Table \ref{ref:EMRI_parameters}. The magnitude of the secondary spin is $\chi = 0.5$, the SNR is $\rho_{AET}\sim 70$ and the time of observation is $T_{\text{obs}} = 1.5$ years. The blue, red, and purple parameter posteriors are generated when recovering with a cir1PA w/ spin model, cir0PA + 1PA-3PN w/ spin model, and cir0PA w/ spin models respectively. The black vertical lines indicate the true parameters. The take-home message here is that the spin of the secondary \emph{can} be constrained using either the cir1PA w/ spin or cir1PA + 1PA-3PN w/ spin model. The cir0PA w/ spin model yields significant biases and \emph{cannot} constrain the spin of the secondary.}
    \label{fig:corner_q_1e4_spin}
\end{figure}

\begin{figure}[hbpt!]
    \centering
    \includegraphics[width = \textwidth]{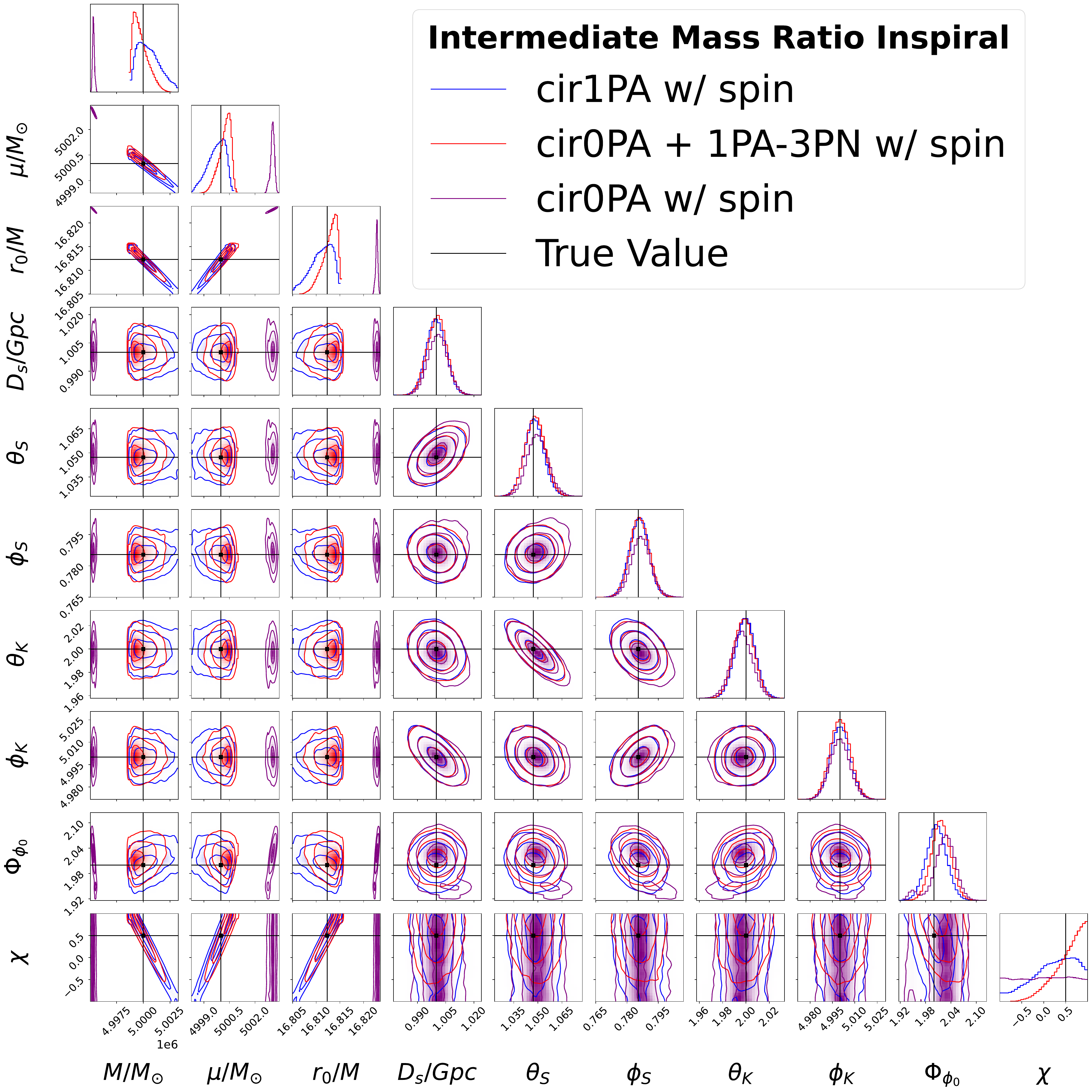}
    \caption{\textbf{(Mass ratio $\eps = 10^{-3}$):} Here we inject an EMRI waveform with \emph{spinning} CO on a circular orbit with parameters $M = 5\cdot 10^{6}M_{\odot}$, $\mu = 5000M_{\odot}$, $r_{0}/M = 16.81230$, $D_{\rm S} = 1\,$Gpc and extrinsic parameters given in the caption of Table~\ref{ref:EMRI_parameters}. The magnitude of the secondary spin is $\chi = 0.5$, the SNR is $\rho_{AET}\sim 340$ and the time of observation is $T_{\text{obs}} = 1$ year. The blue, red, and purple parameter posteriors are generated when recovering with a cir1PA w/ spin model, cir0PA + 1PA-3PN w/ spin model, and cir0PA w/ spin model, respectively. The black vertical lines indicate the true parameters. The take-home message here is that all 1PA terms (including the spin on the secondary) must be included to perform parameter estimation. The PN approximated waveform cir0PA + 1PA-3PN w/ spin shows significant biases and yields biased results on the secondary spin. 
    The cir0PA w/ spin model is unable to account for correlations and constrain the spin on the secondary body.}
    \label{fig:corner_q_1e3_spin}
\end{figure}

We now focus our attention on constraining the spin $\chi$ of the CO. Similar to Sec.~\ref{sec:mismodelling}, we study each mass ratio $\eps = \{10^{-5}, 10^{-4}, 10^{-3}\}$ with parameters given by Table~\ref{ref:EMRI_parameters}, and perform three parameter estimation simulations. The injection is a cir1PA w/ spin model, and approximate waveforms are similar to \eqref{eq:various_cases_no_spin} but with spin included:
\begin{equation}\label{eq:various_cases_spin}
h_{m} =
    \begin{cases}
        \text{cir0PA+1PA-3PN w/ spin}, \\
        \text{cir0PA w/ spin}. 
    \end{cases}
\end{equation}
Our corner plots for each of the $\eps = \{10^{-5}, 10^{-4}, 10^{-3}\}$ are displayed in Figs.~\ref{fig:corner_q_1e5_spin}, ~\ref{fig:corner_q_1e4_spin} and ~\ref{fig:corner_q_1e3_spin}, respectively. The first four rows of Tables \ref{tab:table_eps_1e-5}, \ref{tab:table_eps_1e-4} and \ref{tab:table_eps_1e-3} can be referred to for summary statistics of the intrinsic parameter set $\{M, \mu, r_{0}/M, \chi\}$. We will first discuss the $\eps = 10^{-5}$ case.

From Fig.~\ref{fig:corner_q_1e5_spin}, we see that the parameter $\chi$ \emph{cannot} be constrained for the $\eps = 10^{-5}$ case at $\rho_{AET} \sim 82$. The marginalized posterior distribution for $\chi$ is almost flat. This implies that our posterior information is not dominated by the likelihood (a function of the data), but instead dominated by the prior (a function of the parameters, irrespective of the data). We have tested various values of $\chi = \{-1, 0.5, 0, 0.5, 1\}$, and in no situation can the secondary spin be constrained. The exact model cir1PA w/ spin and approximate cir0PA + 1PA-3PN w/ spin model are indistinguishable. When recovering the exact cir1PA w/ spin with the exact model itself and the approximate cir0PA + 1PA-3PN w/ spin model, statistically insignificant biases to the intrinsic parameters are observed. For the case with mass ratio $\eps = 10^{-5}$, the spin on the secondary can be compensated by the minor tweaking of the intrinsic parameters. Finally, we report statistically significant biases when employing the cir0PA w/ spin model to extract parameters from a cir1PA waveform. These biases are consistent with the top panel and fourth row of Fig.~\ref{fig:Summary_all_mass_ratios_spin}. To conclude, neither the exact model nor approximate model templates are able to detect the presence of the spin on the smaller companion for a mass ratio $\eps = 10^{-5}$ and $\rho_{AET} \sim 82$.

Our results for mass ratio $\eps = 10^{-4}$ with $\rho_{AET} \sim 70$ are given in Fig.~\ref{fig:corner_q_1e4_spin}. The secondary spin has a more noticeable impact, and it can be constrained when using model templates given by cir1PA w/ spin and cir0PA + 1PA-3PN w/ spin. From the 2D marginalized posteriors, it is evident that the distributions on the intrinsic parameters and secondary spin are not Gaussian due to the presence of heavy tails, highlighting strong correlations between these parameters. Recall the red posterior in the middle panel of Fig.~\ref{fig:Summary_all_mass_ratios_spin}, where biases on parameters were observed if we recovered a cir1PA w/ spin template using a cir0PA + 1PA-3PN w/o spin model. 
We see from the red posterior in Fig.~\ref{fig:corner_q_1e4_spin} that including the spin on the cir0PA + 1PA-3PN model eliminates the biases on parameters, making it completely indistinguishable from the true cir1PA w/ spin model. This indicates that the cir0PA + 1PA-3PN would be suitable for parameter estimation, \emph{only if} the secondary spin parameter was included in the approximate model~\cite{Piovano:2020ooe}. 
Finally, we see that the cir0PA w/ spin model fails to constrain the secondary spin, with biases consistent with the second panel and fourth row of Fig.~\ref{fig:Summary_all_mass_ratios_spin}. Thus, neglecting 1PA components of the GSF will have a detrimental effect on recovering the spin of the smaller companion.    

To conclude this section, we now discuss the impact of the secondary spin on IMRIs with a mass ratio $\eps = 10^{-3}$ with $\rho_{AET} \sim 340$. Our results are displayed in Fig.~\ref{fig:corner_q_1e3_spin} for each of the various model templates. For this example at mass-ratio $\epsilon = 10^{-3}$, we find the only waveform suitable for parameter estimation is the exact model cir1PA w/ spin. The cir0PA + 1PA-3PN w/spin model yields biases on the intrinsic parameters and the secondary spin, although it accounts for correlations between the parameters. By contrast, the cir0PA w/ spin model exhibits significantly stronger biases because it does not correctly represent such correlations due to the lack of 1PA information. This can be seen in the 2D marginalized posteriors. The ``hard cut-offs'' observed in the two posteriors for cir1PA w/ spin and cir1PA + 1PA-3PN w/ spin are due to the spin on the secondary reaching the prior bounds. These cut-offs are not physical but merely a sampling artifact. 

In Appendix \ref{app:corner_plot_q1e3_wide_prior}, we have performed the same simulations that resulted in Fig.~\ref{fig:corner_q_1e3_spin}, but with a less informed prior on $\chi$ such that $\chi \sim U[-5,5]$. With a secondary mass $\mu = 10M_{\odot}$, this larger prior allows for more exotic compact objects\footnote{Black holes and neutron stars are the only known astrophysical objects that are compact enough to survive tidal disruption until the last stable orbit. The regularity theorem of the Kerr metric prevents spins $\chi > 1$, whereas neutron stars spin $\chi \lesssim  0.7$ due to the mass shedding limit. 
Other stellar objects (like white and brown dwarfs) can have spins $\chi \gg 1$. However, they are much less compact, and are tidally disrupted well before reaching the ISCO. Moreover, white dwarfs and neutron stars have masses in the narrow range $(1-2)M_\odot$. Thus, a secondary with mass $\mu \sim 10$, or larger, and spin $\chi > 1$ suggest the existence of an exotic compact object (for more details, see also~\cite{Piovano:2020zin,Piovano:2020ooe}).
}
. When injecting and recovering with the cir1PA w/ spin model, the intrinsic parameter posteriors still show non-Gaussian features (they exhibit heavy tails) and the correlation with the secondary spin is still present. The secondary spin is recovered with posterior median $\sim 0.5001$ with 68\% HPDI $\approx (-0.26, 1.12)$. When recovering the cir1PA w/ spin model with the cir0PA + 1PA-3PN w/ spin model, we see that the intrinsic parameters show statistically significant biases. For the secondary spin, we obtain a posterior median of $\sim 1.44$ with corresponding 68\% HPDI $0.5\not\in (0.72, 2.15)$, suggesting evidence that the secondary could be an exotic compact object. Finally, similar to Fig.~\ref{fig:corner_q_1e3_spin}, using the approximate cir0PA w/ spin waveform model to recover the cir1PA w/ spin model offers no information on the secondary spin. The secondary spin prior is recovered and the 68\% HPDIs are around an order of magnitude tighter. Again, this results in a lack of accuracy but a dangerous increase in precision in the parameters.    

These analyses suggests that the bias on the secondary spin is unacceptable for all approximate models and one must be careful, in this case, of using PN results to approximate the 1PA components of the self-force. For IMRIs, it is essential that we have full access to 1PA waveforms when performing parameter estimation. 
  
We conclude this section by briefly discussing the impact of first post-adiabatic effects on small-mass-ratio binaries with $\epsilon < 10^{-5}$. Although the main sources considered in this work are small-mass-ratio binaries with $\epsilon \in [10^{-5},10^{-3}]$, we have also studied a strong-field EMRI with $\epsilon = 10^{-6}$ with SNR $\rho_{AET} \sim 22$. Our analysis indicates that the secondary parameter cannot be constrained and that parameter estimation studies can be conducted with 0PA waveforms. That is, the 1PA components of the self-force are negligible for such small mass-ratios. We do not present our posterior results here, but the result can be extrapolated from 0PA results of Fig.~\ref{fig:Summary_all_mass_ratios_spin}. The primary mass $M$ shows the strongest level of bias, with value approximately $\sim \epsilon M$. For a mass ratio $\epsilon = 10^{-6}$, the bias on the primary mass $M$ is $\mathcal{O}(1M_{\odot})$. This bias is well contained within the statistical error given by the approximate posterior, making 0PA waveforms suitable for parameter estimation. This implies that the true parameter is contained within the 68\% credible interval of the 0PA approximate posterior, satisfying Eq.~\eqref{eq:systematic_ratio} with $\mathcal{C} = 1$. We conclude for this example that adiabatic waveforms would be suitable for \emph{both} search and characterizing 1PA waveforms at $\epsilon = 10^{-6}$, at least for quasi-circular systems with non-spinning primaries. 

In the analysis above we have tested three specific EMRI/IMRI configurations at mass-ratios $\epsilon \in \{10^{-6},10^{-5},10^{-4},10^{-3}\}$ with SNR $\rho_{AET}\in\{22,82,70,340\}$ respectively. We remind the reader that the conclusions in the four paragraphs above are governed by two quantities: (1) the specific configuration of parameters describing the system, the resultant SNR and (2) the \emph{geometry} of the inspiral and resultant waveform. Our Schwarzschild inspirals are strong-field orbits with trajectories evolved to within $\hat{r} \sim 6.27$. However, the orbits could evolve much closer to the horizon for a spinning primary. The increase in the number of orbits, compounded with closer-horizon geometry, would increase significantly the precision in parameters \emph{and} the SNR. This is strongly supported by work in~\cite{Babak:17aa}, see Fig. 6 and Fig. 11. See also  Ref.~\cite{Burke:20aa} for a detailed discussion on enhancements on parameter precision for circular orbits into rotating MBHs. 
For more complex orbits, it is then expected that one could achieve SNRs exceeding those of what we present here. As a result, the accuracy requirements on EMRIs would become more stringent. 

On the other hand, some might view our choices of luminosity distances as overly optimistic. We assume a flat-$\Lambda$CDM cosmological model with matter density $\Omega_{\rm m} = 0.274$, dark energy density $\Omega_{\Lambda} = 0.726$ and Hubble constant $H_{0} = 70.5 \text{km s}^{-1} \, \text{Mpc}$. Our choices for the luminosity distances $D_{\rm S} \in \{1,2\}$Gpc correspond to small redshifts $z_{\text{S}} = \{0.203, 0.371\}$, less than $z_{\rm S} = 1$. By comparison,  $z_{\rm S}=1$  corresponds to a luminosity distance of $D_{\rm S}(z_{\rm S} = 1) = 6.716\,$Gpc. Fig.~9 in Ref.~\cite{Babak:17aa} represents the redshift distributions of detected EMRI events assuming 12 distinct astrophysical models. All such distributions are peaked between $z_{\rm S} \in [1,2]$. Thus, our choices represent \emph{golden} EMRIs: strong-field in their orbital characteristics and placed at low redshifts $z_{S} \ll 1$. Assuming an astrophysically relevant luminosity distance $D_{\rm S} = 6.67$, the SNR of our sources would decrease significantly since $\rho \sim 1/D_{\rm S}$. In fact, our $\epsilon = 10^{-5}$ case in Table~\ref{fig:Summary_all_mass_ratios_spin} would only reach an SNR $\rho_{AET} \sim 11$, which is lower than the detection threshold for EMRIs. 

This highlights a fundamental point. Since the statistical error on the parameters scales with the SNR, one can choose an SNR $\sim 20$ for $\epsilon = 10^{-5}$ such that 0PA waveforms are suitable for parameter estimation of 1PA waveforms. For a more relevant, but still quite small, luminosity distance $D_{\rm S} \sim 3\,$Gpc, corresponding to $z_{\rm S} = 0.52$, $\rho_{AET} \sim 20$ for $\epsilon = 10^{-5}$. The true parameter could then be within the 68\% credible interval of the purple posterior in Fig.~\ref{fig:Summary_all_mass_ratios_spin}. For such a system we could then conclude that adiabatic waveforms are suitable for \emph{both} detection and parameter extraction of 1PA waveforms.  Similarly, one could place the source at exceptionally low redshift $z_{\rm S} = 2.5\times 10^{-4}$, giving a luminosity distance $D_{\rm S} = 0.01\,$~Gpc and increasing the SNR by a factor of 100 in the $\epsilon = 10^{-5}$ case. The spinning secondary may be observable in such a situation, but the probability that such an EMRI will be observed is essentially zero according to~\cite{Babak:17aa}. 

In conclusion, we have chosen parameter sets such that all of our orbits are astrophysically sound, allowing us to draw reasonable conclusions. Within the range of realistic sources, we have focused on ones that are loud enough to be of most interest for high-precision science. The high-accuracy models built by the self-force community naturally aim for these golden EMRIs (and IMRIs), which have high SNRs and are in the strong-field regime.

\onecolumngrid

\newpage
\newpage
\newpage
\newpage

\twocolumngrid

\subsection{Systematic biases --- mismodeling evolution of eccentric orbits}\label{sec:eccentricity}

Our previous analyses focused on the simplest class of black-hole orbits: quasi-circular orbits within Schwarzschild spacetimes. It is not obvious how our results will extend to more generic systems. In this section we begin to explore that question by assessing the potential impact of mismodeling EMRI waveforms for eccentric orbits. 
As 1PA models do not yet exist for eccentric orbits, we inject waveforms with our adiabatic ecc0PA model (see Sec.~\ref{sec:0PAecc}) and attempt to recover them with our approximate ecc0PA-9PN model. We consider injected waveforms with various values of eccentricity $e = \{0, 0.01, 0.1, 0.2\}$.

The trajectories are evolved using a low-eccentricity 9PN expansion, exhibiting slow convergence as the eccentricity increases. We do not consider eccentricities $e > 0.2$ for two reasons. The first is the clear breakdown of the PN expansion, yielding artificially inflated biases that could be less severe in reality. The second reason is the enormous difficulty for the sampler to converge to the point of highest likelihood. For $e = 0.3$, we noticed that the number of secondary maxima in the likelihood surface grew significantly in comparison to $e\in\{0.1,0.2\}$. This means that the individual chains of the sampler got stuck for $\sim \mathcal{O}(10^{3})$ iterations even with $\sim 10$ temperatures (see App.~\ref{appendix:MCMC} for more details).
Whether this is an artifact of eccentricity itself due to the increased mode content, or the PN expansion breaking down, is unclear. For the aforementioned reasons, we consider weak-field orbits with low eccentricities to avoid exaggerating the impact of eccentricity when mismodeling templates.

For all cases presented in this section, the injected waveforms' parameters are set to $M = 10^{6}M_{\odot}$, $\mu = 10M_{\odot}$, and initial semi-latus rectum $p_{0} = 9.75$. The trajectories are evolved for $T_{\text{obs}} = 1$ year until a final semi-latus rectum of $p \sim 7.73$ is reached. We do not evolve the trajectories further than this point due to the breakdown of the PN expansions in the strong-field regime. We choose the same extrinsic parameters as presented in Table~\ref{ref:EMRI_parameters}, but with luminosity distance $D_{\rm S} = 0.8\,$\text{Gpc} and initial radial phase $\Phi_{r_{0}} = 3$ for all cases. For each eccentricity $e = \{0, 0.01, 0.1, 0.2\}$, Eq.~\eqref{eq:dephasing_injected} gives a dephasing on the order $\Delta \Phi^{\text{(inj)}} \approx \{3.52, 3.78, 6.39, 22.37\}$ radians, respectively, between the two models. Finally, the number of modes in the waveform for both models, $\#\text{modes}$, for each chosen eccentricity $e$ is $(e,\#\text{modes}) = \{(0,13), (0.01,24), (0.1,62), (0.2,96)\}$ respectively. Finally, both circular and eccentric waveform models yield similar SNRs on the order of $\rho_{AET} \sim 70$. The details of how we used \texttt{eryn} for our eccentric parameter estimation simulations are presented in App.~\ref{appendix:MCMC}. 

\begin{figure*}[hbpt!]
    \centering
    \includegraphics[width = \textwidth]{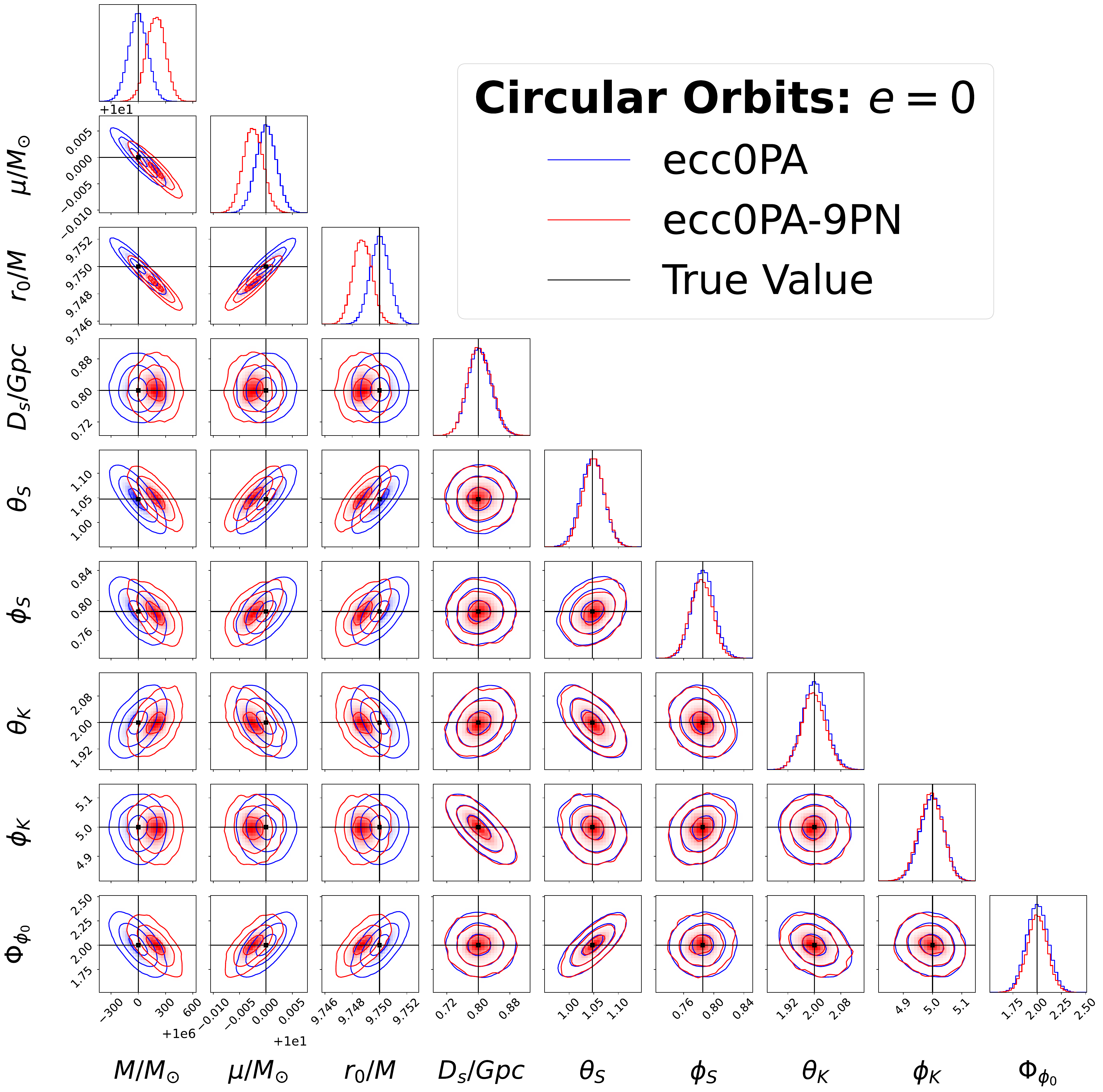}
    \caption{ \textbf{(Circular Orbits:)} Here we consider as injection an ecc0PA model (vanilla adiabatic FEW) with parameters $M = 10^{6}M_{\odot}$, $\mu = 10M_{\odot}$, $r_{0}/M = 9.75$, $e = 0$. The trajectory evolves for one year and terminates at a radial coordinate of $r_{0}/M \approx 7.73$, giving an SNR $\rho_{AET} \sim 70$. We treat parameters related to eccentricity $\{e = 0,\Phi_{r_0} = 3\}$ as known. The blue posterior is generated through recovery using the injected model and the red posterior is built using the approximate ecc0PA-9PN model. The black vertical lines indicate the location of the true parameters.}
    \label{fig:circular_9PN_result_e0_0p0}
\end{figure*}

We begin with the circular orbit case with the result shown in Fig.~\ref{fig:circular_9PN_result_e0_0p0}. The details on the individual runs can be found in the caption. We treat the parameters $e = 0$ and $\Phi_{r_0} = 3$ as known, and we do not sample over them. The intrinsic parameters exhibit statistically significant biases, whereas the extrinsic parameters are unbiased. Figure~\ref{fig:circular_9PN_result_e0_0p0} will be our reference figure when making direct comparisons with eccentric orbits. 

\begin{figure*}[hbpt!]
    \centering
    \includegraphics[width = \textwidth]{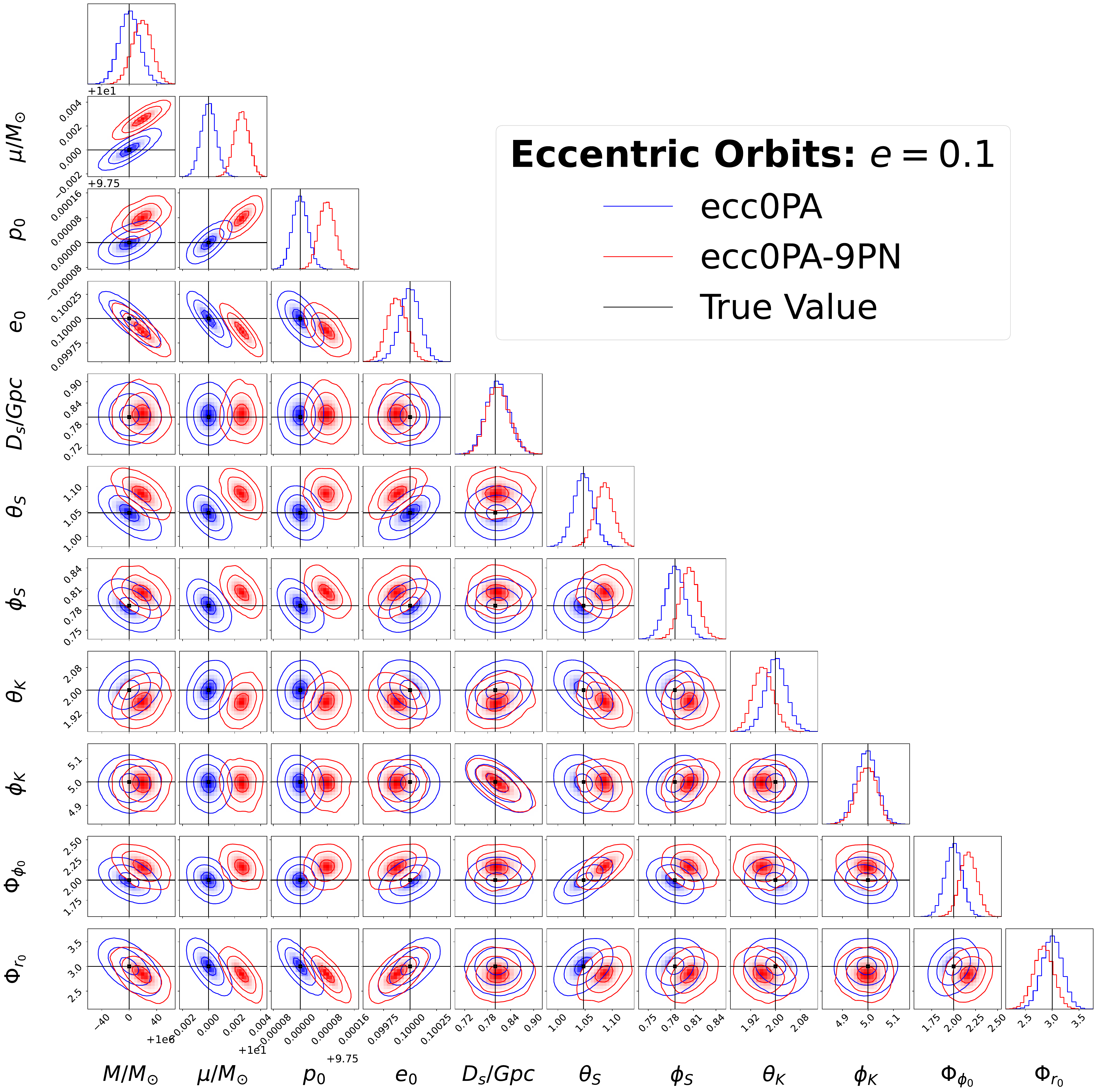}
    \caption{ \textbf{(Eccentric Orbits:)} The same parameters and set up as Fig.~\ref{fig:circular_9PN_result_e0_0p0} but now sampling over eccentric parameters with $e = 0.1$ and $\Phi_{r_{0}} = 3$. Compared to Fig.~\ref{fig:circular_9PN_result_e0_0p0}, the biases on the intrinsic parameters are more severe across the parameter space. We also observe biases on the extrinsic parameters, a feature not seen for the circular orbit case presented in Fig.~\ref{fig:circular_9PN_result_e0_0p0}. }
    \label{fig:eccentric_9PN_result_e0_0p1}
\end{figure*}

\begin{figure*}[hbpt!]
    \centering
    \includegraphics[width = \textwidth]{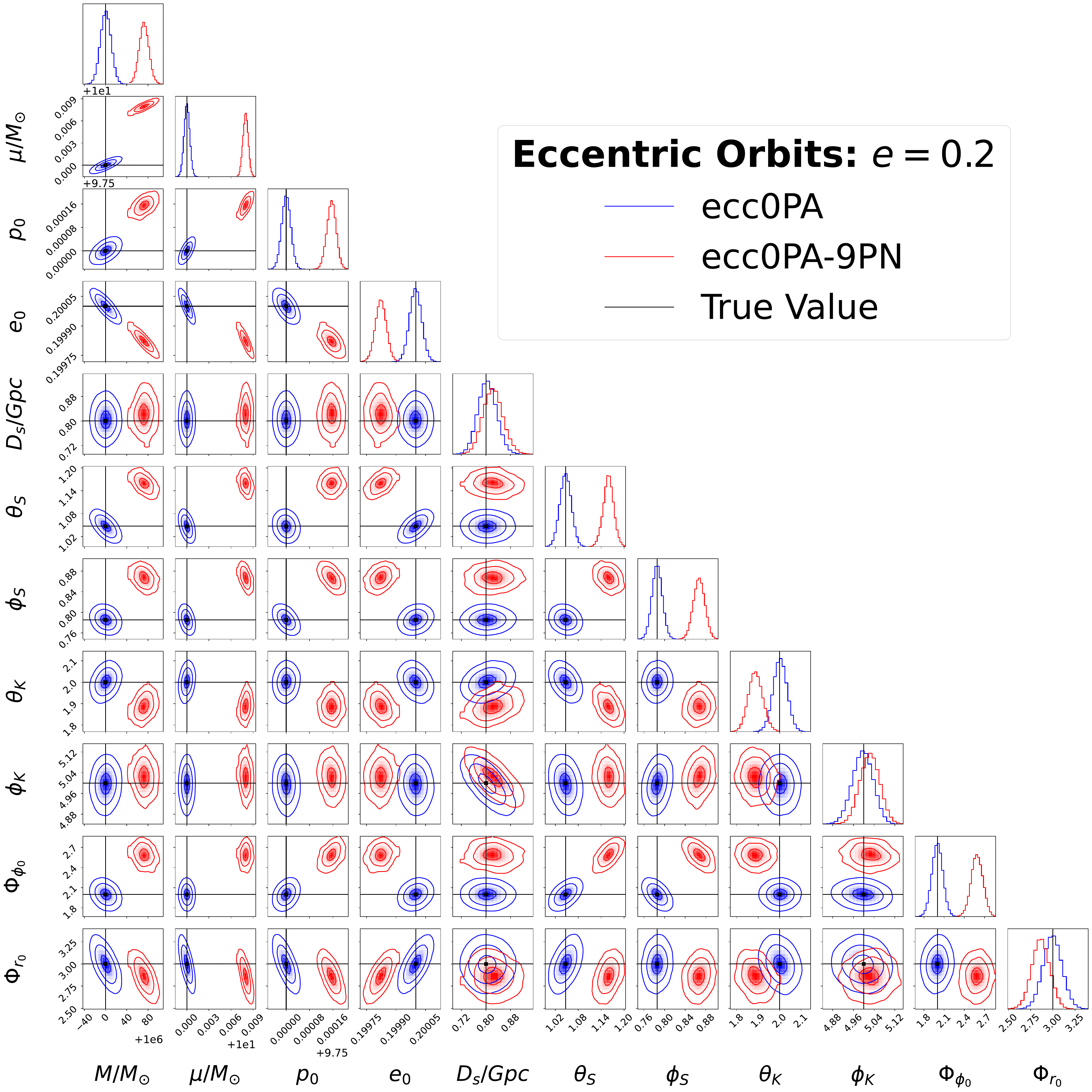}
    \caption{\textbf{(Eccentric orbits:)} 
    The same parameters and set up as Fig.~\ref{fig:eccentric_9PN_result_e0_0p1} but with $e = 0.2$ and $\Phi_{r_{0}} = 3$. Compared to Fig.~\ref{fig:circular_9PN_result_e0_0p0}, the biases on the intrinsic parameters are more dramatic across the parameter space.}
    \label{fig:eccentric_9PN_result_e0_0p2}
\end{figure*}

We perform a series of similar simulations of recovering an exact adiabatic model (ecc0PA), with an approximate (ecc0PA-9PN) model template with small to moderate eccentricities $e = \{0.01, 0.1,0.2\}$.  The case with $e = 0.01$ is qualitatively similar to Fig.~\ref{fig:circular_9PN_result_e0_0p0}, so we will not present it here. The results for $e \in \{0.1,0.2\}$ are shown in Fig.~\ref{fig:eccentric_9PN_result_e0_0p1} and Fig.~\ref{fig:eccentric_9PN_result_e0_0p2}. For both plots, all the intrinsic parameters show severe levels of biases, stronger compared to the circular orbits. Furthermore, unlike the circular case, the angular parameters $\{\theta_{S},\phi_{S},\theta_{K},\phi_{K}\}$ and initial phases $\{\Phi_{\phi_{0}},\Phi_{r_{0}}\}$ show statistically significant levels of bias. It should be noted that the magnitude of the biases on all parameters increases significantly between the $e = 0, e=0.1$ and $e = 0.2$ cases. 

The biases on the extrinsic parameters for eccentric orbits stem from two effects. The first is the correlations between the parameters, which are more pronounced compared to circular orbits. Unlike the circular orbit case, the intrinsic and extrinsic parameter spaces are not orthogonal: minor tweaks to the intrinsic parameters can be compensated by minor tweaks in the extrinsic parameters. The second reason is that the orbital evolution is more complex, resulting in a LISA-responsed waveform with a richer structure in comparison to its circular counterpart. 

The tweaking of the angular parameters in response to the biases in the intrinsic parameters appears to minimize the mismatch between the two signals. This can be explored by comparing the ecc0PA waveform at the true parameters and ecc0PA-9PN waveform at the recovered parameters, but fixing the angular parameters $\{\theta_{S},\phi_{S},\theta_{K},\phi_{K}\}$ to their true values. We obtain mismatches $\mathcal{M} \sim 0.243$ and accumulated SNR $\rho^{\text{(bf w/ inj angles)}}/\rho^{(\text{opt})} \sim 70\%$, indicating that the two waveforms quickly go out of phase. Comparing the true ecc0PA signal with ecc0PA-9PN evaluated at the recovered parameters, including the recovered angular parameters, yields $\mathcal{M} \sim 10^{-2}$ and accumulated SNR $\rho^{\text{(bf)}}/\rho^{\text{(opt)}} \sim 98\%$, indicating that the approximate model template remains in phase with the true reference signal for a much longer duration. 

For circular orbits, there are weak correlations between the intrinsic and extrinsic parameters, demonstrated by the marginalized 2D distributions in Fig.~\ref{fig:circular_9PN_result_e0_0p0}. Minor shifts to the intrinsic parameters will not affect the harmonic structure, and thus there will be no biases across the extrinsic parameters.
\onecolumngrid
\newpage
\newpage

\newpage
\twocolumngrid
\section{Summary}\label{sec:summary}
In this section, we summarize the results presented in Sec.~\ref{sec:results}. For the configuration of parameters in this work (see Table~\ref{ref:EMRI_parameters}), adiabatic templates \emph{are not suitable} for EMRI parameter estimation, where the exact model contains full post-adiabatic information. Adiabatic templates though, are suitable for detection purposes in all cases. We also highlight the performance of suitably re-summed third-order PN expansions when approximating the 1PA components of the self-force. For mass ratios $\eps = 10^{-5}$ and $\eps = 10^{-4}$, such approximate PN-based models are suitable for EMRI data analysis (at least for the simple binary configurations we consider), assuming that the spin on the smaller companion is included. We also found that such PN-based approximate waveforms break down within the IMRI regime $\eps = 10^{-3}$, where full post-adiabatic information on the model templates is required. Neglecting the spin on the secondary and/or the 1PA components of the self-force results in both significant biases and fictitiously tighter constraints on parameters. This is due to neglecting the significant correlations between the secondary spin and intrinsic parameters. We conclude that the spin of the secondary should always be retained and only post-adiabatic templates should be used to characterize post-adiabatic models. 

The feasibility of constraining the secondary spin was then discussed. For $\eps = 10^{-5}$, the secondary spin cannot be constrained for our choice of SNR $\rho_{AET} \sim 82$, whereas, for $\eps = 10^{-4}$ and $\eps = 10^{-3}$, it could be constrained at $\rho_{AET} \sim 70$ and $\rho_{AET} \sim 340$, respectively. In the IMRI regime,
the secondary spin and intrinsic parameters exhibit strong correlations, which significantly degrade the precision measurement of the intrinsic parameters. 

Finally, we investigated the impact of mismodeling eccentric binaries. By injecting an adiabatic eccentric waveform and recovering with a ninth-order PN-based waveform, we observed severe biases across \emph{both} the intrinsic and extrinsic parameters, whereas only the intrinsic parameters are biased for purely circular orbits. It is impossible to say, for now, how this will translate when using 0PA waveforms for inference on 1PA waveforms. However, we expect that the severity of the biases with respect to circular orbits will increase across the full parameter space. We conclude here that care must be taken for both the modeling and parameter inference of eccentric 1PA waveforms.

\section{Discussion}\label{discussion}

This paper presented, for the first time, a detailed Bayesian study using MCMC with state-of-the-art \emph{first post-adiabatic} EMRI and IMRI waveforms for circular orbits in the Schwarzschild spacetime. We showed that neglecting 1PA terms will induce statistically significant biases in the parameters. However, we can still detect and characterize the first post-adiabatic waveform in the data stream using adiabatic waveform models with biased parameters. 
We have confirmed that adiabatic waveforms are only suited for detection purposes (at least for $ 10^{-5}\lesssim \eps \lesssim 10^{-3}$). The systematic errors are subjectively quite small for adiabatic waveforms and small mass-ratios\footnote{For example, the top panel of Fig.~\ref{fig:Summary_all_mass_ratios_spin} shows that we can recover the primary mass $M = 10^{6}M_{\odot}$ with a bias on the order of $\sim 10M_{\odot}$.}, and might not be relevant for some astrophysical applications like population studies~\cite{10.1093/mnras/stad1397,Babak:2017tow}. However, applications within fundamental physics (like testing alternative/modified theories of gravity~\cite{Canizares:2012is,Pani:2011xj,Yunes:2011aa,Blazquez-Salcedo:2016enn}, investigating the nature of massive black holes~\cite{Maggio:2021uge,Barack:2006pq,Barack:2006pq,Datta:2019epe}, the presence of additional fields~\cite{Fell:2023mtf,Maselli:2021men,Barsanti:2022vvl}, and so on) crucially rely on both precise and accurate measurements of the binary parameters. Even relatively small statistically significant systematic errors could spoil the enormous scientific potential of EMRIs. Thus, first post-adiabatic waveforms are \emph{essential} in order to reap the  full scientific rewards of EMRI data analysis.  

Bayesian methods are the gold standard technique when studying waveform systematics because they do not introduce any approximations from a statistical point of view.
Fisher matrices, the Lindblom criterion, overlaps/mismatches, and orbital dephasings must be used with caution. Such systematic tests are suitable for exploration and gaining insight into the accuracy of the waveform models, but conclusions must be taken lightly with respect to their accurate Bayesian counterpart. Fisher matrices are hard to accurately compute for EMRIs (see the works of Refs.~\cite{Burke:20aa,Piovano:2020zin,speri2021assessing,Maselli:2021men}), potentially leading to false conclusions on biases and precision statements of parameters. Mismatches give no information about potential biases on parameters, and fully optimized fitting factors  such as Eq.~\eqref{eq:mismatch_recovered_params} can only be calculated through stochastic sampling algorithms. Similarly, the Lindblom criterion is overly conservative~\cite{purrer2020gravitational}, and if taken at face value could force unreasonably stringent accuracy requirements on waveform templates. Finally, comparisons of the orbital dephasings between two models are performed at the trajectory level, and so neglect the waveform structure, SNR of the source or correlations between parameters. These systematic studies tell an important story, but proper Bayesian inference completes the picture. 
For example, we have shown that the posterior densities of EMRIs and IMRIs can yield non-Gaussian features when the secondary spin is included, becoming ever more dramatic as $\eps \gg 10^{-5}$. No other systematic test can reveal such interesting features. 

This importance of Bayesian methods can be seen in Sec.~\ref{sec:mismodelling}, which presents the parameter distributions from mis-modelling in Fig.~\ref{fig:Summary_all_mass_ratios_spin} and the associated summary statistics in Table~\ref{Tab:Summary_Stats_MCMC}. We highlight from this analysis that statistically significant biases \emph{are not observed} when the orbital dephasing $\Delta\Phi^{\text{(inj)}}\lesssim 1$ radians (see Eq.~\eqref{eq:dephasing_injected}). From the top panel and bottom row of Fig.~\ref{fig:Summary_all_mass_ratios_spin}, for $\epsilon = 10^{-5}$, the orbital dephasing between the 0PA and 1PA waveforms is $\sim 3$ radians. We see that strong biases are observed, yet a concrete detection is made with accumulated SNR at the best-fit parameters $\rho^{\text{(bf)}}/\rho^{\text{(opt)}} \sim 99.8\%$. This is further exemplified in the bottom panel, where there is a $\sim 15$ radian difference between the 0PA and 1PA waveforms and $\rho^{\text{(bf)}}/\rho^{\text{(opt)}} \sim 99.9\%$, resulting in a clear detection. This is comforting to see, as it implies that the requirements often used by the modelling community (e.g., that orbital phase errors should be $\Delta \phi\lesssim 1/{\rm SNR}$~rad ~\cite{Hughes:2016xwf}) may be too stringent. However, the orbital trajectories completely neglect the SNR of the system, a \emph{crucial} ingredient for both detection and parameter estimation. Relying on only orbital dephasing should be taken with caution. Similarly, large mismatches between the approximate and exact waveform model (for identical parameter values) might suggest that the waveform is not detectable with the approximate model. But this is not the case: Table~\ref{Tab:Summary_Stats_MCMC} contradicts it, giving small mismatches $\mathcal{M}\sim 10^{-5}$ at the recovered parameters for the largest of mass ratios $\epsilon = 10^{-3}$ even though there are large mismatches $\mathcal{M} \sim 0.938$ at the injected parameters. In conclusion, our results reinforce that Bayesian inference is key to making significant claims about approximations of waveforms. 

Another outcome of our work is that we have shown the importance of including the secondary spin parameter in our waveform models. Although it appears impossible to constrain at mass ratios $\epsilon \leq 10^{-5}$, a measurement can be made at higher mass ratios $\epsilon \geq 10^{-4}$. Correlations between the intrinsic parameters and the secondary spin are significant, leading to degraded measurements for IMRIs only if the spin parameter is included. We have demonstrated that neglecting the secondary spin causes not only a bias but fictitiously tight constraints on the other intrinsic parameters. This becomes more prominent as the mass ratio increases. Furthermore, we have shown that knowledge of the 1PA components of the GSF is essential when trying to measure the secondary spin. It may not possible to make definite conclusions on the potential constraints on  the secondary spin for genetic orbits without including other post-adiabatic terms.

Finally, we assessed the importance of eccentricity when mismodeling eccentric templates. At the moment, only some self-force contributions at 1PA are known for eccentric orbits, therefore it is not possible yet to make tests similar to Sec.~\ref{sec:mismodelling} and Sec.~\ref{sec:secondary_spin}. Instead, we injected an adiabatic waveform and attempted to recover with an approximate adiabatic waveform with trajectories evolved through 9PN fluxes expanded in eccentricity. For low to moderate eccentricities, $e \in \{0, 0.01, 0.1, 0.2\}$, it is clear that the severity of the biases worsens in comparison to the circular orbit case as $e$ increases (cf. Figs.~\ref{fig:circular_9PN_result_e0_0p0} to \ref{fig:eccentric_9PN_result_e0_0p2}). For moderate eccentricities, biases are observed across the entire parameter space, notably in the angular parameters and initial phases. Such biases in the sky position would be unacceptable for studies within cosmology, where the construction of galaxy catalogs allows one to infer cosmological parameters, such as the Hubble constant~\cite{laghi2021gravitational}. Our analysis show that the inclusion of eccentricity will complicate the picture. More work is required on the self-force and data analysis front to understand the impact of these orbits on EMRI parameter estimation.   

\section{Future Work}\label{sec:Future_Work}

The work presented here has only scratched the surface of EMRI accuracy requirements. Clearly, it is essential to repeat this analysis once second-order self-force results become available for more generic orbits. Indeed, generic orbits may break potential degeneracies with the secondary spin parameter, leading to  improved constraints at smaller mass ratios $\epsilon \sim 10^{-5}$. We have also shown remarkable success with the use of resummed PN expansions when approximating the 1PA components in the context of parameter estimation. For mass ratios $\epsilon \gtrsim 10^{-4}$, one could perform preliminary studies on the secondary spin parameter, assuming suitable 1PA-$n$PN results were available for more general orbits. 
Moreover, we observed that,  due to correlations,  the secondary spin parameter deteriorates the precision with which the intrinsic parameters can be recovered.
It would then be interesting to understand whether there exist degeneracies between the secondary spin and the scalar charge due to extra scalar fields, which may spoil the precision measurements on the latter~\cite{Barsanti:2022vvl,Maselli:2021men,Barsanti:2022ana}. 

The field of EMRI systematics can now answer some crucial practical questions for modeling EMRI waveforms. For instance, calculating the first-order self-force for generic Kerr inspirals is very expensive with just a single point in the parameter space taking $\mathcal{O}(10^4)$ CPU hours \cite{vandeMeent:2017bcc}. One then has to repeat this calculation many times across the 4-dimensional generic Kerr parameter space to produce interpolants for the 0PA equations of motion, and then repeat the process at 1PA. It is important to estimate the required accuracy at each point, the minimum number of points, and their optimal placement, for both the 0PA and 1PA contributions in order to avoid biasing parameters significantly. The choice of a suitable interpolation method is important as well. It is still up for debate whether to use Chebychev-interpolation methods (which have favorable convergence properties) or  less expensive splines. 

We conclude by discussing one last topic that is still relatively untouched: the EMRI search problem. The search problem has been ``solved'' in extremely simplified circumstances by various groups within the LISA community~\cite{gair2008constrained, cornish2011detection, gair2008improved, babak2010mock}. The underlying noise properties were well understood and tight priors were placed on the parameters to recover. Furthermore, many of these groups exploited the analytical features of the injected model (the self-inconsistent ``Analytical Kludge'' waveforms from Ref.~\cite{Barack:2003fp}), and the known structure of the likelihood local maxima to reach the global maximum indicating the true parameters. It is unknown whether fully relativistic waveforms will simplify or complicate the search problem. EMRI search is a difficult open problem, the authors hope that work can restart on this vital research topic thanks to the advent of FEW and access to accurate waveforms. 


\begin{acknowledgments}
O.~Burke thanks both the University College Dublin Relativity Group and the Albert Einstein Institute for hosting him during the preparation of this work. He also gives thanks to Alvin Chua, Christian Chapman-Bird, Leor Barack, Maarten van de Meent, Sylvain Marsat, Michael Katz and Jonathan Gair for various insightful discussions. This project has received financial support from the CNRS through the MITI interdisciplinary programs. 
GAP acknowledges support from an Irish Research Council Fellowship under grant number GOIPD/2022/496. He also thanks Alvin Chua and Enrico Barausse for insightful discussions.
NW acknowledges support from a Royal Society - Science Foundation Ireland University Research Fellowship. 
This publication has emanated from research conducted with the financial support of Science Foundation Ireland under Grant number 22/RS-URF-R/3825.
AP acknowledges the support of a Royal Society University Research Fellowship and a UKRI Frontier Research Grant under the Horizon Europe Guarantee scheme [grant number EP/Y008251/1]. He also thanks Alvin Chua and Leor Barack for helpful discussions. CK acknowledges support from Science Foundation Ireland under Grant number 21/PATH-S/9610. This work has used the following \texttt{python} packages: \texttt{emcee}, \texttt{eryn}, \texttt{lisa-on-gpu}, \texttt{scipy}, \texttt{numpy},  \texttt{matplotlib}, \texttt{corner}, \texttt{cupy}, \texttt{astropy}, \texttt{chainconsumer} and \texttt{lisatools} ~\citep{Foreman-Mackey:2013, Karnesis:2023ras, michael_katz_2023_7705496, harris2020array, Hunter:2007, corner, Hinton2016, astropy:2022, cupy_learningsys2017,KatzTools}. 
This work makes use of the following packages of the Black Hole Perturbation toolkit~\cite{BHPToolkit}: \texttt{FastEMRIWaveforms}~\cite{michael_l_katz_2023_8190418}, \texttt{KerrGeodesics}~\cite{KerrGeodesicsPackage} and \texttt{PostNewtonianSelfForce}~\cite{PostNewtonianSelfForcePackage}. 
\end{acknowledgments}

\section*{Author Contributions}
OB: Conceptualization, Data curation, Formal analysis: MCMC simulations, Investigation, Methodology, Project administration, Resources, Software: All data analysis techniques, Supervision, Validation, Visualization: All plots, Writing – original draft. 

GP: Conceptualization, Data curation: secondary spin fluxes, Methodology, Validation: Fisher based estimates, Investigation, Writing – original draft.

NW: Conceptualization, Data curation: second-order self-force results, Methodology, Software: implementation of the waveform models, Writing - Review \& Editing.

PL: Conceptualization, Methodology, Software: implementation of the waveform models, Writing - Review \& Editing.

LS: Conceptualization, Methodology, Software: implementation of the waveform models, Visualization, Writing - Review \& Editing.

CK: Conceptualization, Methodology, Software: PN expansions, Writing - Review \& Editing.

BW: Conceptualization, Data curation: second-order self-force results, Methodology, Writing - Review \& Editing.

AP: Conceptualization, Data curation: second-order self-force results, Methodology, Writing - Review \& Editing.

LD: Data curation: second-order self force results

JM: Data curation: second-order self force results

\appendix

\section{MCMC implementation details}\label{appendix:MCMC}
\subsection{\texttt{eryn} and \texttt{emcee}}\label{app:eryn_emcee}
As discussed in Sec.~\ref{sec:PE_algorithm}, our analysis employs ensemble-based Markov-Chain Monte-Carlo (MCMC) techniques to sample from the posterior distribution. We use a mixture of \texttt{emcee} and \texttt{eryn}, depending on the complexity of the orbits. Both samplers apply an affine transformation to the posterior distribution to produce a new density that is far easier to sample from. This is very useful for EMRIs since the intrinsic parameters are usually highly correlated. On all runs, we used fifty individual walkers, where each walker represents an individual chain that explores the available prior space set to be uniform over a range of values. One difficulty with this approach, for EMRIs, is that multiple chains could become stuck on secondary maxima of the likelihood surface. These secondary maxima appear due to non-trivial overlap with the underlying signal and the model template. An MCMC technique lessens the impact of these secondary maxima is \emph{parallel-tempering}, which we describe below.

A significant advantage of \texttt{eryn} over \texttt{emcee} is the implementation of parallel tempering, a technique that ``heats'' the likelihood in order to mitigate the secondary maxima structure of the log-likelihood function. Defining a set of $N$ ``temperatures'' $\mathcal{T} = \{1,\ldots,T_{N}\}$, one can deploy a collection of $N$ sets of walkers, one set for each temperature. Each set of walkers explore (in parallel) a \emph{tempered} distribution, given by
\begin{equation}
p(\boldsymbol{\theta}|d) \propto p(d|\boldsymbol{\theta})^{1/T}p(\boldsymbol{\theta}),
\end{equation}
where $T \in \mathcal{T}$ is an individual temperature. When $T = 1$, the (cool) chains would sample from the true posterior density. For intermediate temperatures $1<T<T_{N}$, the likelihood itself is smoothed out so the individual (hot) chains, corresponding to those intermediate temperatures, are less likely to get stuck on local maxima. In the limit as $T \rightarrow \infty$, the walkers would sample from the prior distribution $p(\boldsymbol{\theta})$. 

The individual chains are sent out in parallel, where samples corresponding to that specific chain are either accepted/rejected depending on a specific acceptance criterion (given by Eq.~(19) in \cite{Karnesis:2023ras}). Then, to further encourage mixing, individual points (known as states) of individual chains may also swap between individual temperatures based on a swapping criterion (see Eq.~(20) in \cite{Karnesis:2023ras}). The overall idea of parallel tempering is to lessen the probability of getting stuck on local maxima by encouraging movement between the chains. Once the algorithm is complete, we discard all samples with temperatures $T \neq 1$ and only use samples from the true posterior $\boldsymbol{\theta}^{(i)} \sim p(d|\boldsymbol{\theta})p(\boldsymbol{\theta})$ at $T = 1$. The only disadvantage of using parallel-tempered MCMC over vanilla MCMC is the computational cost. The computational cost of parallel-tempered MCMC algorithms scale linearly with the numbers of temperatures. This was the main reason for our choice of using \texttt{emcee} for the simpler circular orbit runs presented in this paper.  

It is well known that the likelihood surface of EMRIs is notoriously multi-modal (for instance, see Fig.12 in~\cite{chua2022nonlocal}), and it is not uncommon to get ``stuck'' on secondary likelihood maxima during the sampling stage of the posterior. We found that \texttt{emcee} performed fine when dealing with EMRIs in circular orbits. The small parameter space and the small number of modes present in the waveform model ensured that individual chains converged to the maximum likelihood point quickly without getting stuck. Only in the most extreme of runs, where we search for 1PA templates possessing spin using adiabatic templates did we observe individual walkers get stuck on secondary modes. For individual walkers stuck on nuisance modes, those walkers were removed from the set of chains to ensure that we extract summary statistics described by the parameters governing the highest point in likelihood. However, when we focused our attention on eccentric orbits in Schwarzschild, the use of \texttt{emcee} ultimately led to failure. This was due to individual chains remaining stuck on secondary modes for $\sim$~thousands of iterations. We found success using \texttt{eryn}, with five temperatures $T = \{1,\ldots,5\}$ when analyzing the eccentric orbit cases. 

To be clear, we refer to the ``convergence'' of an ensemble sampler when those individual walkers, in our case 50, reach approximately the same value of log-likelihood for at least $\sim 1,000$ iterations. For the simulations present in Fig.~\ref{fig:Summary_all_mass_ratios_spin}, the maximum log-likelihood values attained are given by the last column of Table~\ref{Tab:Summary_Stats_MCMC}. If the log-likelihood does not climb in value past 1,000 iterations, we deem the sampler to have converged to the correct stationary distribution. In other words, we apply the ``thick-pen'' procedure; we refer the reader to~\cite{cowles1996markov} for discussion of this procedure (and a plethora of other convergence diagnostic tests).

\subsection{Starting points, prior choices, burn-in and computational cost}\label{app:mcmc_start_prior_burnin}

\textbf{Starting points:} For each walker, the initial states (samples) were chosen such that $\boldsymbol{\theta}^{(0)} \approx \boldsymbol{\theta}_{\text{tr}}$. Starting this close to the true parameters is reasonable: our goal is to identify the impact on parameter estimation when inaccurate EMRI/IMRI waveforms are employed, rather than perform a search. 

\textbf{Prior choices:} We determine the prior ranges as follows. We first perform inference on the exact waveform model using an exact model template. With knowledge of the true parameters, we can set vague prior bounds as our starting coordinates $\boldsymbol{\theta}^{(0)}$ are very close to the true parameters. This will result in a reference exact posterior that can be used for comparison between approximate posteriors computed through the use of inaccurate model templates. From the exact posterior samples, we can identify how well we can constrain parameters given Monte-Carlo estimates of the posterior standard deviation $\Delta\theta^{i}_{\text{bf}}$ for  $\theta^{i} \in \mathbf{\Theta}$, the parameter space of the studied EMRI.  

The statistical uncertainty of parameters is dominated by the leading-order piece of the self-force, the adiabatic component. It is expected in the high-SNR regime, to a reasonable approximation, that the statistical uncertainty of waveforms exhibiting full post-adiabatic information should be similar to that of adiabatic waveforms. With this in mind, when performing inference with approximate model templates, we set restrictive prior bounds 
\begin{equation}\label{eq:prior_distribution}
    \theta^{i} \sim \theta^{i}_{\text{tr}} + 25U[-\Delta\theta^{i}_{\text{bf}}, \Delta\theta^{i}_{\text{bf}}]\,.
\end{equation}
We found the prior~\eqref{eq:prior_distribution} to be suitably wide enough to allow for proper sampling of approximate distributions, but narrow enough to stop the sampler from sampling enormous regions of parameter space where it may get lost/stuck. The prior ranges given by Eq.~\eqref{eq:prior_distribution} were essential when performing the eccentric runs presented in Sec.~\ref{sec:eccentricity}. When performing inference on the secondary spin of the compact object, we set a uniform prior $\chi \sim U[-1,1]$ and only focus on secondary black hole counterparts.

\textbf{Burn-in:} Once (all) the chains had converged to a point of sufficiently high likelihood for a sufficient number of iterations, the earlier, exploratory, samples from the chains are discarded. This is the process of ``burn-in'', which ensures that the only samples that remain are those from the correct stationary distribution. The number of samples that were discarded as burn-in was usually set as a post-processing step. After each run terminated, we would analyze each chain's log-likelihood, apply the ``thick-pen'' procedure, and discard early iterations that were used to explore the posterior. From the final set of samples, it is then possible to extract summary statistics such as the point maximum a-posteriori estimate $\boldsymbol{\theta}_{\text{bf}}$, marginalised credible intervals, posterior variance etc. 

\textbf{Computational cost:} All simulations were performed on NVIDIA a100 or a30 GPUs, whichever were available at the time. The computational cost of the algorithm largely depended on two factors: the complexity of the orbit and whether \texttt{eryn} or \texttt{emcee} was used. The circular orbit runs presented here, using \texttt{emcee}, usually finished in $\sim 6\,$hours. The eccentric runs using \texttt{eryn} took around $\sim 2\,$ days. This extra cost was due to the number of temperatures required (in our case five) for the eccentric case and also the number of modes present in the eccentric waveform summation stage of FEW.

\begin{figure*}
    \centering
    \includegraphics[width = 0.95\textwidth]{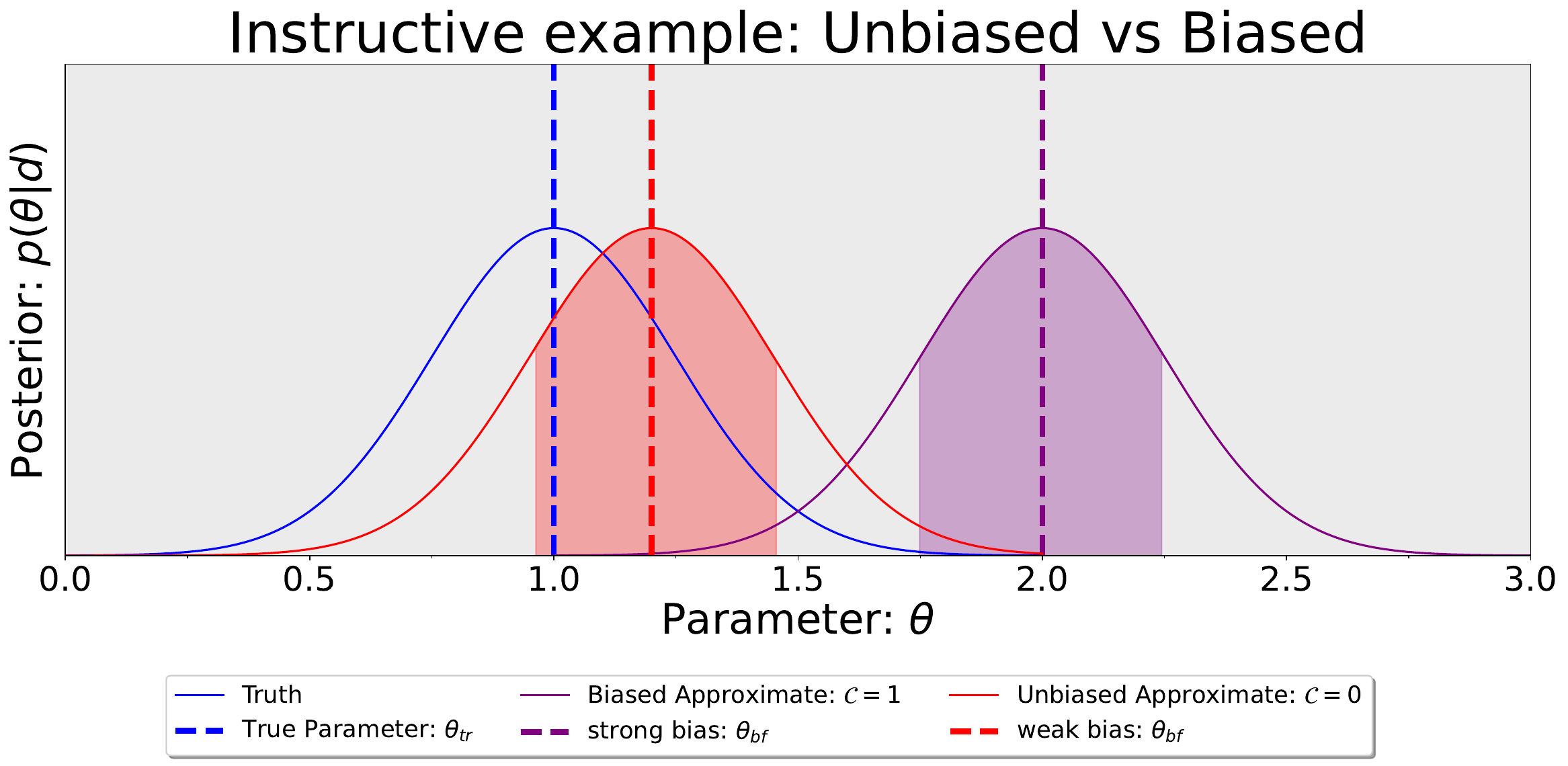}
    \caption{A simple example showing the biased criterion Eq.~\eqref{eq:systematic_ratio} in use. The blue Gaussian represents the true model posterior, where an exact waveform $h_{e}$ (with zero-noise injection) has been used to infer the true parameter $\theta = 1$.  Red and purple Gaussians represent the posterior distributions of two approximate waveforms $h_{m \text{ red}}(\theta_{tr})$ and  $h_{m \text{ purple}}(\theta_{tr})$, respectively,  with different levels of bias. The red Gaussian envelopes the true parameter within the $1\sigma$ width, or 68\% credible interval (formally the 68\% highest posterior density interval) $C_{p(\boldsymbol{\theta}|d)}$, given by the red shaded region. Hence the recovered parameter is \emph{consistent} with statistical shifts due to the noise, and the $h_{m \text{ red}}(\theta_{tr})$ waveform is suitable for data analysis. 
    On the other hand, the true parameter is not contained within the $1\sigma$ or 68\% HPDI of the purple posterior. This means that the approximate model waveform $h_{m \text{ purple}}(\theta_{tr})$ is not suitable for parameter estimation.
    }
    \label{fig:easy_example_biases}
\end{figure*}%
\section{Post-Newtonian expressions for 1PA forcing term}\label{appendix:PN}

In the 3PN model of the 1PA forcing term we use post-Newtonian expressions for the fluxes at leading and next-to-leading order in the mass-ratio. The explicit expressions we use are     
\begin{widetext}
\begin{align}
    \mathcal{F}_0&=\frac{32 x^5}{5}\Bigg[1-\frac{1247 }{336}x+4 \pi  x^{3/2}-\frac{44711 }{9072}x^2-\frac{8191}{672} \pi  x^{5/2}+ \nonumber\\ 
    &\qquad\left(\frac{6643739519}{69854400}+\frac{16 \pi ^2}{3}-\frac{1712 \gamma
   }{105}-\frac{856}{105}\log(x)-\frac{3424 \log (2)}{105}\right)x^3\Bigg], \\
   \mathcal{F}_1&=\frac{32 x^5}{5}\Bigg[-\frac{35 }{12}x+\frac{9271 }{504}x^2-\frac{583}{24} \pi  x^{5/2}+\left(\frac{41 \pi ^2}{48}-\frac{134543}{7776}\right) x^3+\frac{214745 \pi }{1728} x^{7/2}+ \nonumber \\
   &\qquad\left(-\frac{1452202403629}{1466942400}-\frac{267127 \pi ^2}{4608}+\frac{41478 \gamma }{245}+\frac{23693 \log (3)}{196}+\frac{479062 \log (2)}{2205}+\frac{20739
   }{245}\log(x)\right)x^4 \Bigg],
\end{align}
\end{widetext}
where $x=\frac{1}{\hat{r}}$ and $\gamma$ is the Euler–Mascheroni constant. Note that these expressions are only used in the 1PA forcing term $F_1$ in Eq.~\eqref{eq:radius_ev}, as written in Eq.~\eqref{eq:fix_freq_force_term}, not in the 0PA forcing term $F_0$.

\section{Detectability by LISA  of the corrections to the EMRI waveform amplitudes}
\label{appendix:importance_amplitudes}
The complex amplitudes $\mathcal{A}_{\ell m} = \mathcal{A}_{\ell m}^{(0)} + \eps \mathcal{A}_{\ell m}^{(1)}+ \mathcal{O}(\eps^2)$ in the waveform~\eqref{eq:Teukolskycircwave} can be separated in their real $\mathcal X_{\ell m}$ and imaginary part $\mathcal Y_{\ell m}$
\begin{align*}
 \mathcal X^{(0)}_{\ell m}(t) &= \Re \mathcal A^{(0)}_{\ell m}(t) \qquad \mathcal X^{(1)}_{\ell m}(t) = \Re \mathcal A^{(1)}_{\ell m}(t) \ , \\
\mathcal  Y^{(0)}_{\ell m}(t) &= \Im \mathcal A^{(0)}_{\ell m}(t) \qquad \mathcal Y^{(1)}_{\ell m}(t) = \Im \mathcal A^{(1)}_{\ell m}(t) \ . 
\end{align*}

Thus, the waveform in Eq.~\eqref{eq:Teukolskycircwave} can be expanded as
\begin{align*}
 h_+- i h_\times &= \frac{\mu}{D_{\rm S}} \displaystyle \sum_{\ell,m} \mathcal C_{\ell m}(t)Y_{\ell m}(\vartheta, \varphi)  e^{i \Psi_{\ell m}(t)}  \ , 
\end{align*}
where 
\begin{align*}
\mathcal C_{\ell m}(t) &= \mathcal C^{(0)}_{\ell m}(t) +\epsilon \mathcal C^{(1)}_{\ell m}(t) \ , \\
\mathcal C^{(0)}_{\ell m}(t)  &= \sqrt{\big(\mathcal X^{(0)}_{\ell m}(t)\big)^2 + \big(\mathcal Y^{(0)}_{\ell m}(t)\big)^2} \ , \\
\mathcal C^{(1)}_{\ell m}(t)  &=\frac{\mathcal X^{(0)}_{\ell m}(t)\mathcal X^{(1)}_{\ell m}(t) + \mathcal Y^{(0)}_{\ell m}(t) \mathcal Y^{(1)}_{\ell m}(t) }{\mathcal A^{(0)}_{\ell m}(t)}
\end{align*}
while the phase $\Psi_{\ell m}(t)$ is
\begin{equation*}
\Psi_{\ell m} = -m \Phi_{\phi}(t) + \beta^{(0)}_{\ell m}(t) + \epsilon \beta^{(1)}_{\ell m}(t)
\end{equation*}
with
\begin{align*}
\beta^{(0)}_{\ell m}(t) &= \arctan \!\Bigg(\frac{\mathcal Y^{(0)}_{\ell m}(t)}{\mathcal X^{(0)}_{\ell m}(t)}\Bigg) \ , \\
\beta^{(1)}_{\ell m}(t) &= \frac{\mathcal X^{(0)}_{\ell m}(t)\mathcal Y^{(1)}_{\ell m}(t) - \mathcal X^{(1)}_{\ell m}(t) \mathcal Y^{(0)}_{\ell m}(t) }{\big(\mathcal A^{(0)}_{\ell m}(t) \big)^2} \ . 
\end{align*}
It is then easy to see that the phase term $\beta^{(1)}_{\ell m}(t)$ of the complex amplitude correction $\mathcal{A}_{\ell m}^{(1)}$ is a 2PA term contribution to the overall waveform phase $\Psi_{\ell m}(t)$. By expanding $\Psi_{\ell m}(t)$ in a two-time scale expansion~\cite{Pound:2021qin}
\begin{equation*}
\Psi_{\ell m}(t) = -\frac{m}{\epsilon}\Phi^{(0)}_{\phi}(\tilde t) - m\Phi^{(1)}_{\phi}(\tilde t) + \beta^{(0)}_{\ell m}(\tilde t) + \epsilon \beta^{(1)}_{\ell m}(\tilde t)
\end{equation*}
with $\bar t = \epsilon t$, while $\Phi^{(0)}_{\phi}(\bar t)$ and $\Phi^{(1)}_{\phi}(\bar t)$ are respectively the 0PA and 1PA contributions from the orbital phase. We can substitute $t \to \bar t$ in  $\beta^{(0)}_{\ell m}(t)$ and  $\beta^{(1)}_{\ell m}(t)$ in the previous expression since the complex amplitudes $\mathcal A_{\ell m}$ are slow varying quantities:
\begin{align}
 \bigg|\totder{\mathcal X^{(0)}_{\ell m}(t)}{t}  \bigg| &=  \bigg|\frac{\partial \mathcal X_{\ell m}}{\partial r} \totder{r}{t} \bigg| \sim \epsilon\bigg|\frac{\partial \mathcal X_{\ell m}}{\partial r} F_0(\hat{r} )  \bigg| \ll \Big| \totder{\Phi_\phi}{t} \Big|  \ , \label{eq:slowevolve1} \\
\bigg|\totder{\mathcal Y^{(0)}_{\ell m}(t)}{t}  \bigg| &=  \bigg|\frac{\partial \mathcal Y_{\ell m}}{\partial r} \totder{r}{t} \bigg| \sim \epsilon\bigg|\frac{\partial \mathcal Y_{\ell m}}{\partial r} F_0(r)  \bigg| \ll \Big|\totder{\Phi_\phi}{t} \Big|  \ , \label{eq:slowevolve2}
\end{align}
Thus, the phase terms $\beta^{(0)}_{\ell m}(t)$ and $\beta^{(1)}_{\ell m}(t)$ enter at 1PA and 2PA order, respectively, in the waveform phase.

Moreover, using the Lindblom criterion~\cite{lindblom2008model}, we can now show that the real amplitudes $C^{(1)}_{\ell m}(t)$ in EMRIs (and IMRIs) are practically undetectable. With the aid of the Stationary Phase Approximation (SPA)~\cite{Hughes:2021exa} together with the conditions~\eqref{eq:slowevolve1} and~\eqref{eq:slowevolve2}, we can approximate the Fourier transform of the time-domain waveform~\eqref{eq:Teukolskycircwave}, schematically, as
\begin{equation}
\tilde h = \frac{\mu}{D_{\rm S}}\sum_{\ell, m}^{}Y_{\ell m}(\vartheta, \varphi)e^{ i (2\pi f  \tilde t(f) - \Psi_{\ell m}(f))} \mathcal B_{\ell m}(f) \mathcal C_{\ell m}(f) \label{eq:1PASPAwaveform} \ ,
\end{equation}
where $\mathcal B_{\ell m}(f)$ are real functions due to the SPA~\cite{Hughes:2021exa}.
Let us now define two waveforms:
\begin{align}
\tilde h_0 &= \frac{\mu}{D_{\rm S}}\sum_{\ell, m}^{}Y_{\ell m}(\vartheta, \varphi)e^{ i (2\pi f  \tilde t(f) - \Phi_{\ell m}(f))} \mathcal B_{\ell m}(f)\mathcal C^{(0)}_{\ell m}(f)  \label{eq:1PASPAwaveform0} \ , \\
\tilde h_1 &= \tilde h_0 + \epsilon \delta \tilde h
\end{align}
where
\begin{equation}
 \delta \tilde h = \frac{\mu}{D} \sum_{\ell m}^{}Y_{\ell m}(\vartheta, \varphi)e^{ i (2\pi f  \tilde t(f) - \Phi_{\ell m}(f))} \mathcal B_{\ell m}(f) \mathcal C^{(1)}_{\ell m}(f)  \label{eq:1PASPAwaveform1}    
\end{equation}

We can then write
\begin{align}
\mathcal M &= 1 - \frac{\langle  h_1 |  h_0 \rangle}{\sqrt{\langle  h_0 |  h_0 \rangle \langle  h_1 | h_1 \rangle}} \nonumber \\ 
 &= 1 - \frac{\langle  h_0 |  h_0 \rangle+ \epsilon\langle  h_0 |  \delta h \rangle}{\sqrt{\langle  h_0 |  h_0 \rangle} \sqrt{\langle  h_0 |  h_0 \rangle + 2 \epsilon \langle  \delta h | h_0 \rangle +  \epsilon^2 \langle  \delta h |  \delta h \rangle)}}   \nonumber \\
& \simeq \epsilon^2 \frac{ \langle  h_0 |  h_0 \rangle  \langle \delta h |  \delta h\rangle - \langle  \delta h | h_0 \rangle^2}{2 \langle  h_0 |  h_0 \rangle^2 } + \mathcal O(\epsilon^3) \nonumber \\
&\simeq \frac{\epsilon^2}{2} + \mathcal O(\epsilon^3)
\end{align}
Thus, the mismatch $\mathcal M$ between $\tilde h_0(f)$  and $\tilde h_1(f)$ is
\begin{equation}
\mathcal M = 1 -  \frac{\langle  h_1 |  h_0 \rangle}{\sqrt{\langle  h_0 |  h_0 \rangle \langle  h_1 | h_1 \rangle}}  \simeq \frac{\epsilon^2}{2}
\end{equation}
and according to the Lindblom criterion , the required SNR to observe the real amplitude $\mathcal A^{(1)}_{\ell m}$ is
\begin{equation}
\rho \geq \frac{1}{\sqrt{2 \mathcal M}} \simeq \frac{1}{\epsilon}
\end{equation}
For EMRIs, the largest expected $\epsilon$ is $\sim 10^{-5}$, while the loudest SNR for a ``golden'' source is $\sim 100$. It is then unlikely to detect $\mathcal C^{(1)}_{\ell m}$ for astrophysically relevant distances.

\begin{widetext}

\section{Summary statistics}

\begin{table}[ht!]
\begin{tabular}{l|c|c|c}
\multicolumn{1}{c|}{\textbf{Model Waveform: $\epsilon = 10^{-5}$}} & \textbf{Parameter} & \textbf{Posterior Median} & \textbf{68\% HPDI} \\ \hline \hline
                                                                   & $M/M_{\odot}$      & 1000001.39                & (999997.74, 1000005.30)                          \\
\multicolumn{1}{c|}{Cir1PA w/ spin}                                & $\mu/M_{\odot}$    & 10.000000842              & (9.999937 10.000062)                             \\
                                                                   & $r_{0}/M$          & 10.602490                 & (10.602461, 10.602518)    \\
                                                                   & $\chi$             & 0.016568    & (-0.675632, 0.681355)
                                                                   \\\hline\hline
                                                                   & $M/M_{\odot}$      & 1000001.77                & (999998.55, 1000005.05)                          \\ 
\multicolumn{1}{c|}{Cir1PA w/o spin}                               & $\mu/M_{\odot}$    & 9.999995                  & (9.999930, 10.000057)                            \\
                                                                   & $r_{0}/M$          & 10.602488                 & (10.602462, 10.602512)                           \\ \hline\hline
                                                                   & $M/M_{\odot}$      & 1000000.90                & (999997.73, 1000004.19)                          \\
\multicolumn{1}{c|}{Cir0PA + 1PA-3PN w/o spin}                     & $\mu/M_{\odot}$    & 10.00000878               & (9.999944, 10.0000719)                           \\
                                                                   & $r_{0}/M$          & 10.602493                 & (10.6024682, 10.602518)                          \\ \hline\hline
                                                                   & $M/M_{\odot}$      & 999989.96                 & (999986.61, 999993.32)                           \\
\multicolumn{1}{c|}{Cir0PA w/o spin}                               & $\mu/M_{\odot}$    & 10.00007160               & (10.000006, 10.000137)                           \\
                                                                   & $r_{0}/M$          & 10.602571                 & (10.602545, 10.602596)                          
\end{tabular}
\caption{ Summary statistics for Config.(1) in Table \ref{ref:EMRI_parameters}. The  posterior distributions are shown in the first row of Fig. \ref{fig:Summary_all_mass_ratios_spin}. The first column present waveform models, the second column parameter choices, the third column estimates of the posterior median and the final column the 68\% highest posterior density interval. Notice here that the true parameters are contained in all the 68\% HPDIs apart from the cir0PA w/o spin waveform model. This means that the purely adiabatic model w/o spin is unsuitable for parameter estimation purposes. For all choices of waveform models, the true extrinsic parameters within their approximate posterior 68\% HPDI.}
\label{tab:table_eps_1e-5}
\end{table}

\begin{table}[hb!]
\begin{tabular}{l|c|c|c}
\multicolumn{1}{c|}{\textbf{Model Waveform: $\epsilon = 10^{-4}$}} & \textbf{Parameter} & \textbf{Posterior Median} & \textbf{68\% HPDI} \\ \hline \hline
                                                                   & $M/M_{\odot}$      & 1000001.11               & (999995.09, 1000016.51)                        \\
\multicolumn{1}{c|}{Cir1PA w/ spin}                                & $\mu/M_{\odot}$    & 99.999856             & (99.998832, 100.000625)                            \\
                                                                   & $r_{0}/M$          & 15.790489                &  (15.790330, 15.790555)                          \\ 
                                                                   & $\chi$             & 0.449103    & (0.0500912, 0.692787) \\
                                                                   \hline\hline
                                                                   & $M/M_{\odot}$      & 1000014.32               & (1000006.43, 1000022.81)                         \\ 
\multicolumn{1}{c|}{Cir1PA w/o spin}                               & $\mu/M_{\odot}$    & 99.999749                 & (99.998684, 100.000757)                           \\
                                                                   & $r_{0}/M$          & 15.790353                 & (15.790258, 15.790445)                        \\ \hline\hline
                                                                   & $M/M_{\odot}$      & 1000012.55               & (1000003.67, 1000021.45)                        \\
\multicolumn{1}{c|}{Cir0PA + 1PA-3PN w/o spin}                     & $\mu/M_{\odot}$    & 100.000098               & (99.999049, 100.001168)                          \\
                                                                   & $r_{0}/M$          & 15.790375                 & (15.790276, 15.790475)                        \\ \hline\hline
                                                                   & $M/M_{\odot}$      & 999908.159826                & (999901.34, 999916.69)                          \\
\multicolumn{1}{c|}{Cir0PA w/o spin}                               & $\mu/M_{\odot}$    & 100.005805             & (100.004816, 100.006748)                         \\
                                                                   & $r_{0}/M$          & 15.791467                & (15.791373, 15.791547)                       
\end{tabular}
\caption{The same as Table \ref{tab:table_eps_1e-5} but for Config.(2) of Table \ref{ref:EMRI_parameters}. The  posterior distributions are shown in the second row of Fig. \ref{fig:Summary_all_mass_ratios_spin}. Notice here that no approximate waveform models are suitable for parameter estimation. The true intrinsic parameters under Config.(2) lie outside the approximate 68\% HPDI.}
\label{tab:table_eps_1e-4}
\end{table}

\begin{table*}[]
\begin{tabular}{l|c|c|c}
\multicolumn{1}{c|}{\textbf{Model Waveform: $\epsilon = 10^{-3}$}} & \textbf{Parameter} & \textbf{Posterior Median} & \textbf{68\% HPDI} \\ \hline \hline
                                                                   & $M/M_{\odot}$      & 5000415.91              & (4999481.91, 5001638.03)                       \\
\multicolumn{1}{c|}{Cir1PA w/ spin}                                & $\mu/M_{\odot}$    & 4999.802             & (4999.201, 5000.253)                           \\
                                                                   & $r_{0}/M$          &  16.811376               &  (16.808657, 16.813453)   \\                     
                                                                   & $\chi$             & 0.297799                    & $(-0.288256, 0.749646)$ \\ \hline\hline
                                                                   & $M/M_{\odot}$      & 5001035.32              & (5000965.94, 5001105.69)                        \\ 
\multicolumn{1}{c|}{Cir1PA w/o spin}                               & $\mu/M_{\odot}$    &  4999.499             &  (4999.411, 4999.589)                           \\
                                                                   & $r_{0}/M$          &  16.810000                & (16.809826, 16.810171)                        \\ \hline\hline
                                                                   & $M/M_{\odot}$      & 5000958.72               & (5000889.88, 5001026.58)                       \\
\multicolumn{1}{c|}{Cir0PA + 1PA-3PN w/o spin}                     & $\mu/M_{\odot}$    & 4999.651              & (4999.564, 4999.738)                         \\
                                                                   & $r_{0}/M$          & 16.810197              & (16.810031, 16.810366)                      \\ \hline\hline
                                                                   & $M/M_{\odot}$      & 4995324.90               & (4995254.22, 4995394.23)           \\
\multicolumn{1}{c|}{Cir0PA w/o spin}                               & $\mu/M_{\odot}$    &  5003.051           & (5002.96, 5003.14)                       \\
                                                                   & $r_{0}/M$          & 16.822834              &  (16.822662 16.823007)                       
\end{tabular}
\caption{The same as Table \ref{tab:table_eps_1e-5} but for Config.(3) of Table \ref{ref:EMRI_parameters}. The  posterior distributions are shown in the third row of Fig. \ref{fig:Summary_all_mass_ratios_spin}. Making reference to the first row of Fig.\ref{fig:Summary_all_mass_ratios_spin}, we see that the distribution is no longer normal. Such feature is liekly due to correlations with the secondary spin, where the excess Kurtosis of the $\{M, \mu, r_{0}, \chi\}$ distributions are each $\approx -0.5$ with other (extrinsic parameters) $\sim \pm 0.01$. The quoted 68\% HPDI demonstrates a significant posterior widening when sampling over the secondary spin. We can also see that the approximate (w/o spin) models 68\% HPDIs are $\sim 1$ order of magnitude tighter than the true 68\% HPDI given by the cir1PA w/ spin model. } 
\label{tab:table_eps_1e-3}
\end{table*}

\end{widetext}

\clearpage
\newpage

\begin{widetext}
\section{Wide prior on $\chi$}\label{app:corner_plot_q1e3_wide_prior}

\begin{figure*}[hbpt!]
    \centering
    \includegraphics[width = \textwidth]{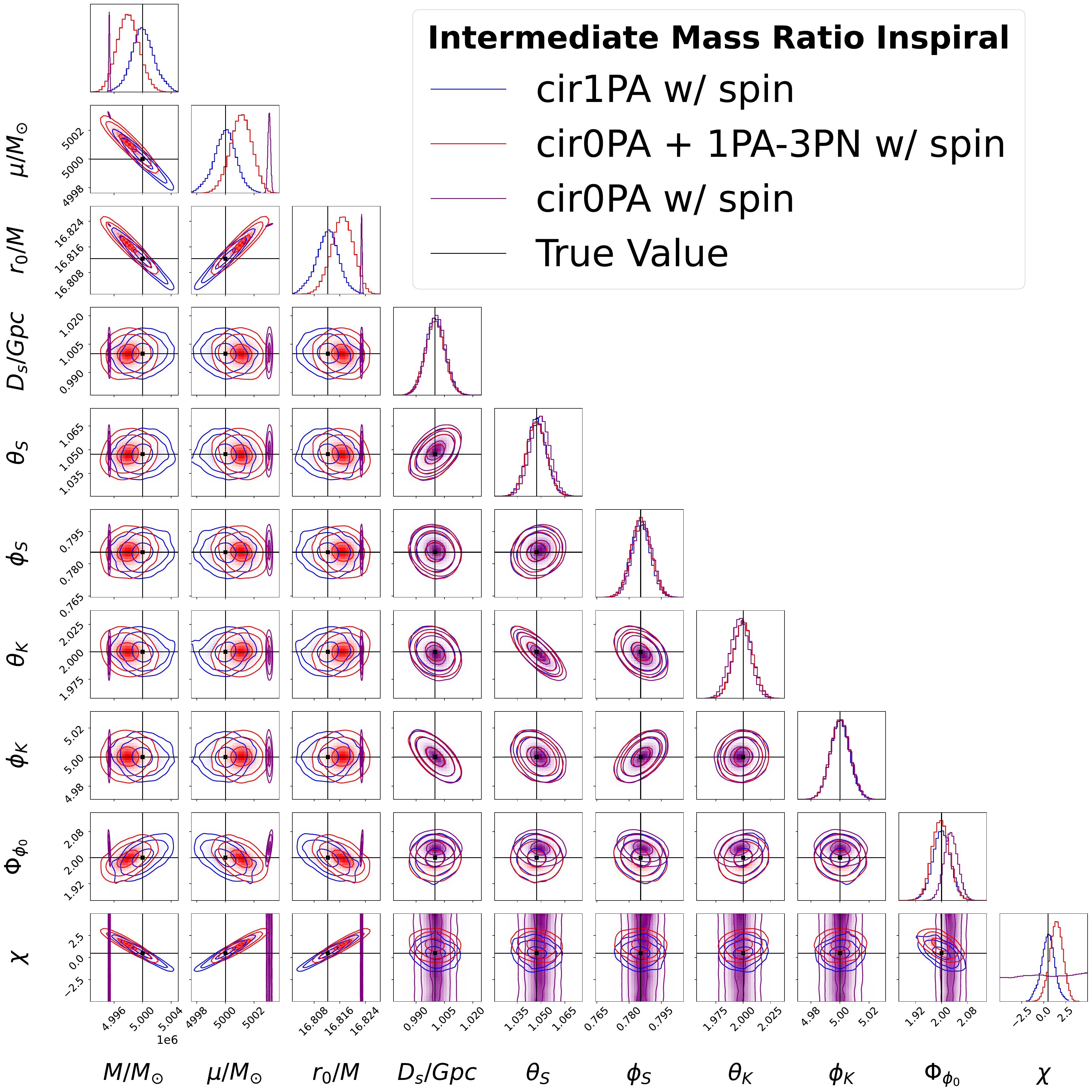}
    \caption{\textbf{(Mass ratio: $\epsilon = 10^{-3}$
):} The same set up as Fig.~\ref{fig:corner_q_1e3_spin} but with a wider prior on $\chi \sim U[-5,5]$.}
    \label{fig:circular_wide_prior_q1e3}
\end{figure*}
\end{widetext}

\bibliographystyle{IEEEtran}
\bibliography{1PA_project}

\begin{thebibliography}{100}
\providecommand{\url}[1]{#1}
\csname url@samestyle\endcsname
\providecommand{\newblock}{\relax}
\providecommand{\bibinfo}[2]{#2}
\providecommand{\BIBentrySTDinterwordspacing}{\spaceskip=0pt\relax}
\providecommand{\BIBentryALTinterwordstretchfactor}{4}
\providecommand{\BIBentryALTinterwordspacing}{\spaceskip=\fontdimen2\font plus
\BIBentryALTinterwordstretchfactor\fontdimen3\font minus \fontdimen4\font\relax}
\providecommand{\BIBforeignlanguage}[2]{{%
\expandafter\ifx\csname l@#1\endcsname\relax
\typeout{** WARNING: IEEEtran.bst: No hyphenation pattern has been}%
\typeout{** loaded for the language `#1'. Using the pattern for}%
\typeout{** the default language instead.}%
\else
\language=\csname l@#1\endcsname
\fi
#2}}
\providecommand{\BIBdecl}{\relax}
\BIBdecl

\bibitem{barack2007using}
L.~Barack and C.~Cutler, ``Using lisa extreme-mass-ratio inspiral sources to test off-kerr deviations in the geometry of massive black holes,'' \emph{Physical Review D}, vol.~75, no.~4, p. 042003, 2007.

\bibitem{gair2013testing}
J.~R. Gair, M.~Vallisneri, S.~L. Larson, and J.~G. Baker, ``Testing general relativity with low-frequency, space-based gravitational-wave detectors,'' \emph{Living Reviews in Relativity}, vol.~16, pp. 1--109, 2013.

\bibitem{Maselli:2020zgv}
A.~Maselli, N.~Franchini, L.~Gualtieri, and T.~P. Sotiriou, ``{Detecting scalar fields with Extreme Mass Ratio Inspirals},'' \emph{Phys. Rev. Lett.}, vol. 125, no.~14, p. 141101, 2020.

\bibitem{Maselli:2021men}
A.~Maselli, N.~Franchini, L.~Gualtieri, T.~P. Sotiriou, S.~Barsanti, and P.~Pani, ``{Detecting fundamental fields with LISA observations of gravitational waves from extreme mass-ratio inspirals},'' \emph{Nature Astron.}, vol.~6, no.~4, pp. 464--470, 2022.

\bibitem{Barsanti:2022ana}
S.~Barsanti, N.~Franchini, L.~Gualtieri, A.~Maselli, and T.~P. Sotiriou, ``{Extreme mass-ratio inspirals as probes of scalar fields: eccentric equatorial orbits around Kerr black holes},'' 3 2022.

\bibitem{barack2018self}
L.~Barack and A.~Pound, ``Self-force and radiation reaction in general relativity,'' \emph{Reports on Progress in Physics}, vol.~82, no.~1, p. 016904, 2018.

\bibitem{Pound:2021qin}
A.~Pound and B.~Wardell, ``Black hole perturbation theory and gravitational self-force,'' \emph{Handbook of Gravitational Wave Astronomy}, pp. 1--119, 2022.

\bibitem{Speri:2023jte}
L.~Speri, M.~L. Katz, A.~J. Chua, S.~A. Hughes, N.~Warburton, J.~E. Thompson, C.~E. Chapman-Bird, and J.~R. Gair, ``Fast and fourier: Extreme mass ratio inspiral waveforms in the frequency domain,'' \emph{arXiv preprint arXiv:2307.12585}, 2023.

\bibitem{katz2021fast}
M.~L. Katz, A.~J. Chua, L.~Speri, N.~Warburton, and S.~A. Hughes, ``Fast extreme-mass-ratio-inspiral waveforms: New tools for millihertz gravitational-wave data analysis,'' \emph{Physical Review D}, vol. 104, no.~6, p. 064047, 2021.

\bibitem{chua2021rapid}
A.~J. Chua, M.~L. Katz, N.~Warburton, and S.~A. Hughes, ``Rapid generation of fully relativistic extreme-mass-ratio-inspiral waveform templates for lisa data analysis,'' \emph{Physical Review Letters}, vol. 126, no.~5, p. 051102, 2021.

\bibitem{VanDeMeent:2018cgn}
M.~Van De~Meent and N.~Warburton, ``{Fast Self-forced Inspirals},'' \emph{Class. Quant. Grav.}, vol.~35, no.~14, p. 144003, 2018.

\bibitem{McCart:2021upc}
J.~McCart, T.~Osburn, and J.~Y.~J. Burton, ``{Highly eccentric extreme-mass-ratio-inspiral waveforms via fast self-forced inspirals},'' \emph{Phys. Rev. D}, vol. 104, no.~8, p. 084050, 2021.

\bibitem{Lynch:2021ogr}
{Lynch, Philip and van de Meent, Maarten and Warburton, Niels}, ``{Eccentric self-forced inspirals into a rotating black hole},'' \emph{Class. Quant. Grav.}, vol.~39, no.~14, p. 145004, 2022.

\bibitem{Lynch:2023gpu}
P.~Lynch, M.~van~de Meent, and N.~Warburton, ``Self-forced inspirals with spin-orbit precession,'' \emph{arXiv preprint arXiv:2305.10533}, 2023.

\bibitem{Isoyama:2021jjd}
S.~Isoyama, R.~Fujita, A.~J.~K. Chua, H.~Nakano, A.~Pound, and N.~Sago, ``{Adiabatic Waveforms from Extreme-Mass-Ratio Inspirals: An Analytical Approach},'' \emph{Phys. Rev. Lett.}, vol. 128, no.~23, p. 231101, 2022.

\bibitem{Gupta:2022fbe}
P.~Gupta, L.~Speri, B.~Bonga, A.~J.~K. Chua, and T.~Tanaka, ``{Modeling transient resonances in extreme-mass-ratio inspirals},'' \emph{Phys. Rev. D}, vol. 106, no.~10, p. 104001, 2022.

\bibitem{speri2021assessing}
L.~Speri and J.~R. Gair, ``Assessing the impact of transient orbital resonances,'' \emph{Physical Review D}, vol. 103, no.~12, p. 124032, 2021.

\bibitem{Lynch2022}
P.~{Lynch}, ``{Efficient trajectory calculations for extreme mass-ratio inspirals using near-identity (averaging) transformations},'' Ph.D. dissertation, University College Dublin, 2022, available electronically at http://hdl.handle.net/10197/13347.

\bibitem{Akcay:2019bvk}
S.~Akcay, S.~R. Dolan, C.~Kavanagh, J.~Moxon, N.~Warburton, and B.~Wardell, ``{Dissipation in extreme-mass ratio binaries with a spinning secondary},'' \emph{Phys. Rev. D}, vol. 102, no.~6, p. 064013, 2020.

\bibitem{Piovano:2021iwv}
G.~A. Piovano, R.~Brito, A.~Maselli, and P.~Pani, ``{Assessing the detectability of the secondary spin in extreme mass-ratio inspirals with fully relativistic numerical waveforms},'' \emph{Phys. Rev. D}, vol. 104, no.~12, p. 124019, 2021.

\bibitem{Skoupy:2021asz}
V.~Skoup\'y and G.~Lukes-Gerakopoulos, ``{Spinning test body orbiting around a Kerr black hole: Eccentric equatorial orbits and their asymptotic gravitational-wave fluxes},'' \emph{Phys. Rev. D}, vol. 103, no.~10, p. 104045, 2021.

\bibitem{Mathews:2021rod}
J.~Mathews, A.~Pound, and B.~Wardell, ``{Self-force calculations with a spinning secondary},'' \emph{Phys. Rev. D}, vol. 105, no.~8, p. 084031, 2022.

\bibitem{Drummond:2022xej}
L.~V. Drummond and S.~A. Hughes, ``{Precisely computing bound orbits of spinning bodies around black holes. I. General framework and results for nearly equatorial orbits},'' \emph{Phys. Rev. D}, vol. 105, no.~12, p. 124040, 2022.

\bibitem{Drummond:2022efc}
{Drummond, Lisa V. and Hughes, Scott A.}, ``{Precisely computing bound orbits of spinning bodies around black holes. II. Generic orbits},'' \emph{Phys. Rev. D}, vol. 105, no.~12, p. 124041, 2022.

\bibitem{Skoupy:2023lih}
V.~Skoup{\`y}, G.~Lukes-Gerakopoulos, L.~V. Drummond, and S.~A. Hughes, ``Asymptotic gravitational-wave fluxes from a spinning test body on generic orbits around a kerr black hole,'' \emph{arXiv preprint arXiv:2303.16798}, 2023.

\bibitem{Hughes:2021exa}
S.~A. Hughes, N.~Warburton, G.~Khanna, A.~J.~K. Chua, and M.~L. Katz, ``{Adiabatic waveforms for extreme mass-ratio inspirals via multivoice decomposition in time and frequency},'' \emph{Phys. Rev. D}, vol. 103, no.~10, p. 104014, 2021, [Erratum: Phys.Rev.D 107, 089901 (2023)].

\bibitem{vandeMeent:2017bcc}
M.~van~de Meent, ``{Gravitational self-force on generic bound geodesics in Kerr spacetime},'' \emph{Phys. Rev. D}, vol.~97, no.~10, p. 104033, 2018.

\bibitem{Wardell:2021fyy}
B.~Wardell, A.~Pound, N.~Warburton, J.~Miller, L.~Durkan, and A.~Le~Tiec, ``{Gravitational Waveforms for Compact Binaries from Second-Order Self-Force Theory},'' \emph{Phys. Rev. Lett.}, vol. 130, no.~24, p. 241402, 2023.

\bibitem{MockLISADataChallengeTaskForce:2009wir}
S.~Babak \emph{et~al.}, ``{The Mock LISA Data Challenges: From Challenge 3 to Challenge 4},'' \emph{Class. Quant. Grav.}, vol.~27, p. 084009, 2010.

\bibitem{Chua:2019wgs}
A.~J.~K. Chua, N.~Korsakova, C.~J. Moore, J.~R. Gair, and S.~Babak, ``{Gaussian processes for the interpolation and marginalization of waveform error in extreme-mass-ratio-inspiral parameter estimation},'' \emph{Phys. Rev. D}, vol. 101, no.~4, p. 044027, 2020.

\bibitem{chua2022nonlocal}
A.~J. Chua and C.~J. Cutler, ``Nonlocal parameter degeneracy in the intrinsic space of gravitational-wave signals from extreme-mass-ratio inspirals,'' \emph{Physical Review D}, vol. 106, no.~12, p. 124046, 2022.

\bibitem{gair2004event}
J.~R. Gair, L.~Barack, T.~Creighton, C.~Cutler, S.~L. Larson, E.~S. Phinney, and M.~Vallisneri, ``Event rate estimates for lisa extreme mass ratio capture sources,'' \emph{Classical and Quantum Gravity}, vol.~21, no.~20, p. S1595, 2004.

\bibitem{Babak:17aa}
S.~{Babak}, J.~{Gair}, A.~{Sesana}, E.~{Barausse}, C.~F. {Sopuerta}, C.~P.~L. {Berry}, E.~{Berti}, P.~{Amaro-Seoane}, A.~{Petiteau}, and A.~{Klein}, ``{Science with the space-based interferometer LISA. V. Extreme mass-ratio inspirals},'' \emph{prd}, vol.~95, no.~10, p. 103012, May 2017.

\bibitem{burke2021extreme}
A.~I. Burke and O.~Burke, ``Extreme precision and extreme complexity: source modelling and data analysis development for the laser interferometer space antenna,'' Ph.D. dissertation, The University of Edinburgh, 2022, available electronically at https://era.ed.ac.uk/handle/1842/38372.

\bibitem{Hinderer:2008dm}
T.~Hinderer and E.~E. Flanagan, ``{Two timescale analysis of extreme mass ratio inspirals in Kerr. I. Orbital Motion},'' \emph{Phys. Rev. D}, vol.~78, p. 064028, 2008.

\bibitem{Fujita:2009us}
R.~Fujita, W.~Hikida, and H.~Tagoshi, ``{An Efficient Numerical Method for Computing Gravitational Waves Induced by a Particle Moving on Eccentric Inclined Orbits around a Kerr Black Hole},'' \emph{Prog. Theor. Phys.}, vol. 121, pp. 843--874, 2009.

\bibitem{Flanagan:2012kg}
E.~E. Flanagan, S.~A. Hughes, and U.~Ruangsri, ``{Resonantly enhanced and diminished strong-field gravitational-wave fluxes},'' \emph{Phys. Rev. D}, vol.~89, no.~8, p. 084028, 2014.

\bibitem{LISA:2022kgy}
K.~G. Arun \emph{et~al.}, ``{New horizons for fundamental physics with LISA},'' \emph{Living Rev. Rel.}, vol.~25, no.~1, p.~4, 2022.

\bibitem{LISA:2022yao}
P.~A. Seoane \emph{et~al.}, ``{Astrophysics with the Laser Interferometer Space Antenna},'' \emph{Living Rev. Rel.}, vol.~26, no.~1, p.~2, 2023.

\bibitem{Miller:2020bft}
J.~Miller and A.~Pound, ``{Two-timescale evolution of extreme-mass-ratio inspirals: waveform generation scheme for quasicircular orbits in {S}chwarzschild spacetime},'' \emph{Phys. Rev. D}, vol. 103, no.~6, p. 064048, 2021.

\bibitem{Pound:2019lzj}
A.~Pound, B.~Wardell, N.~Warburton, and J.~Miller, ``{Second-Order Self-Force Calculation of Gravitational Binding Energy in Compact Binaries},'' \emph{Phys. Rev. Lett.}, vol. 124, no.~2, p. 021101, 2020.

\bibitem{Warburton:2021kwk}
N.~Warburton, A.~Pound, B.~Wardell, J.~Miller, and L.~Durkan, ``{Gravitational-Wave Energy Flux for Compact Binaries through Second Order in the Mass Ratio},'' \emph{Phys. Rev. Lett.}, vol. 127, no.~15, p. 151102, 2021.

\bibitem{Albertini:2022rfe}
A.~Albertini, A.~Nagar, A.~Pound, N.~Warburton, B.~Wardell, L.~Durkan, and J.~Miller, ``{Comparing second-order gravitational self-force, numerical relativity, and effective one body waveforms from inspiralling, quasicircular, and nonspinning black hole binaries},'' \emph{Phys. Rev. D}, vol. 106, no.~8, p. 084061, 2022.

\bibitem{barack2004lisa}
L.~Barack and C.~Cutler, ``Lisa capture sources: Approximate waveforms, signal-to-noise ratios, and parameter estimation accuracy,'' \emph{Physical Review D}, vol.~69, no.~8, p. 082005, 2004.

\bibitem{babak2007kludge}
S.~Babak, H.~Fang, J.~R. Gair, K.~Glampedakis, and S.~A. Hughes, ``“kludge” gravitational waveforms for a test-body orbiting a kerr black hole,'' \emph{Physical Review D}, vol.~75, no.~2, p. 024005, 2007.

\bibitem{chua2017augmented}
A.~J. Chua, C.~J. Moore, and J.~R. Gair, ``Augmented kludge waveforms for detecting extreme-mass-ratio inspirals,'' \emph{Physical Review D}, vol.~96, no.~4, p. 044005, 2017.

\bibitem{lindblom2008model}
L.~Lindblom, B.~J. Owen, and D.~A. Brown, ``Model waveform accuracy standards for gravitational wave data analysis,'' \emph{Physical Review D}, vol.~78, no.~12, p. 124020, 2008.

\bibitem{vallisneri2008use}
M.~Vallisneri, ``Use and abuse of the fisher information matrix in the assessment of gravitational-wave parameter-estimation prospects,'' \emph{Physical Review D}, vol.~77, no.~4, p. 042001, 2008.

\bibitem{Piovano:2022ojl}
G.~A. Piovano, A.~Maselli, and P.~Pani, ``{Constraining the tidal deformability of supermassive objects with extreme mass ratio inspirals and semianalytical frequency-domain waveforms},'' \emph{Phys. Rev. D}, vol. 107, no.~2, p. 024021, 2023.

\bibitem{gupta2022modeling}
P.~Gupta, L.~Speri, B.~Bonga, A.~J. Chua, and T.~Tanaka, ``Modeling transient resonances in extreme-mass-ratio inspirals,'' \emph{Physical Review D}, vol. 106, no.~10, p. 104001, 2022.

\bibitem{moore2017gravitational}
C.~J. Moore, A.~J. Chua, and J.~R. Gair, ``Gravitational waves from extreme mass ratio inspirals around bumpy black holes,'' \emph{Classical and Quantum Gravity}, vol.~34, no.~19, p. 195009, 2017.

\bibitem{maselli2022detecting}
A.~Maselli, N.~Franchini, L.~Gualtieri, T.~P. Sotiriou, S.~Barsanti, and P.~Pani, ``Detecting fundamental fields with lisa observations of gravitational waves from extreme mass-ratio inspirals,'' \emph{Nature Astronomy}, vol.~6, no.~4, pp. 464--470, 2022.

\bibitem{cutler2007lisa}
C.~Cutler and M.~Vallisneri, ``Lisa detections of massive black hole inspirals: Parameter extraction errors due to inaccurate template waveforms,'' \emph{Physical Review D}, vol.~76, no.~10, p. 104018, 2007.

\bibitem{Osburn:2015duj}
T.~Osburn, N.~Warburton, and C.~R. Evans, ``{Highly eccentric inspirals into a black hole},'' \emph{Phys. Rev. D}, vol.~93, no.~6, p. 064024, 2016.

\bibitem{Warburton:2017sxk}
N.~Warburton, T.~Osburn, and C.~R. Evans, ``{Evolution of small-mass-ratio binaries with a spinning secondary},'' \emph{Phys. Rev. D}, vol.~96, no.~8, p. 084057, 2017.

\bibitem{Drummond:2023loz}
L.~V. Drummond, A.~G. Hanselman, D.~R. Becker, and S.~A. Hughes, ``Extreme mass-ratio inspiral of a spinning body into a kerr black hole i: Evolution along generic trajectories,'' \emph{arXiv preprint arXiv:2305.08919}, 2023.

\bibitem{katz2022assessing}
M.~L. Katz, J.-B. Bayle, A.~J. Chua, and M.~Vallisneri, ``Assessing the data-analysis impact of lisa orbit approximations using a gpu-accelerated response model,'' \emph{Physical Review D}, vol. 106, no.~10, p. 103001, 2022.

\bibitem{martens2021trajectory}
W.~Martens and E.~Joffre, ``Trajectory design for the esa lisa mission,'' \emph{The Journal of the Astronautical Sciences}, vol.~68, no.~2, pp. 402--443, 2021.

\bibitem{LISAsr:18aa}
\BIBentryALTinterwordspacing
{LISA Science Study Team}, ``Lisa science requirements document, esa-l3-est-sci-rs-001. technical report 1.0,'' ESA, Tech. Rep., May 2018. [Online]. Available: \url{https://www.cosmos.esa.int/web/lisa/ lisa-documents/}
\BIBentrySTDinterwordspacing

\bibitem{mandel2009can}
I.~Mandel and J.~R. Gair, ``Can we detect intermediate mass ratio inspirals?'' \emph{Classical and Quantum Gravity}, vol.~26, no.~9, p. 094036, 2009.

\bibitem{mandel2008rates}
I.~Mandel, D.~A. Brown, J.~R. Gair, and M.~C. Miller, ``Rates and characteristics of intermediate mass ratio inspirals detectable by advanced ligo,'' \emph{The Astrophysical Journal}, vol. 681, no.~2, p. 1431, 2008.

\bibitem{vandeMeent:2020xgc}
M.~van~de Meent and H.~P. Pfeiffer, ``{Intermediate mass-ratio black hole binaries: Applicability of small mass-ratio perturbation theory},'' \emph{Phys. Rev. Lett.}, vol. 125, no.~18, p. 181101, 2020.

\bibitem{BHPToolkit}
``Black hole perturbation toolkit,'' (\href{http://bhptoolkit.org/}{bhptoolkit.org}), 2018.

\bibitem{Munna:2020juq}
C.~Munna, C.~R. Evans, S.~Hopper, and E.~Forseth, ``{Determination of new coefficients in the angular momentum and energy fluxes at infinity to 9PN order for eccentric Schwarzschild extreme-mass-ratio inspirals using mode-by-mode fitting},'' \emph{Phys. Rev. D}, vol. 102, no.~2, p. 024047, 2020.

\bibitem{Chua:2020stf}
A.~J.~K. Chua, M.~L. Katz, N.~Warburton, and S.~A. Hughes, ``{Rapid generation of fully relativistic extreme-mass-ratio-inspiral waveform templates for LISA data analysis},'' \emph{Phys. Rev. Lett.}, vol. 126, no.~5, p. 051102, 2021.

\bibitem{Katz:2021yft}
M.~L. Katz, A.~J.~K. Chua, L.~Speri, N.~Warburton, and S.~A. Hughes, ``{Fast extreme-mass-ratio-inspiral waveforms: New tools for millihertz gravitational-wave data analysis},'' \emph{Phys. Rev. D}, vol. 104, no.~6, p. 064047, 2021.

\bibitem{Harte:2014wya}
A.~I. Harte, ``{Motion in classical field theories and the foundations of the self-force problem},'' \emph{Fund. Theor. Phys.}, vol. 179, pp. 327--398, 2015.

\bibitem{Tanaka:1996ht}
T.~Tanaka, Y.~Mino, M.~Sasaki, and M.~Shibata, ``{Gravitational waves from a spinning particle in circular orbits around a rotating black hole},'' \emph{Phys. Rev. D}, vol.~54, pp. 3762--3777, 1996.

\bibitem{Dixon:1964NCim}
W.~Dixon, ``{A covariant multipole formalism for extended test bodies in general relativity},'' \emph{Il Nuovo Cimento}, vol.~34, no.~2, pp. 317--339, Oct 1964.

\bibitem{Dixon:1970I}
W.~G. Dixon, ``{Dynamics of extended bodies in general relativity. I. Momentum and angular momentum},'' \emph{Proc. Roy. Soc. Lond.}, vol. A314, pp. 499--527, 1970.

\bibitem{Dixon:1970II}
------, ``{Dynamics of extended bodies in general relativity. II. Moments of the charge-current vector},'' \emph{Proc. Roy. Soc. Lond.}, vol. A319, pp. 509--547, 1970.

\bibitem{Piovano:2020zin}
G.~A. Piovano, A.~Maselli, and P.~Pani, ``{Extreme mass ratio inspirals with spinning secondary: a detailed study of equatorial circular motion},'' \emph{Phys. Rev. D}, vol. 102, no.~2, p. 024041, 2020.

\bibitem{Rudiger:1981a}
\BIBentryALTinterwordspacing
R.~R{\"u}diger, ``Conserved quantities of spinning test particles in general relativity. i,'' \emph{Proceedings of the Royal Society of London. Series A, Mathematical and Physical Sciences}, vol. 375, no. 1761, pp. 185--193, 1981. [Online]. Available: \url{http://www.jstor.org/stable/2990231}
\BIBentrySTDinterwordspacing

\bibitem{Rudiger:1981b}
\BIBentryALTinterwordspacing
------, ``Conserved quantities of spinning test particles in general relativity. ii,'' \emph{Proceedings of the Royal Society of London. Series A, Mathematical and Physical Sciences}, vol. 385, no. 1788, pp. 229--239, 1983. [Online]. Available: \url{http://www.jstor.org/stable/2397480}
\BIBentrySTDinterwordspacing

\bibitem{Witzany:2023bmq}
V.~Witzany and G.~A. Piovano, ``{Analytic Solutions for the Motion of Spinning Particles near Spherically Symmetric Black Holes and Exotic Compact Objects},'' \emph{Phys. Rev. Lett.}, vol. 132, no.~17, p. 171401, 2024.

\bibitem{Jefremov:2015gza}
P.~I. Jefremov, O.~{\relax Yu}. Tsupko, and G.~S. Bisnovatyi-Kogan, ``{Innermost stable circular orbits of spinning test particles in Schwarzschild and Kerr space-times},'' \emph{Phys. Rev.}, vol. D91, no.~12, p. 124030, 2015.

\bibitem{LeTiec:2011dp}
A.~Le~Tiec, E.~Barausse, and A.~Buonanno, ``{Gravitational Self-Force Correction to the Binding Energy of Compact Binary Systems},'' \emph{Phys. Rev. Lett.}, vol. 108, p. 131103, 2012.

\bibitem{Martel:2003jj}
K.~Martel, ``{Gravitational wave forms from a point particle orbiting a Schwarzschild black hole},'' \emph{Phys. Rev. D}, vol.~69, p. 044025, 2004.

\bibitem{Hughes:2018qxz}
S.~A. Hughes, ``{Bound orbits of a slowly evolving black hole},'' \emph{Phys. Rev. D}, vol. 100, no.~6, p. 064001, 2019.

\bibitem{Blanchet:2023sbv}
L.~Blanchet, G.~Faye, Q.~Henry, F.~Larrouturou, and D.~Trestini, ``{Gravitational-wave flux and quadrupole modes from quasicircular nonspinning compact binaries to the fourth post-Newtonian order},'' \emph{Phys. Rev. D}, vol. 108, no.~6, p. 064041, 2023.

\bibitem{Fujita:2012cm}
R.~Fujita, ``{Gravitational Waves from a Particle in Circular Orbits around a Schwarzschild Black Hole to the 22nd Post-Newtonian Order},'' \emph{Prog. Theor. Phys.}, vol. 128, pp. 971--992, 2012.

\bibitem{ori2000transition}
A.~Ori and K.~S. Thorne, ``Transition from inspiral to plunge for a compact body in a circular equatorial orbit around a massive, spinning black hole,'' \emph{Physical Review D}, vol.~62, no.~12, p. 124022, 2000.

\bibitem{o2003transition}
R.~O’Shaughnessy, ``Transition from inspiral to plunge for eccentric equatorial kerr orbits,'' \emph{Physical Review D}, vol.~67, no.~4, p. 044004, 2003.

\bibitem{kesden2011transition}
M.~Kesden, ``Transition from adiabatic inspiral to plunge into a spinning black hole,'' \emph{Physical Review D}, vol.~83, no.~10, p. 104011, 2011.

\bibitem{apte2019exciting}
A.~Apte and S.~A. Hughes, ``Exciting black hole modes via misaligned coalescences. i. inspiral, transition, and plunge trajectories using a generalized ori-thorne procedure,'' \emph{Physical Review D}, vol. 100, no.~8, p. 084031, 2019.

\bibitem{burke2020transition}
O.~Burke, J.~Gair, and J.~Sim{\'o}n, ``Transition from inspiral to plunge: A complete near-extremal trajectory and associated waveform,'' \emph{Physical Review D}, vol. 101, no.~6, p. 064026, 2020.

\bibitem{compere2020transition}
G.~Comp{\`e}re, K.~Fransen, and C.~Jonas, ``Transition from inspiral to plunge into a highly spinning black hole,'' \emph{Classical and Quantum Gravity}, vol.~37, no.~9, p. 095013, 2020.

\bibitem{compere2021self}
G.~Comp{\`e}re and L.~K{\"u}chler, ``Self-consistent adiabatic inspiral and transition motion,'' \emph{Physical review letters}, vol. 126, no.~24, p. 241106, 2021.

\bibitem{Cutler:1994pb}
C.~Cutler, D.~Kennefick, and E.~Poisson, ``{Gravitational radiation reaction for bound motion around a Schwarzschild black hole},'' \emph{Phys. Rev. D}, vol.~50, pp. 3816--3835, 1994.

\bibitem{Fujita:2009bp}
R.~Fujita and W.~Hikida, ``{Analytical solutions of bound timelike geodesic orbits in Kerr spacetime},'' \emph{Class. Quant. Grav.}, vol.~26, p. 135002, 2009.

\bibitem{KerrGeodesicsPackage}
\BIBentryALTinterwordspacing
N.~Warburton, B.~Wardell, O.~Long, S.~Upton, P.~Lynch, Z.~Nasipak, and L.~C. Stein, ``Kerrgeodesics,'' Jul. 2023. [Online]. Available: \url{https://doi.org/10.5281/zenodo.8108265}
\BIBentrySTDinterwordspacing

\bibitem{Kennefick:1995za}
D.~Kennefick and A.~Ori, ``{Radiation reaction induced evolution of circular orbits of particles around Kerr black holes},'' \emph{Phys. Rev. D}, vol.~53, pp. 4319--4326, 1996.

\bibitem{Munna:2023vds}
C.~Munna, C.~R. Evans, and E.~Forseth, ``{Tidal heating and torquing of the primary black hole in eccentric-orbit, nonspinning, extreme-mass-ratio inspirals to 22PN order},'' \emph{Phys. Rev. D}, vol. 108, no.~4, p. 044039, 2023.

\bibitem{PostNewtonianSelfForcePackage}
\BIBentryALTinterwordspacing
N.~Warburton, B.~Wardell, C.~Munna, and C.~Kavanagh, ``Postnewtonianselfforce,'' Jul. 2023. [Online]. Available: \url{https://doi.org/10.5281/zenodo.8112975}
\BIBentrySTDinterwordspacing

\bibitem{Goldberg:1967}
J.~N. {Goldberg}, A.~J. {Macfarlane}, E.~T. {Newman}, F.~{Rohrlich}, and E.~C.~G. {Sudarshan}, ``{Spin-s Spherical Harmonics and {\dh}},'' \emph{Journal of Mathematical Physics}, vol.~8, no.~11, pp. 2155--2161, Nov. 1967.

\bibitem{Barack:2003fp}
L.~Barack and C.~Cutler, ``{LISA capture sources: Approximate waveforms, signal-to-noise ratios, and parameter estimation accuracy},'' \emph{Phys. Rev. D}, vol.~69, p. 082005, 2004.

\bibitem{armstrong1999time}
J.~Armstrong, F.~Estabrook, and M.~Tinto, ``Time-delay interferometry for space-based gravitational wave searches,'' \emph{The Astrophysical Journal}, vol. 527, no.~2, p. 814, 1999.

\bibitem{Tinto:2004wu}
M.~Tinto and S.~V. Dhurandhar, ``{TIME DELAY},'' \emph{Living Rev. Rel.}, vol.~8, p.~4, 2005.

\bibitem{vallisneri2005synthetic}
M.~Vallisneri, ``Synthetic lisa: Simulating time delay interferometry in a model lisa,'' \emph{Physical Review D}, vol.~71, no.~2, p. 022001, 2005.

\bibitem{vallisneri2012non}
M.~Vallisneri and C.~R. Galley, ``Non-sky-averaged sensitivity curves for space-based gravitational-wave observatories,'' \emph{Classical and Quantum Gravity}, vol.~29, no.~12, p. 124015, 2012.

\bibitem{talbot2021inference}
C.~Talbot, E.~Thrane, S.~Biscoveanu, and R.~Smith, ``Inference with finite time series: Observing the gravitational universe through windows,'' \emph{Physical Review Research}, vol.~3, no.~4, p. 043049, 2021.

\bibitem{edy2021issues}
O.~Edy, A.~Lundgren, and L.~K. Nuttall, ``Issues of mismodeling gravitational-wave data for parameter estimation,'' \emph{Physical Review D}, vol. 103, no.~12, p. 124061, 2021.

\bibitem{wiener1930generalized}
N.~Wiener \emph{et~al.}, ``Generalized harmonic analysis,'' \emph{Acta mathematica}, vol.~55, pp. 117--258, 1930.

\bibitem{khintchine1934korrelationstheorie}
A.~Khintchine, ``Korrelationstheorie der station{\"a}ren stochastischen prozesse,'' \emph{Mathematische Annalen}, vol. 109, no.~1, pp. 604--615, 1934.

\bibitem{LISAScienceRequirementsDocument}
``{LISA Science Study Team, “LISA Science Requirements Document,”},'' 2018.

\bibitem{KatzTools}
\BIBentryALTinterwordspacing
{Michael Katz}, ``Lisaanalysistools,'' AEI, Tech. Rep., May 2020. [Online]. Available: \url{https://github.com/mikekatz04/LISAanalysistools}
\BIBentrySTDinterwordspacing

\bibitem{whittle:1957}
P.~Whittle, ``Curve and periodogram smoothing,'' \emph{Journal of the Royal Statistical Society: Series B (Statistical Methodology)}, vol.~19, pp. 38--63, 1957.

\bibitem{finn1992detection}
L.~S. Finn, ``Detection, measurement, and gravitational radiation,'' \emph{Physical Review D}, vol.~46, no.~12, p. 5236, 1992.

\bibitem{woodward2014probability}
P.~M. Woodward, \emph{Probability and information theory, with applications to radar: international series of monographs on electronics and instrumentation}.\hskip 1em plus 0.5em minus 0.4em\relax Elsevier, 2014, vol.~3.

\bibitem{babak2009algorithm}
S.~Babak, J.~R. Gair, and E.~K. Porter, ``An algorithm for the detection of extreme mass ratio inspirals in lisa data,'' \emph{Classical and quantum gravity}, vol.~26, no.~13, p. 135004, 2009.

\bibitem{Karnesis:2023ras}
N.~Karnesis, M.~L. Katz, N.~Korsakova, J.~R. Gair, and N.~Stergioulas, ``Eryn: A multi-purpose sampler for bayesian inference,'' \emph{arXiv preprint arXiv:2303.02164}, 2023.

\bibitem{michael_katz_2023_7705496}
M.~Katz, N.~Karnesis, and N.~Korsakova, ``mikekatz04/eryn: first full release,'' Mar. 2023, available at https://doi.org/10.5281/zenodo.7705496.

\bibitem{Foreman-Mackey:2013}
D.~{Foreman-Mackey}, D.~W. {Hogg}, D.~{Lang}, and J.~{Goodman}, ``{emcee: The MCMC Hammer},'' \emph{Astronomical Society of the Pacific}, vol. 125, no. 925, p. 306, Mar. 2013.

\bibitem{goodman2010ensemble}
J.~Goodman and J.~Weare, ``Ensemble samplers with affine invariance,'' \emph{Communications in applied mathematics and computational science}, vol.~5, no.~1, pp. 65--80, 2010.

\bibitem{Cornish_2017}
\BIBentryALTinterwordspacing
N.~Cornish and T.~Robson, ``Galactic binary science with the new lisa design,'' \emph{Journal of Physics: Conference Series}, vol. 840, no.~1, p. 012024, may 2017. [Online]. Available: \url{https://dx.doi.org/10.1088/1742-6596/840/1/012024}
\BIBentrySTDinterwordspacing

\bibitem{moore2014gravitational}
C.~J. Moore, R.~H. Cole, and C.~P. Berry, ``Gravitational-wave sensitivity curves,'' \emph{Classical and Quantum Gravity}, vol.~32, no.~1, p. 015014, 2014.

\bibitem{babak2021lisa}
S.~Babak, M.~Hewitson, and A.~Petiteau, ``Lisa sensitivity and snr calculations,'' \emph{arXiv preprint arXiv:2108.01167}, 2021.

\bibitem{robson2019construction}
T.~Robson, N.~J. Cornish, and C.~Liu, ``The construction and use of lisa sensitivity curves,'' \emph{Classical and Quantum Gravity}, vol.~36, no.~10, p. 105011, 2019.

\bibitem{bayle2019simulation}
J.-B. Bayle, ``Simulation and data analysis for lisa (instrumental modeling, time-delay interferometry, noise-reduction performance study, and discrimination of transient gravitational signals),'' Ph.D. dissertation, Universit{\'e} de Paris; Universit{\'e} Paris Diderot; Laboratoire Astroparticules~…, 2019.

\bibitem{Piovano:2020ooe}
G.~A. Piovano, A.~Maselli, and P.~Pani, ``{Model independent tests of the Kerr bound with extreme mass ratio inspirals},'' \emph{Phys. Lett. B}, vol. 811, p. 135860, 2020.

\bibitem{Burke:20aa}
O.~{Burke}, J.~R. {Gair}, J.~{Sim{\'o}n}, and M.~C. {Edwards}, ``{Constraining the spin parameter of near-extremal black holes using LISA},'' \emph{prd}, vol. 102, no.~12, p. 124054, Dec. 2020.

\bibitem{10.1093/mnras/stad1397}
\BIBentryALTinterwordspacing
C.~E.~A. Chapman-Bird, C.~P.~L. Berry, and G.~Woan, ``{Rapid determination of LISA sensitivity to extreme mass ratio inspirals with machine learning},'' \emph{Monthly Notices of the Royal Astronomical Society}, vol. 522, no.~4, pp. 6043--6054, 05 2023. [Online]. Available: \url{https://doi.org/10.1093/mnras/stad1397}
\BIBentrySTDinterwordspacing

\bibitem{Babak:2017tow}
S.~Babak, J.~Gair, A.~Sesana, E.~Barausse, C.~F. Sopuerta, C.~P.~L. Berry, E.~Berti, P.~Amaro-Seoane, A.~Petiteau, and A.~Klein, ``{Science with the space-based interferometer LISA. V: Extreme mass-ratio inspirals},'' \emph{Phys. Rev. D}, vol.~95, no.~10, p. 103012, 2017.

\bibitem{Canizares:2012is}
P.~Canizares, J.~R. Gair, and C.~F. Sopuerta, ``{Testing Chern-Simons Modified Gravity with Gravitational-Wave Detections of Extreme-Mass-Ratio Binaries},'' \emph{Phys. Rev. D}, vol.~86, p. 044010, 2012.

\bibitem{Pani:2011xj}
P.~Pani, V.~Cardoso, and L.~Gualtieri, ``{Gravitational waves from extreme mass-ratio inspirals in Dynamical Chern-Simons gravity},'' \emph{Phys. Rev. D}, vol.~83, p. 104048, 2011.

\bibitem{Yunes:2011aa}
N.~Yunes, P.~Pani, and V.~Cardoso, ``{Gravitational Waves from Quasicircular Extreme Mass-Ratio Inspirals as Probes of Scalar-Tensor Theories},'' \emph{Phys. Rev. D}, vol.~85, p. 102003, 2012.

\bibitem{Blazquez-Salcedo:2016enn}
J.~L. Bl\'azquez-Salcedo, C.~F.~B. Macedo, V.~Cardoso, V.~Ferrari, L.~Gualtieri, F.~S. Khoo, J.~Kunz, and P.~Pani, ``{Perturbed black holes in Einstein-dilaton-Gauss-Bonnet gravity: Stability, ringdown, and gravitational-wave emission},'' \emph{Phys. Rev. D}, vol.~94, no.~10, p. 104024, 2016.

\bibitem{Maggio:2021uge}
E.~Maggio, M.~van~de Meent, and P.~Pani, ``{Extreme mass-ratio inspirals around a spinning horizonless compact object},'' \emph{Phys. Rev. D}, vol. 104, no.~10, p. 104026, 2021.

\bibitem{Barack:2006pq}
L.~Barack and C.~Cutler, ``{Using LISA EMRI sources to test off-Kerr deviations in the geometry of massive black holes},'' \emph{Phys. Rev. D}, vol.~75, p. 042003, 2007.

\bibitem{Datta:2019epe}
S.~Datta, R.~Brito, S.~Bose, P.~Pani, and S.~A. Hughes, ``{Tidal heating as a discriminator for horizons in extreme mass ratio inspirals},'' \emph{Phys. Rev. D}, vol. 101, no.~4, p. 044004, 2020.

\bibitem{Fell:2023mtf}
S.~D.~B. Fell, L.~Heisenberg, and D.~Veske, ``{Detecting Fundamental Vector Fields with LISA},'' 4 2023.

\bibitem{Barsanti:2022vvl}
S.~Barsanti, A.~Maselli, T.~P. Sotiriou, and L.~Gualtieri, ``{Detecting Massive Scalar Fields with Extreme Mass-Ratio Inspirals},'' \emph{Phys. Rev. Lett.}, vol. 131, no.~5, p. 051401, 2023.

\bibitem{purrer2020gravitational}
M.~P{\"u}rrer and C.-J. Haster, ``Gravitational waveform accuracy requirements for future ground-based detectors,'' \emph{Physical Review Research}, vol.~2, no.~2, p. 023151, 2020.

\bibitem{Hughes:2016xwf}
S.~A. Hughes, ``{Adiabatic and post-adiabatic approaches to extreme mass ratio inspiral},'' in \emph{{14th Marcel Grossmann Meeting on Recent Developments in Theoretical and Experimental General Relativity, Astrophysics, and Relativistic Field Theories}}, vol.~2, 2017, pp. 1953--1959.

\bibitem{laghi2021gravitational}
D.~Laghi, N.~Tamanini, W.~Del~Pozzo, A.~Sesana, J.~Gair, S.~Babak, and D.~Izquierdo-Villalba, ``Gravitational-wave cosmology with extreme mass-ratio inspirals,'' \emph{Monthly Notices of the Royal Astronomical Society}, vol. 508, no.~3, pp. 4512--4531, 2021.

\bibitem{gair2008constrained}
J.~R. Gair, E.~Porter, S.~Babak, and L.~Barack, ``A constrained metropolis--hastings search for emris in the mock lisa data challenge 1b,'' \emph{Classical and Quantum Gravity}, vol.~25, no.~18, p. 184030, 2008.

\bibitem{cornish2011detection}
N.~J. Cornish, ``Detection strategies for extreme mass ratio inspirals,'' \emph{Classical and Quantum Gravity}, vol.~28, no.~9, p. 094016, 2011.

\bibitem{gair2008improved}
J.~R. Gair, I.~Mandel, and L.~Wen, ``Improved time--frequency analysis of extreme-mass-ratio inspiral signals in mock lisa data,'' \emph{Classical and Quantum Gravity}, vol.~25, no.~18, p. 184031, 2008.

\bibitem{babak2010mock}
S.~Babak, J.~G. Baker, M.~J. Benacquista, N.~J. Cornish, S.~L. Larson, I.~Mandel, S.~T. McWilliams, A.~Petiteau, E.~K. Porter, E.~L. Robinson \emph{et~al.}, ``The mock lisa data challenges: from challenge 3 to challenge 4,'' \emph{Classical and Quantum Gravity}, vol.~27, no.~8, p. 084009, 2010.

\bibitem{harris2020array}
\BIBentryALTinterwordspacing
C.~R. Harris, K.~J. Millman, S.~J. van~der Walt, R.~Gommers, P.~Virtanen, D.~Cournapeau, E.~Wieser, J.~Taylor, S.~Berg, N.~J. Smith, R.~Kern, M.~Picus, S.~Hoyer, M.~H. van Kerkwijk, M.~Brett, A.~Haldane, J.~F. del R{\'{i}}o, M.~Wiebe, P.~Peterson, P.~G{\'{e}}rard-Marchant, K.~Sheppard, T.~Reddy, W.~Weckesser, H.~Abbasi, C.~Gohlke, and T.~E. Oliphant, ``Array programming with {NumPy},'' \emph{Nature}, vol. 585, no. 7825, pp. 357--362, Sep. 2020. [Online]. Available: \url{https://doi.org/10.1038/s41586-020-2649-2}
\BIBentrySTDinterwordspacing

\bibitem{Hunter:2007}
J.~D. Hunter, ``Matplotlib: A 2d graphics environment,'' \emph{Computing in Science \& Engineering}, vol.~9, no.~3, pp. 90--95, 2007.

\bibitem{corner}
\BIBentryALTinterwordspacing
D.~Foreman-Mackey, ``corner.py: Scatterplot matrices in python,'' \emph{The Journal of Open Source Software}, vol.~1, no.~2, p.~24, jun 2016. [Online]. Available: \url{https://doi.org/10.21105/joss.00024}
\BIBentrySTDinterwordspacing

\bibitem{Hinton2016}
S.~R. {Hinton}, ``{ChainConsumer},'' \emph{The Journal of Open Source Software}, vol.~1, p. 00045, Aug. 2016.

\bibitem{astropy:2022}
{Astropy Collaboration}, A.~M. {Price-Whelan}, P.~L. {Lim}, N.~{Earl}, N.~{Starkman}, L.~{Bradley}, D.~L. {Shupe}, A.~A. {Patil}, L.~{Corrales}, C.~E. {Brasseur}, M.~{N{"o}the}, A.~{Donath}, E.~{Tollerud}, B.~M. {Morris}, A.~{Ginsburg}, E.~{Vaher}, B.~A. {Weaver}, J.~{Tocknell}, W.~{Jamieson}, M.~H. {van Kerkwijk}, T.~P. {Robitaille}, B.~{Merry}, M.~{Bachetti}, H.~M. {G{"u}nther}, T.~L. {Aldcroft}, J.~A. {Alvarado-Montes}, A.~M. {Archibald}, A.~{B{'o}di}, S.~{Bapat}, G.~{Barentsen}, J.~{Baz{'a}n}, M.~{Biswas}, M.~{Boquien}, D.~J. {Burke}, D.~{Cara}, M.~{Cara}, K.~E. {Conroy}, S.~{Conseil}, M.~W. {Craig}, R.~M. {Cross}, K.~L. {Cruz}, F.~{D'Eugenio}, N.~{Dencheva}, H.~A.~R. {Devillepoix}, J.~P. {Dietrich}, A.~D. {Eigenbrot}, T.~{Erben}, L.~{Ferreira}, D.~{Foreman-Mackey}, R.~{Fox}, N.~{Freij}, S.~{Garg}, R.~{Geda}, L.~{Glattly}, Y.~{Gondhalekar}, K.~D. {Gordon}, D.~{Grant}, P.~{Greenfield}, A.~M. {Groener}, S.~{Guest}, S.~{Gurovich}, R.~{Handberg}, A.~{Hart}, Z.~{Hatfield-Dodds}, D.~{Homeier},
  G.~{Hosseinzadeh}, T.~{Jenness}, C.~K. {Jones}, P.~{Joseph}, J.~B. {Kalmbach}, E.~{Karamehmetoglu}, M.~{Ka{l}uszy{'n}ski}, M.~S.~P. {Kelley}, N.~{Kern}, W.~E. {Kerzendorf}, E.~W. {Koch}, S.~{Kulumani}, A.~{Lee}, C.~{Ly}, Z.~{Ma}, C.~{MacBride}, J.~M. {Maljaars}, D.~{Muna}, N.~A. {Murphy}, H.~{Norman}, R.~{O'Steen}, K.~A. {Oman}, C.~{Pacifici}, S.~{Pascual}, J.~{Pascual-Granado}, R.~R. {Patil}, G.~I. {Perren}, T.~E. {Pickering}, T.~{Rastogi}, B.~R. {Roulston}, D.~F. {Ryan}, E.~S. {Rykoff}, J.~{Sabater}, P.~{Sakurikar}, J.~{Salgado}, A.~{Sanghi}, N.~{Saunders}, V.~{Savchenko}, L.~{Schwardt}, M.~{Seifert-Eckert}, A.~Y. {Shih}, A.~S. {Jain}, G.~{Shukla}, J.~{Sick}, C.~{Simpson}, S.~{Singanamalla}, L.~P. {Singer}, J.~{Singhal}, M.~{Sinha}, B.~M. {Sip{H{o}}cz}, L.~R. {Spitler}, D.~{Stansby}, O.~{Streicher}, J.~{Sumak}, J.~D. {Swinbank}, D.~S. {Taranu}, N.~{Tewary}, G.~R. {Tremblay}, M.~d. {Val-Borro}, S.~J. {Van Kooten}, Z.~{Vasovi{'c}}, S.~{Verma}, J.~V. {de Miranda Cardoso}, P.~K.~G. {Williams}, T.~J. {Wilson},
  B.~{Winkel}, W.~M. {Wood-Vasey}, R.~{Xue}, P.~{Yoachim}, C.~{Zhang}, A.~{Zonca}, and {Astropy Project Contributors}, ``{The Astropy Project: Sustaining and Growing a Community-oriented Open-source Project and the Latest Major Release (v5.0) of the Core Package},'' \emph{apj}, vol. 935, no.~2, p. 167, Aug. 2022.

\bibitem{cupy_learningsys2017}
\BIBentryALTinterwordspacing
R.~Okuta, Y.~Unno, D.~Nishino, S.~Hido, and C.~Loomis, ``Cupy: A numpy-compatible library for nvidia gpu calculations,'' in \emph{Proceedings of Workshop on Machine Learning Systems (LearningSys) in The Thirty-first Annual Conference on Neural Information Processing Systems (NIPS)}, 2017. [Online]. Available: \url{http://learningsys.org/nips17/assets/papers/paper_16.pdf}
\BIBentrySTDinterwordspacing

\bibitem{michael_l_katz_2023_8190418}
\BIBentryALTinterwordspacing
M.~L. Katz, L.~Speri, A.~J.~K. Chua, C.~E.~A. Chapman-Bird, N.~Warburton, and S.~A. Hughes, ``{BlackHolePerturbationToolkit/FastEMRIWaveforms: Frequency Domain Waveform Added!}'' Jul. 2023. [Online]. Available: \url{https://doi.org/10.5281/zenodo.8190418}
\BIBentrySTDinterwordspacing

\bibitem{cowles1996markov}
M.~K. Cowles and B.~P. Carlin, ``Markov chain monte carlo convergence diagnostics: a comparative review,'' \emph{Journal of the American Statistical Association}, vol.~91, no. 434, pp. 883--904, 1996.

\end{thebibliography}
\newpage\hbox{}\thispagestyle{empty}{\newpage}

\end{document}